\documentclass[a4paper]{article_saj}
\usepackage{graphicx,saj,multicol,subeqnarray}
\usepackage{cuted}
\usepackage{natbib}
\usepackage{float}
\usepackage{xcolor}
\usepackage{widetext}
\usepackage{url}
\usepackage{bm}
\usepackage{tikz} %
\usepackage{pifont} %
\usepackage{amsfonts}
\usepackage{amssymb}
\usepackage{amsmath,upgreek}
\usepackage{titlesec}

\def\point#1{\hbox{\setbox7=\hbox to0.6em{\hfil.\hfil}%
\setbox8=\hbox to0.5em{\hfil$^{#1}$\hfil}%
\box7\kern-0.5em\box8}}

\def\pointmin#1{\hbox{\setbox2=\hbox to0.8em{\hfil.\hfil}%
\setbox3=\hbox to0.6em{\hfil$^{#1}$\hfil}%
\box2\kern-.7em\box3}}

\def\mmm{\pointmin{\mathrm{m}}\kern.15em}

\titlelabel{\thetitle.\quad}
\definecolor{xlinkcolor}{cmyk}{1,0.6,0,0}
\usepackage{amsmath}
\usepackage[bookmarks=false,         %
     pdfnewwindow=true,      %
     colorlinks=true,    %
     linkcolor=xlinkcolor,     %
     citecolor=xlinkcolor,     %
     filecolor=xlinkcolor,  %
     urlcolor=xlinkcolor,      %
final=true
]{hyperref}
\usepackage{xspace}
\usepackage[nolist]{acronym}
\usepackage{multicol}
\usepackage{supertabular}
\usepackage{multirow}
\usepackage{comment}
\makeatletter
\@ifundefined{@setmarks}{\let\@setmarks\relax}{}
\makeatother

\newcommand{\ergcm}[1]{erg~cm$^{-2}$ s$^{-1}$}

\newcommand{\HI}{H{\sc i}}

\newcommand{\SII}{[S{\sc ii}]}
\newcommand{\OIII}{[O{\sc iii}]}
\newcommand{\Halpha}{H${\alpha}$}

\newcommand{\D}{$^\circ$}

\newcommand{\snr}{Veliki}
\newcommand{\snrold}{J0450.4$-$7050}

\def\arcmin{\hbox{$^\prime$}}
\def\arcsec{\hbox{$^{\prime\prime}$}}
\def\kms{km\,s$^{-1}$}

\newcommand{\farcm}{\mbox{\ensuremath{.\mkern-4mu^\prime}}}%
\newcommand{\fdg}{\mbox{\ensuremath{.\!\!^\circ}}}%

\begin{acronym}[AWGN]
\acro{2MASS}{Two Micron All Sky Survey}
\acro{2MASX}{Two Micron All Sky Survey Extended Source Catalogue}
\acro{30Dor}[30 Dor]{30~Doradus}
\acro{AeReS}{Aegean Residual}
\acro{AGN}{active galactic nuclei}
\acro{ARC}{Australian Research Council}
\acro{ATCA}{Australia Telescope Compact Array}
\acro{ATESP}{Australia Telescope ESO Slice Project}
\acro{ATNF}{Australia Telescope National Facility}
\acro{ATOA}{Australia Telescope Online Archive}
\acro{AT20G}{Australia Telescope 20 GHz Survey}
\acro{ASKAP}{Australian Square Kilometre Array Pathfinder}
\acro{BETA}{Boolardy Engineering Test Array}
\acro{BGG}{Brightest Galaxy Group}
\acro{BL}[BL Lac]{BL Lacertae Objects}
\acro{CASS}{CSIRO Astronomy and Space Science}
\acro{CABB}{Compact Array Broadband Back-end}
\acro{CC}{core-collapse}
\acro{CHII}[{\sc CHii}]{Compact \textsc{Hii}}
\acro{CIE}{collisional ionisation equilibrium}
\acro{CMB}{Cosmic Microwave Background}
\acro{CSIRO}{Australian Commonwealth Scientific and Industrial Research Organisation}
\acro{CSS}{Compact Steep Spectrum}
\acro{DESI}{Dark Energy Spectroscopic Instrument}
\acro{DS9}[\textsc{DS9}]{\textsc{SAOImage DS9}}
\acro{DSS}{Digital Sky Survey}
\acro{DSA}{diffuse shock acceleration}
\acro{EBHIS}{Effelsberg--Bonn H\,{\sc i} Survey}
\acro{EM}{Electromagnetic}
\acro{EMU}{Evolutionary Map of the Universe}
\acro{ev}[eV]{electronvolt\acroextra{: 1 eV $\approx 1.6 \times 10^{-19}$ J}}
\acro{eROSITA}{extended R\"Ontgen Survey with an Imaging Telescope Array}
\acro{Fermi-LAT}{Fermi Large Area Telescope}
\acro{FIR}{Far Infrared}
\acro{FITS}[\textsc{Fits}]{Flexible Image Transport System}
\acro{FRB}{Fast Radio Bursts}
\acro{FSRQ}{Flat Spectrum Radio Quasars}
\acro{FWHM}{Full Width at Half-Maximum}
\acro{GASS}{Galactic All-Sky Survey}
\acro{GLEAM}{GaLactic and Extragalactic All-sky MWA}
\acro{GPS}{Gigahertz Peak Spectrum}
\acro{HEMT}{High Electron Mobility Transistor}
\acro{HERITAGE}{HERschel Inventory of The Agents of Galaxy Evolution}
\acro{H.E.S.S.}{High Energy Stereoscopic System}
\acro{HFP}[HFP]{High-Frequency Peaker}
\acro{HPBW}{Half Power Beam Width} 
\acro{HzRGs}{High Redshift Radio Galaxies}
\acro{pHFP}[pHFP]{Potential High Frequency Peaker}
\acro{HI}[H{\sc i}]{Neutral Atomic Hydrogen} 
\acro{HST}{\textit{Hubble Space Telescope}}
\acro{ICM}{intracluster medium}
\acro{IAU}{International Astronomical Union}
\acro{IFRSs}{Infrared Faint Radio Sources}
\acro{ISM}{interstellar medium}
\acro{IFRS}{Infrared Faint Radio Source}
\acro{IR}{Infrared}
\acro{JY}[Jy]{Jansky\acroextra{, 1 Jy = $10^{-26} \times \mathrm{W~ m}^{-2}~\mathrm{Hz}^{-1}$}} 
\acro{LLS}{Largest Linear Size}
\acro{LFAA}{Low-Frequency Aperture Array}
\acro{LMC}{Large Magellanic Cloud}
\acro{LSO}[LSOs]{Large Scale Objects}
\acro{MACHO}{Massive Astrophysical Compact Halo Objects}
\acro{MAGMA}{Magellanic Mopra Assessment}
\acro{MC}[MC]{Magellanic Cloud}
\acro{MS}[MS]{Magellanic Stream} 
\acro{MB}[MB]{Magellanic Bridge} 
\acro{MCELS}{Magellanic Cloud Emission Line Survey}
\acro{MW}{Milky Way}
\acro{MIPS}{Multiband Imaging Photometer}
\acro{MIRIAD}[\textsc{Miriad}]{Multichannel Image Reconstruction, Image Analysis and Display}
\acro{MIT}{Massachusetts Institute of Technology}
\acro{MM}{mixed-morphology}
\acro{MOST}{Molonglo Observatory Synthesis Telescope}
\acro{MQS}{Magellanic Quasars Survey}
\acro{MRC}{Molonglo Reference Catalogue of Radio Sources}
\acro{MWA}{Murchison Widefield Array}
\acro{NED}{NASA/IPAC Extragalactic Database}
\acro{NRAO}{National Radio Astronomy Observatory}
\acro{NVSS}{NRAO VLA Sky Survey}
\acro{OPAL}{Online Proposal Applications \& Links}
\acrodefplural{ORC}{Odd Radio Circles}
\acro{ORC}{Odd Radio Circle}
\acro{OVV}{Optically Violent Variable Quasars}
\acro{PACS}{Photodetector Array Camera and Spectrometer}
\acro{PAF}{Phased Array Feed}
\acro{pc}{parsec\acroextra{: 1 pc $\simeq 3.09 \times 10^{16}$ m}}
\acro{PMN}{Parkes-MIT-NRAO}
\acro{PNe}{Planetary Nebulae}
\acro{PWN}{Pulsar Wind Nebulae}
\acro{QSO}{Quasi-Stellar Object}
\acro{RA}{Right Ascension}
\acro{RFI}{Radio-Frequency Interference}
\acro{RMS}[rms]{Root Mean Squared}
\acro{SAGE}{Surveying the Agents of a Galaxy’s Evolution}
\acro{SARAO}{South African Radio Astronomy Observatory}
\acro{SDSS}{Sloan Digital Sky Survey}
\acro{SED}{spectral energy distribution}
\acro{SI}[$\alpha$]{Spectral Index\acroextra{, $S \propto \nu^\alpha$}}
\acro{SHS}{SuperCOSMOS H$\alpha$ Survey}
\acro{SKA}{Square Kilometre Array}
\acro{SMB}{Super Massive Blackholes}
\acro{SMC}{Small Magellanic Cloud}
\acro{SN}{Supernova}
\acro{SPIRE}{Spectral and Photometric Imaging Receiver}
\acro{SUMSS}{Sydney University Molonglo Sky Survey}
\acro{TOPCAT}[\textsc{Topcat}]{Tool for OPerations on Catalogues And Tables}
\acro{USS}{Ultra Steep Spectrum}
\acro{YSOs}{young stellar objects}
\acro{WBAC}{Wide-Band Analogue Correlator}
\acro{WIFES}[WiFeS]{Wide-Field Spectrograph}
\acro{WISE}{Wide-Field Infrared Survey Explorer}
\acro{VLBI}{Very Long Baseline Interferometry} 
\acro{VLSR}[\textbf{$v_{lsr}$}]{Velocity in the Line of Sight}
\acro{SMASH}{Survey of the MAgellanic Stellar History}
\acro{SNR}{Supernova Remnant}
\acrodefplural{SNR}{Supernova Remnants}
\acro{SD}{single-degenerate}
\acro{SUMSS}{Sydney University Molonglo Sky Survey}
\acro{SMBH}{Super Massive Black Hole}
\acro{CARTA}{Cube Analysis and Rendering Tool for Astronomy}
\acro{NEI}{non-equilibrium ionisation}
\acro{UV}{ultra-violet}
\acro{FUSE}{Far Ultraviolet Spectroscopic Explorer}
\acro{RM}{rotation measure}
\acro{H.E.S.S.}{High Energy Spectroscopic System}
\acro{CTA}{Cherenkov Telescope Array}
\end{acronym}

\def\papertype{\ \hfill\ Editorial}

\def\udc{52}
\begin{document}
\parindent=.5cm
\baselineskip=3.8truemm
\columnsep=.5truecm
\newenvironment{lefteqnarray}{\arraycolsep=0pt\begin{eqnarray}}
{\end{eqnarray}\protect\aftergroup\ignorespaces}
\newenvironment{lefteqnarray*}{\arraycolsep=0pt\begin{eqnarray*}}
{\end{eqnarray*}\protect\aftergroup\ignorespaces}
\newenvironment{leftsubeqnarray}{\arraycolsep=0pt\begin{subeqnarray}}
{\end{subeqnarray}\protect\aftergroup\ignorespaces}

\markboth{\eightrm Radio analysis of \snr} 
{\eightrm Z. J. SMEATON {\lowercase{\eightit{et al.}}}}

\begin{strip}

{\ }

\vskip-1cm

\publ

\type

{\ }

\title{Study of a giant Large Magellanic Cloud Supernova Remnant, Veliki (J0450.4$-$7050)}

\authors{
Z. J. Smeaton$^1$,
M. D. Filipovi\'c$^1$,
R. Z. E. Alsaberi$^{2,1}$,
B. Arbutina$^{3}$,
W. D. Cotton$^{4, 5}$,}
\authors{
E. J. Crawford$^{1}$,
A. M. Hopkins$^6$, 
R. Kothes$^7$,
D. Leahy$^8$,
J. L. Payne$^1$,
N. Rajabpour$^1$,
}
\authors{
H. Sano$^{2, 9}$,
M. Sasaki$^{10}$, 
D. Uro\v sevi\'c$^{3}$,
and J. Th. van Loon$^{11}$
}

\vskip3mm

\address{$^1$Western Sydney University, Locked Bag 1797, Penrith South DC, NSW 2751, Australia}
\address{$^2$Faculty of Engineering, Gifu University, 1-1 Yanagido, Gifu 501-1193, Japan}
\address{$^3$Department of Astronomy, Faculty of Mathematics, University of Belgrade, Studentski trg 16, 11000 Belgrade, Serbia}
\address{$^4$National Radio Astronomy Observatory, 520 Edgemont Road, Charlottesville, VA 22903, USA}
\address{$^5$South African Radio Astronomy Observatory
Liesbeek House, River Park, Gloucester Road
Cape Town, 7700, South Africa}
\address{$^6$School of Mathematical and Physical Sciences, 12 Wally’s Walk, Macquarie University, NSW 2109, Australia}
\address{$^7$Dominion Radio Astrophysical Observatory, Herzberg Astronomy \& Astrophysics, National Research Council Canada, P.O. Box 248, Penticton}
\address{$^8$Department of Physics and Astronomy, University of Calgary, Calgary, Alberta, T2N IN4, Canada}
\address{$^9$Center for Space Research and Utilization Promotion (c-SRUP), Gifu University, 1-1 Yanagido, Gifu 501-1193, Japan}
\address{$^{10}$Dr Karl Remeis Observatory, Erlangen Centre for Astroparticle Physics, Friedrich-Alexander-Universit\"{a}t Erlangen-N\"{u}rnberg, Sternwartstra{\ss}e 7, 96049 Bamberg, Germany}
\address{$^{11}$Lennard-Jones Laboratories, Keele University, ST5 5BG, UK}

\Email{19594271@student.westernsydney.edu.au}

\dates{XXX}{XXX}

\abstract{We present a high-resolution radio-continuum view and a multi-frequency analysis of the \ac{LMC} \ac{SNR} \snrold, which we give the nickname \snr. These high-resolution observations reveal a larger extent than previously measured, making \snrold\ one of the largest \acp{SNR} that we know of. Additionally, we observe a higher than expected radio surface brightness and an unusually flat spectral index ($\alpha\,=\,-0.26\pm0.02$), with little spectral variation over the remnant. We observe a bright \Halpha\ shell indicating significant cooling over the remnant, but also an excess of \OIII\ on the eastern shock front%
. We investigate several theoretical scenarios to explain the emission and radio evolution of \snrold\ in the context of the \ac{LMC} environment, and determine that this is most likely an older, predominantly radiative, SNR with a higher shock compression ratio, which gives a flatter non-thermal spectrum, in combination with a thermal (bremsstrahlung) emission contribution.}


\end{strip}

\acresetall

\section{Introduction}
\label{sec:intro}

\acp{SNR} have been extensively studied in terms of their evolution and impact on stellar and galactic evolution. Extensive studies have been conducted on the Galactic \ac{SNR} population and on the \ac{SNR} populations of nearby galaxies to understand their full evolution and physical properties. Several galaxies have had their \ac{SNR} populations studied in detail, such as the \ac{LMC}~\citep{1996ASPC..112...91F,Filipovic1998a, Payne2008, 2015PKAS...30..149B, Maggi2016, Bozzetto2017, Yew2021, Bozzetto2023,2021MNRAS.507.2885F, Zangrandi2024}, the \ac{SMC}~\citep{Payne2007, Filipovic2008, Maggi2019, Cotton2024}, M31~\citep{Galvin2014M31, Kavanagh2020}, as well as a number of nearby galaxies outside of the Local Group~\citep{Millar2011, Pannuti2011, Galvin2012, Millar2012, OBrien2013, Galvin2014Sculpt, Pannuti2015, Yew2018}, which provides a unique opportunity to study entire galactic \ac{SNR} populations in detail. 

The newest generation of radio telescopes, such as \ac{ASKAP} and MeerKAT, are ideal for analysing \acp{SNR}. With a higher resolution and sensitivity than previously available, they allow in-depth radio-continuum studies to be conducted. This has been demonstrated by the discovery of several new individual \acp{SNR}, such as J0624$-$6948~\citep{Filipovic2022}, G288.8--6.3~\citep[Ancora;][]{Filipovic2023,BurgerSchiedlin2024}, G181.1$-$9.5~\citep{Kothes2017}, G121.1$-$1.9~\citep{Khabibullin2024}, G329.9$-$0.5~\citep[Perun;][]{Smeaton2024}, and G305.4$-$2.2~\citep[Teleios;][]{Filipovic2025Arxiv}) and \ac{SNR} candidates, such as G308.73+1.38~\citep[Raspberry;][]{Lazarevic2024} and G312.65+2.87~\citep[Unicycle;][]{Smeaton2024}, as well as recent studies which uncover new details about previously known \acp{SNR}, such as G278.94+1.35 \citep[Diprotodon;][]{Filipovic2024}. It has also been demonstrated by the large-scale surveys conducted with these telescopes, which have resulted in several discoveries and analysis of several \acp{SNR} and \ac{SNR} candidates~\citep[][Ball et al. 2025, in press]{Ball2023, Cotton2024SMC, Cotton2024, Goedhart2024, Anderson2025}. These highlight the importance of radio-continuum observations in discovering and analysing \ac{SNR} populations.

Due to the \ac{LMC}'s nearby location~\citep[$\sim$50\,kpc;][]{Pietrzynski2019} in an area with relatively weak Galactic emission interference, we are able to fully resolve the \acp{SNR} and study their physical properties in detail. Thus, we can map the evolution of an entire population and determine how the properties of the galaxy influence their evolution. To this end, we present a detailed study of one of the \ac{LMC} \acp{SNR}, \snrold\ (also known as J0450--709 and MC~11), which we give the nickname \snr\footnote{\snr\ ({\rrm Veliki}) meaning large in Serbian.}. \snr\ displays unusual properties which may be attributed, in part, to its location within the different environment of the \ac{LMC}. %

\begin{figure*}
		\centering
			\includegraphics[width=\textwidth]{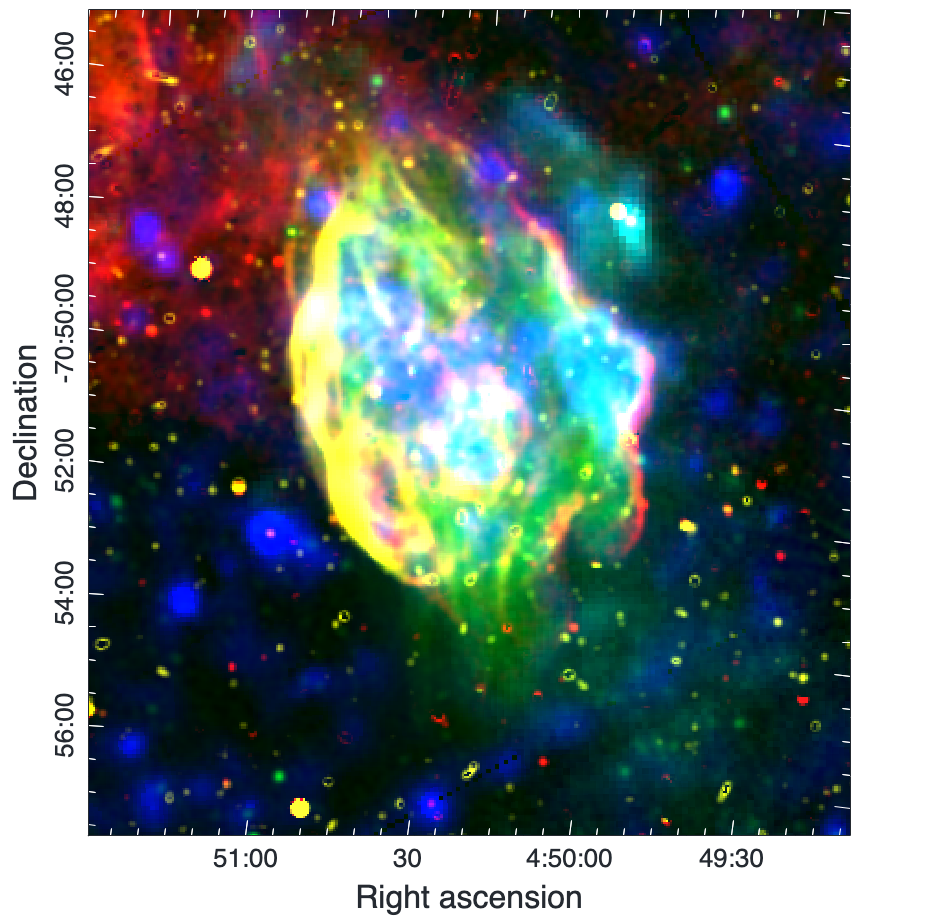}
        \caption{Five colour multi-frequency image of \snr. Red is optical \ac{MCELS} \Halpha, green is 1.3\,GHz MeerKAT radio, blue is {\it XMM-Newton} soft X-ray (0.3$-$1\,keV), yellow is optical \ac{MCELS} \OIII, and cyan is 250\,$\mu$m {\it Herschel} IR.}
        \label{fig:RGBYM}
\end{figure*}

As one of the largest known \acp{SNR} in the \ac{LMC}, \snr\ has been extensively studied at multiple frequencies. It was discovered at radio frequencies as an unclassified radio source at 4800\,MHz using the Parkes radio telescope~\citep{McGee1972}, where it was named MC~11. It appears in several subsequent large-scale radio surveys with radio measurements ranging from 408\,MHz to 8.85\,GHz~\citep{Clarke1976, Wright1994, Filipovic1995, Filipovic1996, Filipovic1998a}. \citet{Mathewson1985} were the first to classify \snr\ as an \ac{SNR} using narrow-band optical observations to measure a [SII]/H$\alpha$ ratio of $\sim$0.7, which classified it as an evolved \ac{SNR} using the criteria of \citet{Mathewson1973}. 

\snr\ was then studied using X-ray and optical data by \citet{Williams2004}. Their analysis revealed a complex filamentary optical structure with distinct inner and outer shell structure, with different expansion velocities; $\sim$120\,\kms\ for the outer shell and $\sim$220\,\kms\ for the inner. They detect interior X-ray emission and estimate a minimum age of $\sim$45,000 years, making \snr\ an evolved \ac{MM} remnant.

\snr\ has also been observed in the Far \ac{UV}~\citep{Blair2006}, and also appears in the \ac{IR} study of \ac{LMC} \acp{SNR} by \citet{Lakicevic2015}. The most recent radio analysis was performed by \citet{Cajko2009}. Their radio analysis showed a spectral index $\alpha\,=\,-0.43\pm0.06$, slightly flatter than expected and attributed to thermal contributions, as well as a physical size of 102$\times$75$\pm$1\,pc and a peak radio polarisation of $\sim$40\%. \snr\ also appears in several other \ac{LMC} \ac{SNR} analyses~\citep{Maggi2016, Bozzetto2017, Zangrandi2024}. \citet{Maggi2016} gives X-ray properties of 0.24$\pm$0.02\,keV plasma temperature and 33.6$^{+36.8}_{-24.1}\times10^{58}$\,cm$^{-3}$ emission measure, which are best fit by a one-component \ac{CIE} model. This suggests that the inner plasma has reached thermal equilibrium between electrons and ions ($kTe=kTi$), consistent with an evolved \ac{SNR} where sufficient time has passed to achieve electron-ion temperature equilibrium through Coulomb collisions~\citep{Vink2012}. The analysis of \citet{Bozzetto2017} also state a possible \ac{CC} origin for \snr.

There have been $\gamma$-ray surveys conducted of the \ac{LMC}, and there are currently only a handful of $\gamma$-ray emitting \ac{LMC} \acp{SNR} known. \snr\ is not detected in the current $\gamma$-ray studies, including the recent Fermi-LAT survey~\citep{Tramacere2025} which detected several $\gamma$-ray emitting \acp{SNR}, but did not cover \snr's location. These detections are expected to increase when \ac{CTA} observations commence, which should achieve higher sensitivities than previously obtained~\citep{Acharyya2023}.

In this paper, we present new radio observations of \snr\ using data from the \ac{ASKAP} and MeerKAT radio telescopes (see Section~\ref{sec:obs and data reduction}) with higher resolution and sensitivity than previous data. We also use archival data at other frequencies to perform a multi-frequency analysis of the properties. In Section~\ref{sec:obs and data reduction} we detail the data used, in Section~\ref{sec:results} we show results of the analysis, in Section~\ref{sec:discussion} we discuss theoretical interpretations, and in Section~\ref{sec:conclusion} we present the final conclusions.%

\section{Observations and Data Reduction}
\label{sec:obs and data reduction}

The continuum data used consists of radio data from the \ac{ASKAP}, MeerKAT, \ac{MWA}, and \ac{MOST} radio telescopes, \ac{IR} data from the {\it Herschel} telescope, optical data from the \ac{MCELS} survey, and X-ray data from the {\it XMM-Newton} telescope. We also use spectral line data, specifically \HI\ data from the \ac{ATCA} and Parkes telescopes, and CO data from the \ac{MAGMA} survey.

\subsection{Radio-continuum data}
\label{subsec:radio}

\subsubsection{ASKAP}
\label{subsubsec:ASKAP}

\snr's location in the \ac{LMC} has been observed with the \ac{ASKAP} telescope~\citep{Johnston2008, Hotan2021} as part of the large-scale \ac{EMU} survey~\citep{Norris2011, Norris2021,2025arXiv250508271H}. This area was covered at 888\,MHz as part of the \ac{ASKAP} commissioning and early science (ACES, project code AS033, scheduling block SB~8532)~\citep{Pennock2021, Bozzetto2023}, and again in the subsequent main survey at 943\,MHz (project code AS201).

Both observations were conducted using the entire 36 antenna \ac{ASKAP} array with a bandwidth of 288\,MHz. The early science 888\,MHz data has a restoring beam of  13.9\arcsec$\times$12.1\arcsec\ (position angle\ PA=--84.4\D) and the main survey 944\,MHz data is convolved to a common beam size of 15\arcsec$\times$15\arcsec. All \ac{EMU} data was reduced using the standard ASKAPSoft pipeline which involves mutli-scale cleaning, self-calibration, and multi-frequency
synthesis imaging~\citep{Guzman2019}. \snr\ appears in three separate \ac{EMU} main scheduling blocks, SB~46957, SB~48589, and SB~46962, however in SB~48589 and SB~46962, \snr\ is located at the edge of the tile where sensitivity drops exponentially \citep{2025arXiv250508271H}. Due to this, only SB~46957 is used in this analysis.

\subsubsection{MeerKAT}
\label{subsubsec:MeerKAT}

The MeerKAT data used comes from the recent MeerKAT survey of the \ac{LMC} under project code SSV-20180505-FC-02. The observations were conducted using the MeerKAT L-band split into 14 sub-band channels from 856$-$1712\,MHz. Two sub-bands (centred at 1198.9 and 1255.8\,MHz) were blanked due to \ac{RFI}. Observations were conducted for all Stokes I, Q, U, and V parameters, and the data consists of all sub-bands as well as an additional broadband image centred at 1295\,MHz across the whole bandwidth. Details of the image generation, calibration, data reduction, and final survey will be presented in Cotton et al. (in prep) and Rajabpour et al. (in prep). We use the Stokes~I data products for our radio-continuum analysis, and the Stokes~I, Q, and U products for our polarisation analysis. 

\subsubsection{MWA}
\label{subsubsec:MWA}

We use radio data from the \ac{MWA} telescope~\citep{Tingay2013}. Specifically, we use the radio-continuum maps of the \ac{LMC} from the \ac{GLEAM}~\citep{Wayth2015, HurleyWalker2017} survey at 88, 118, 155, and 200\,MHz as described in \citet{For2018}. %

\subsubsection{MOST}
\label{subsubsec:MOST}

We also use the more recent 843\,MHz \ac{MOST} radio survey of the \ac{LMC}~\citep{Turtle1991} to remeasure the prior \ac{MOST} measurement of \citet{Mathewson1973}. The \citet{Mathewson1973} 843\,MHz measurement significantly differed from our new \ac{ASKAP} 888\,MHz measurement, and thus we remeasured from the more recent 843\,MHz \ac{MOST} survey data (see Section.~\ref{subsec:flux densities} for more detail). 

\subsection{Optical data}
\label{subsec: other data}

We use optical data from the \ac{MCELS} optical survey~\citep{Smith2000} of the \acp{MC}. The survey imaged the central 8\D$\times$8\D\ of the \ac{LMC} using the UM/CTIO Curtis Schmidt telescope and achieved a resolution of 3-4\arcsec. We use the narrow-band observations of \Halpha\ (centred at 656.3\,nm with a bandwidth of 3\,nm), \SII\ (centred at 672.4\,nm with a bandwidth of 5\,nm), and \OIII (centred at 500.7\,nm with a bandwidth of 4\,nm). 

\subsection{IR}
\label{subsubsec:IR}

We use \ac{FIR} data from the \ac{HERITAGE}~\citep{Meixner2006} survey conducted using the {\it Herschel} Space telescope~\citep{Pilbratt2010}. Specifically, we use the longest wavelengths from the \ac{SPIRE} instrument at 250, 350, and 500\,$\mu$m.

\subsection{X-ray}
\label{subsubsec:X-ray}

We use X-ray data from the {\it XMM-Newton} telescope. We use data from the \ac{MC} {\it XMM-Newton} survey~\citep{Haberl2014, Maggi2016, Maggi2019}, which surveyed the \ac{LMC} and \ac{SMC} with the full {\it XMM-Newton} bandwidth (0.2$-$10.0\,keV) with a sensitivity of $F_\mathrm{X}$ (0.3$-$8\,keV) $\approx 10^{-14}$ erg\,cm$^{-2}$\,s$^{-1}$. We use only the soft X-ray band (0.3$-$1\,keV) in this analysis.

\subsection{\HI}
\label{subsubsec:HI}
We also used archival \HI\ data published by \cite{2003ApJS..148..473K}. The \HI\ data were obtained by combining the Australia Telescope Compact Array and the Parkes 64-m radio telescope observations. The angular resolution of the \HI\ data is 1\farcm0, corresponding to the spatial resolution of $\sim$15~pc at the distance of the LMC. The typical rms noise is $\sim$2.5~K at the velocity resolution of $\sim$1.6~km~s$^{-1}$.

\subsection{CO}
\label{subsubsec:CO}
The $^{12}$CO($J$~=~1--0) data were obtained from the Magellanic Mopra Assessment \citep[MAGMA DR3,][]{2011ApJS..197...16W,2017ApJ...850..139W}. The angular resolution is $\sim$45$''$, corresponding to a spatial resolution of $\sim$11~pc at the distance of the LMC. The typical rms noise is $\sim$0.3~K at a velocity resolution of $\sim$0.5~km~s$^{-1}$.

\section{Results}
\label{sec:results}

\subsection{Morphology}
\label{subsec:morphology}

\begin{figure*}
		\centering
			\includegraphics[width=\textwidth]{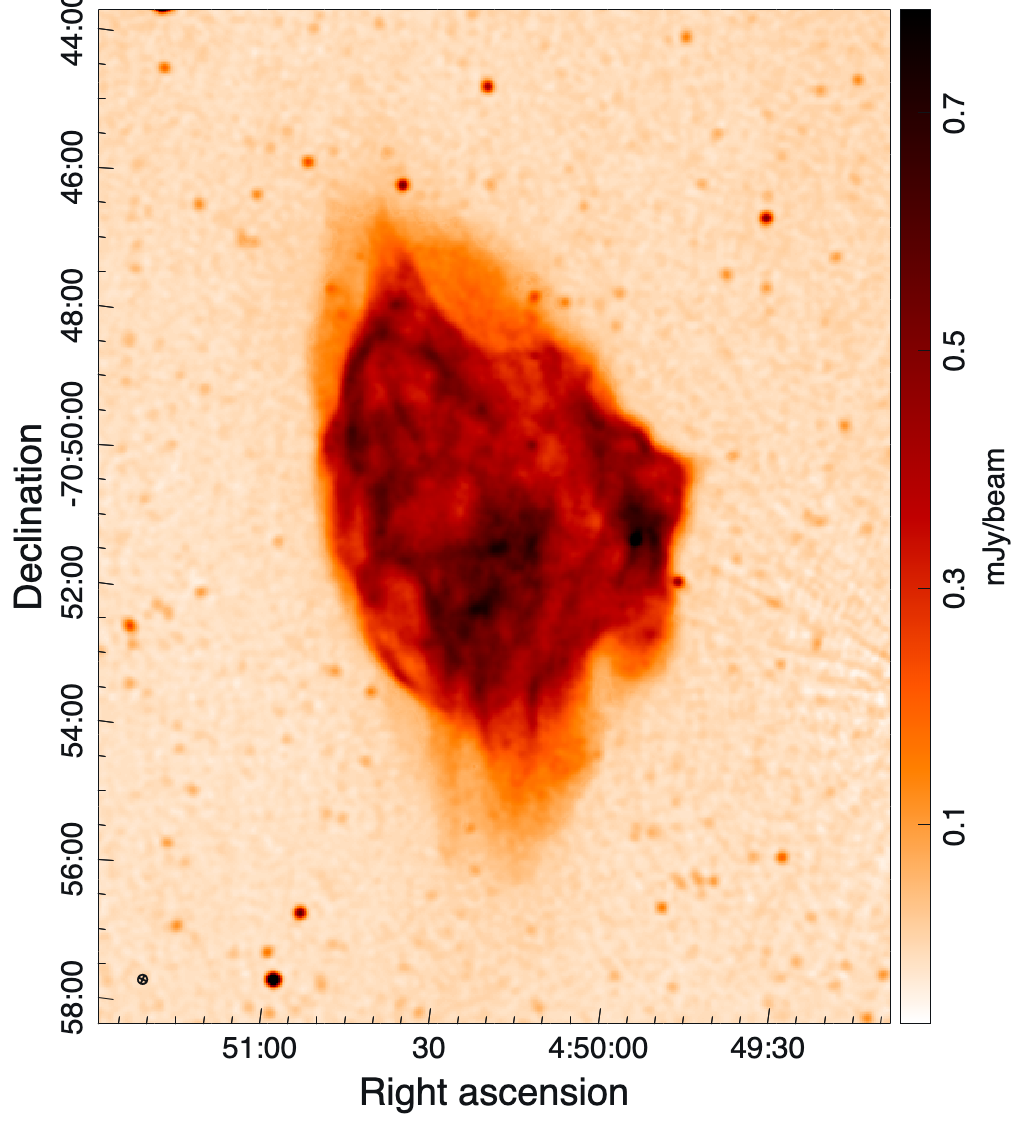}
        \caption{MeerKAT 1.3\,GHz view of LMC SNR J0450--709. The image is linearly scaled, has a synthesised beam of 8\arcsec$\times$8\arcsec shown in the bottom left corner and a measured local \ac{RMS} noise level of 10\,$\mu$Jy beam$^{-1}$.}
        \label{fig:SNR}
\end{figure*}

We analyse the morphology of \snr\ at several different frequencies (see Fig.~\ref{fig:RGBYM}). %
\snr\ has a filamentary shell-like morphology at optical and radio frequencies, elongated along the north-south axis and with a brighter clump of emission on the western edge (see Fig.~\ref{fig:SNR}). From the MeerKAT data, we measure the centre of the remnant to be RA(J2000)\,=\,4$^{\rm h}$50$^{\rm m}$26.8$^{\rm s}$, Dec(J2000)\,=\,$-$70$^{\rm d}$50$^{\rm m}$45.5$^{\rm s}$, with angular diameters 10.3\arcmin$\times$5.6\arcmin\ and position angle 17.6\D\ (measured counterclockwise from north axis), corresponding to a physical size of 150\,pc$\times$81\,pc at the \ac{LMC}~\citep{Pietrzynski2019}. This size is significantly larger than previously reported in radio~\citep[102\,pc$\times$75\,pc;][]{Cajko2009}, and this is due to the faint filamentary structure visible in the north and south, which was not visible in the previous radio images.%

\begin{figure*}
		\centering
			\includegraphics[width=\textwidth]{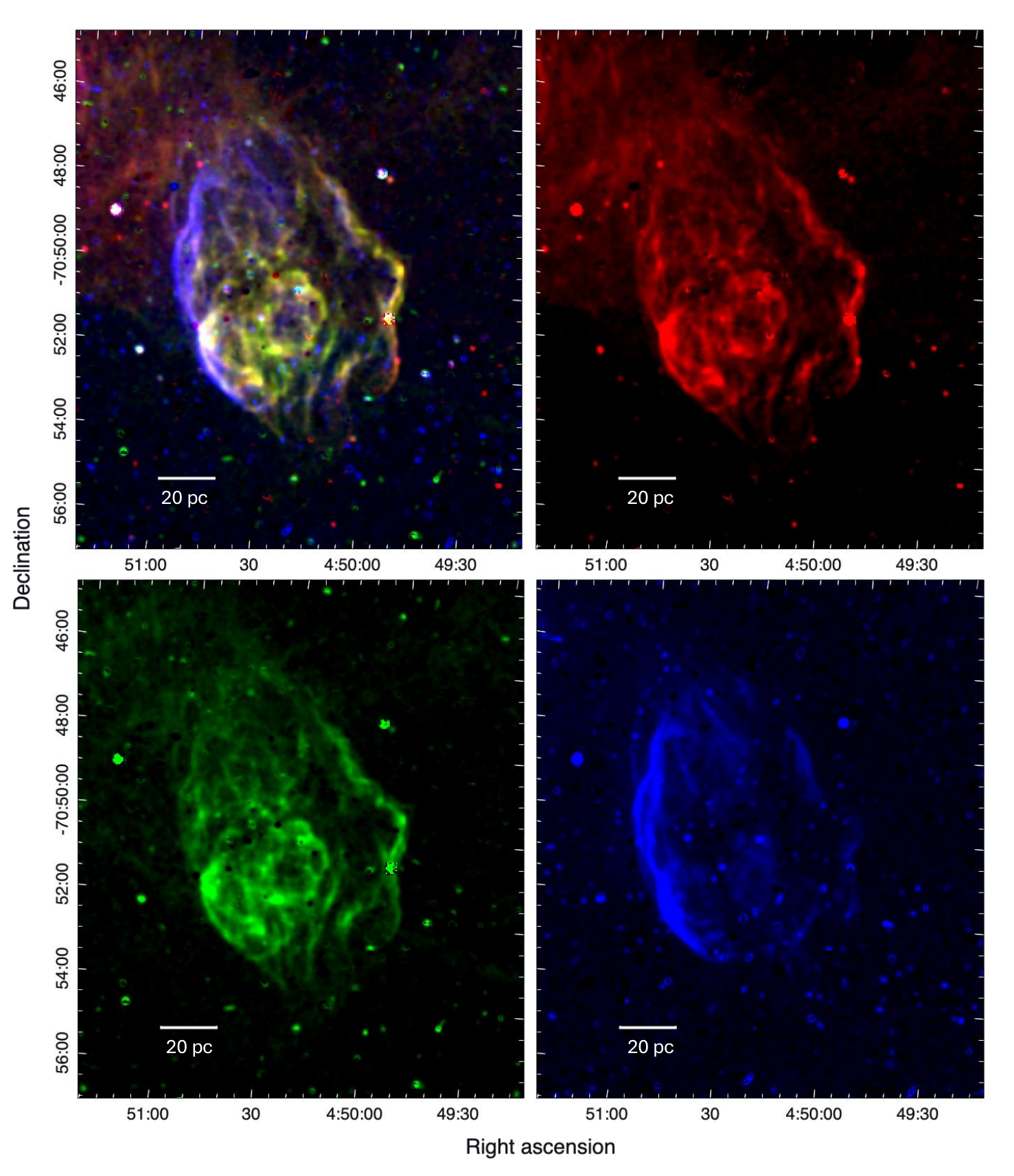}
        \caption{4-panel image of \snr\ at optical wavelengths using \ac{MCELS} data. {\bf Top left:} Optical RGB using three \ac{MCELS} filters. The other three panels show the individual filters in their respective colours which make up the RGB. Red is \Halpha ({\bf top right}), green is \SII ({\bf bottom left}), and blue is \OIII ({\bf bottom right}). All images are linearly scaled and have a 20\,pc scale bar shown in the bottom left corner.}
        \label{fig:opticalRGB}
\end{figure*}

The \Halpha\ and \SII\ emission are closely correlated and form an optical filamentary shell across the entire remnant (see Fig.~\ref{fig:opticalRGB}, first, second, and third panels). %
 The optical and radio emission are closely correlated with the optical leading the radio along the majority of the shell; the only exception being the southern filamentary structure which is only visible in radio. %
 We note an excess of \OIII\ emission in the eastern rim of the remnant, leading both the radio and \SII/\Halpha\ shell (see Fig.~\ref{fig:opticalRGB}, fourth panel).%

\snr\ has interior soft X-ray (0.3$-$1.0\,keV) emission which is not visible at higher frequencies%
($>$1\,keV for {\it XMM-Newton}). This emission is confined to the interior of the \ac{SNR}, with bright patches in the middle eastern section and in the western protruding clump. The exact X-ray properties of these clumps could be better determined with in-depth X-ray analysis, such as by {\it XMM-Newton} or {\it Chandra}, which could potentially reveal the chemical composition of the ejecta with X-ray spectroscopy. There is a complex filamentary \Halpha\ structure near the eastern edge identified by \citet{Williams2004} and measured to be expanding faster ($\sim$220\,km s$^{-1}$) than the outer shell ($\sim$120\,km s$^{-1}$). There are no obvious radio or X-ray structures which correlate with this inner \Halpha\ shell.

\begin{figure}
\centerline{\includegraphics[width=0.5\textwidth]{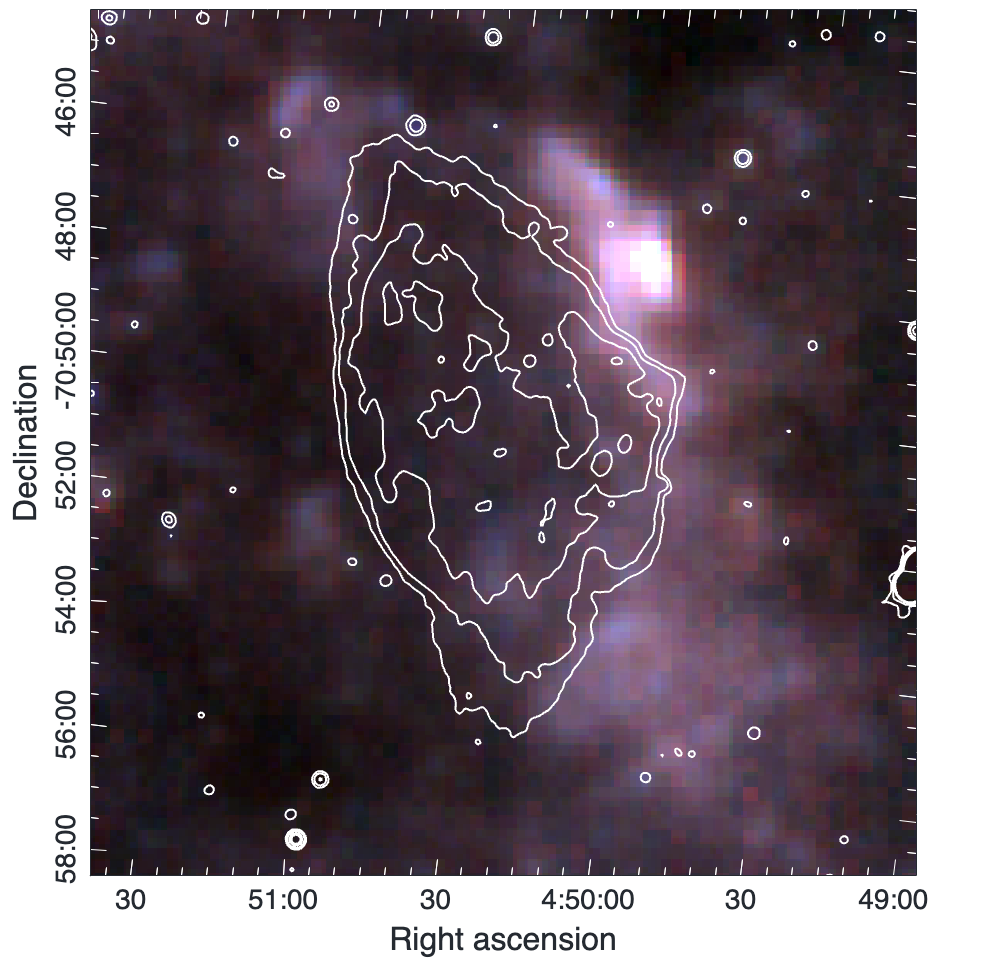}}
\caption{
FIR RGB of \snr\ and surrounding area using {\it Herschel} FIR data. Red is 500\,$\mu$m, green is 350\,$\mu$m, and blue is 250\,$\mu$m. All images are linearly scaled. The contours are from the 1.3\,GHz MeerKAT radio image at levels of 0.05, 0.15, 0.40, and 0.70\,mJy beam$^{-1}$.
}
\label{fig:IR}
\end{figure}

\snr\ itself does not appear at \ac{IR} frequencies (see Fig.~\ref{fig:IR}). There is a bright patch of \ac{IR} emission to the north-east of the shell, which appears to correspond with a catalogued molecular cloud PGCC~G272.62-35.35~\citep{Planck2016}. The cloud is not within the \ac{SNR} extent, however the \ac{IR} emission closely traces the north-eastern radio and \SII/\Halpha\ rim, directly above the western bulge. The findings of \citet{Lakicevic2015} conclude that interaction between \snr\ and a surrounding cloud is in progress or imminent, (see Sec.~\ref{subsec:environment}).%

\begin{figure*}
\centerline{\includegraphics[width=\textwidth]{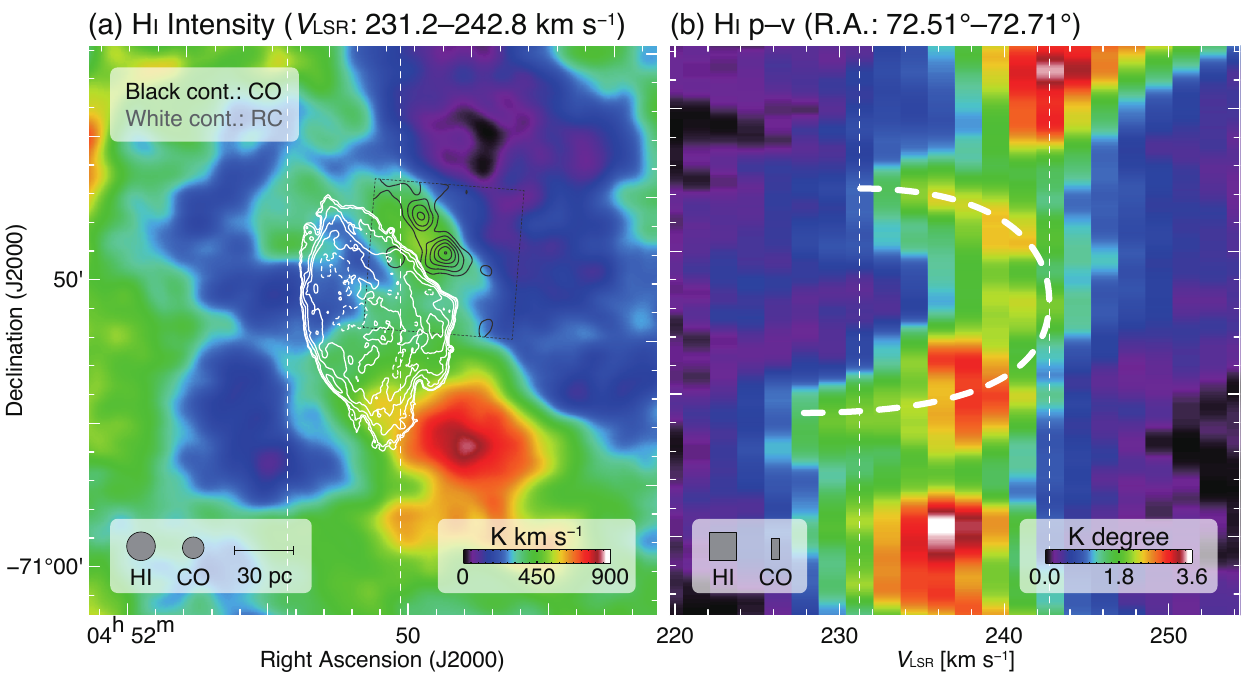}}
\caption{\HI\ and CO maps of Veliki and surrounding environment. Panel (a): \HI\ intensity map of Veliki and surrounding environment. \HI\ intensity is integrated over the $V_\mathrm{LSR}$ velocity range of 231.2--242.8\,km\,s$^{-1}$. White contours are from MeerKAT 1.3\,GHz radio-continuum map and black contours are from Mopra CO map. The contour levels are 0.10, 0.12, 0.18, 0.28, 0.42, and 0.60~mJy~beam$^{-1}$ for radio-continuum, and 1.2, 2.1, 3.0, 3.9, 4.8 and 5.7~K~km~s$^{-1}$ for CO. The dashed black square shows the extent of the CO image which the contours are taken from as it covers only a portion of the north-western shell. The vertical dashed white lines correspond with panel (b) and show the extent of the possible \HI\ cavity. The beam sizes are shown in the bottom left corner along with a scale bar. Panel (b): \HI\ $p$--$v$ diagram, integrated over Veliki's RA location (72\fdg51--72\fdg71). The beam sizes are shown in the bottom left corner. The thin white dashed correspond with panel (a) showing the location relative to Veliki, and the thick dashed white line traces the possible \HI\ cavity.}
\label{fig:HICO}
\end{figure*}

\snr\ itself does not appear in the \HI\ and CO observations (see Fig.~\ref{fig:HICO}). 
The integrated \HI\ covers the entire area, however the CO observations are restricted to a small portion of the north-western rim. %
We detect two distinct peaks in the CO emission corresponding with the observed \ac{IR} patch and the molecular cloud. We also observe a possible \HI\ cavity at the velocity range 231.2~km\,s$^{-1} < V_\mathrm{LSR} <$ 242.8~km\,s$^{-1}$ which spatially corresponds with \snr's location (see Fig.~\ref{fig:HICO}, panel (b)). The \HI, CO, and \ac{IR} observations are discussed in more detail in Sec.~\ref{subsec:environment}, in relation to \snr's environment.

\subsection{Radio flux densities}
\label{subsec:flux densities}

We fit a region around \snr\ using the \ac{CARTA} imaging software~\citep{Comrie2021} to extract flux density measurements. We use an elliptical region based on the remnant centre and angular diameters as described in Sec.~\ref{subsec:morphology}. We measure flux densities from the 88, 118, 155, and 200\,MHz \ac{MWA} data, the 888 and 944\,MHz \ac{ASKAP} data and the 1.3\,GHz MeerKAT data (see Table~\ref{tab:fluxes}). We also remeasured the 843\,MHz flux density measurement from \ac{MOST}. The previously measured value was 837\,mJy from \citet{Mathewson1985}; at odds with our most recent value of 696$\pm$70\,mJy at 888\,MHz from \ac{ASKAP}. We thus used the more recent \ac{MOST} images of the \ac{LMC} by \citet{Turtle1991} to remeasure and verify this data point, and measured a value of 671$\pm$67\,mJy, more consistent with expectations from \ac{ASKAP}. We estimate a 10\% error in our \ac{ASKAP}, MeerKAT, and \ac{MOST} flux density measurements, similar to the methodology used in \citet{Filipovic2022, Filipovic2023, Filipovic2024, Smeaton2024} and \citet{Bradley2025}. Due to the significantly poorer resolution of the \ac{MWA} images caused by the lower frequency, we assume a 20\% error for these measurements. We list these new measurements and previously catalogued flux density measurements for \snr\ %
in Table~\ref{tab:fluxes}.

\begin{table}
    \centering
    \begin{tabular}{l|c|l}
        $\nu$ (MHz) & $S\pm\Delta S$ (mJy) & Reference \\ \hline
        88 & 1580$\pm$316 & This work \\
        118 & 1159$\pm$232 & This work \\
        155 & 1082$\pm$216 & This work \\
        200 & 918$\pm$184 & This work \\
        408 & 610$\pm$150 & \citep{Clarke1976} \\
        843 & 671$\pm$67 & This work \\
        888 & 696$\pm$70 & This work \\
        944 & 779$\pm$78 & This work \\
        1295 & 615$\pm$62 & This work \\
        1400 & 644$\pm$64$^\dag$ & \citep{Cajko2009} \\
        2300 & 540$\pm$47$^\ddag$ & \citep{Filipovic1996} \\
        2450 & 514$\pm$55$^\ddag$ & \citep{Filipovic1995IV} \\
        2700 & 470$\pm$47$^\dag$  & \citep{Filipovic1998a} \\
        4750 & 496$\pm$52$^\ddag$ & \citep{Filipovic1995IV} \\
        4800 & 448$\pm$45$^\dag$ & \citep{Cajko2009} \\
        4850 & 479$\pm$47$^\ddag$ & \citep{Filipovic1995IV} \\
        5010 & 410$\pm$150 & \citep{McGee1972} \\
        8640 & 360$\pm$36$^\dag$ & \citep{Cajko2009} \\
        8850 & 367$\pm$57 & \citep{Filipovic1995IV} \\
    \end{tabular}
    \caption{Measured and catalogued flux densities for \snr. Uncertainties for \ac{ASKAP}, MeerKAT, and MOST are taken as 10\% and uncertainties for \ac{MWA} fluxes are taken as 20\%. Other uncertainties are taken as listed in the catalogue, or are assumed to be 10\% if a value is not given (indicated by $^\dag$). The $^\ddag$ indicates values where an uncertainty was not listed in the catalogue, but an uncertainty equation was given in the listed reference, and this was used to calculate the uncertainty.}
    \label{tab:fluxes}
\end{table}

\subsection{Radio spectral index}
\label{subsec:spectral index}

\begin{figure*}
\centerline{\includegraphics[width=\textwidth]{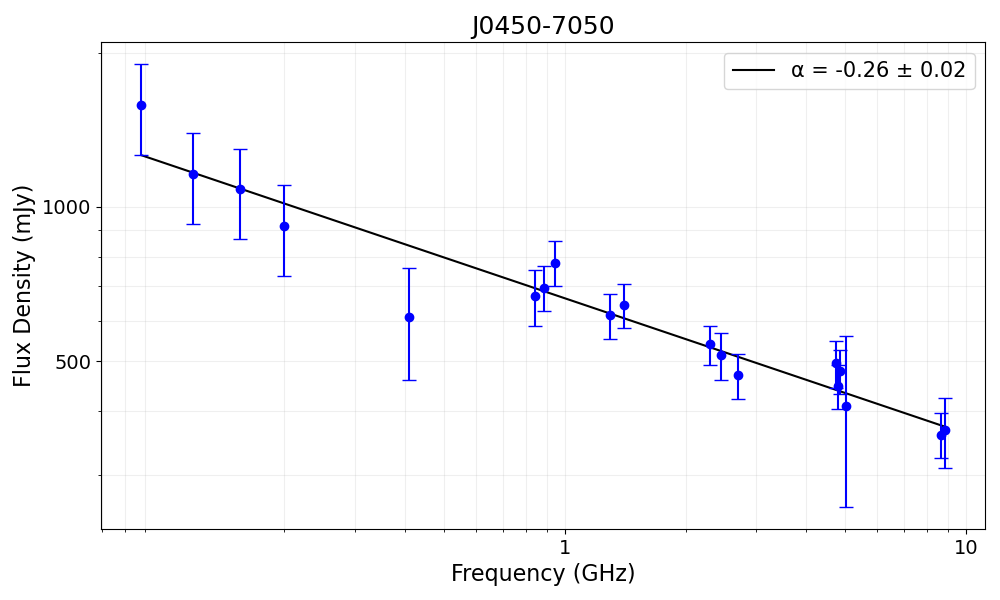}}
\caption{
Spectral index graph of \snr\ using flux density values from Table~\ref{tab:fluxes}. A linear power law is fit and the slope is taken as the spectral index. The quoted error only includes the statistical error of the fit.
}
\label{fig:specindexgraph}
\end{figure*}

We use the flux density values (see  Table~\ref{tab:fluxes}) to calculate the overall spectral index of \snr, defined as $S\propto\nu^\alpha$, where $S$\,=\,integrated flux density, $\nu$\,=\,observing frequency, and $\alpha$\,=\,spectral index~\citep{book1}. We use all available measurements to calculate the total spectral index, and the available high-resolution images to generate a spectral index map. 

\subsubsection{Total spectral index}
\label{subsubsec: total spec index}

The linear fit was conducted using the \textsc{linregress}\footnote{\url{https://docs.scipy.org/doc/scipy/reference/generated/scipy.stats.linregress.html}} function in the Python library \textsc{scipy}~\citep{Virtanen2020} which uses a linear least-squares regression method to find the line of the best fit to the data. The quoted uncertainty is defined as the standard error of the fit~\citep{Theil1950}, and thus likely underestimates the true uncertainty. We calculate a spectral index value of $\alpha$\,=\,$-$0.26$\pm$0.02, with a reduced $\chi^2$ value of 0.53 and 17 degrees of freedom. In order to verify this value, we also fit the data points with a weighted fit using the \textsc{curve\_fit} function in the \textsc{scipy} library, and get similar results of $\alpha$\,=\,$-0.27\pm0.03$, and a reduced $\chi^2$ value of 0.50 with the same degrees of freedom. These reduced $\chi^2$ values and the consistency between the two methods indicate a good fit for the data and provide confidence for the measured spectral index value. 

We note the largest outlier is the 408\,MHz measurement from \citet{Clarke1976}. %
They note that \snr\ appears extended in their observations and this may have caused erroneous measurements due to the older telescope's insensitivity to extended emission. While this outlier could indicate spectral curvature, its inconsistency with recent measurements and high uncertainty prevents us from claiming spectral curvature based solely on this point. %
Even excluding this point, we calculate similar spectral index values as above, with $\alpha$\,=\,$-$0.27$\pm$0.01 (reduced $\chi^2$\,=\,0.42, degrees of freedom\,=\,16) for linear least-squared regression, and $\alpha$\,=\,$-$0.27$\pm$0.03 (reduced $\chi^2$\,=\,0.41, degrees of freedom\,=\,16) for the weighted fit. We therefore determine that this point of higher uncertainty is not significantly impacting the results, and that a linear fit is accurate with no obvious curvature in the spectrum.

We use this total spectral index value to scale the flux to 1\,GHz, and obtain $S_{1\,\text{GHz}}$\,=\,664\,mJy. We calculate a surface brightness, defined as $\Sigma_{1\,\text{GHz}}$\,=\,$S_{1\,\text{GHz}}/\Omega$ where $\Omega$\,=\,angular area. Using the region defined in Sec.~\ref{subsec:morphology} as 10.3\arcmin$\times$5.6\arcmin, we calculate an angular size of $\Omega\,=\,\pi ab\,=\,163086$\,arcsec$^2$ and $\Sigma_{1\,\text{GHz}}$\,=\,1.7$\times$10$^{-21}$\,W\,m$^{-2}$\,Hz$^{-1}$\,sr$^{-1}$. We calculate a radio luminosity defined as $L_{1\,\text{GHz}}$\,=\,4$\pi d^2S$, where $d$\,=\,distance to the object. Using a distance of 50\,kpc, we calculate $L_{1\,\text{GHz}}$\,=\,1.9$\times$10$^{17}$\,W\,Hz$^{-1}$.

\subsubsection{Spectral index map}
\label{subsubsec:spec index map}

\begin{figure*}
\vskip-3mm
\centerline{\includegraphics[width=\textwidth]{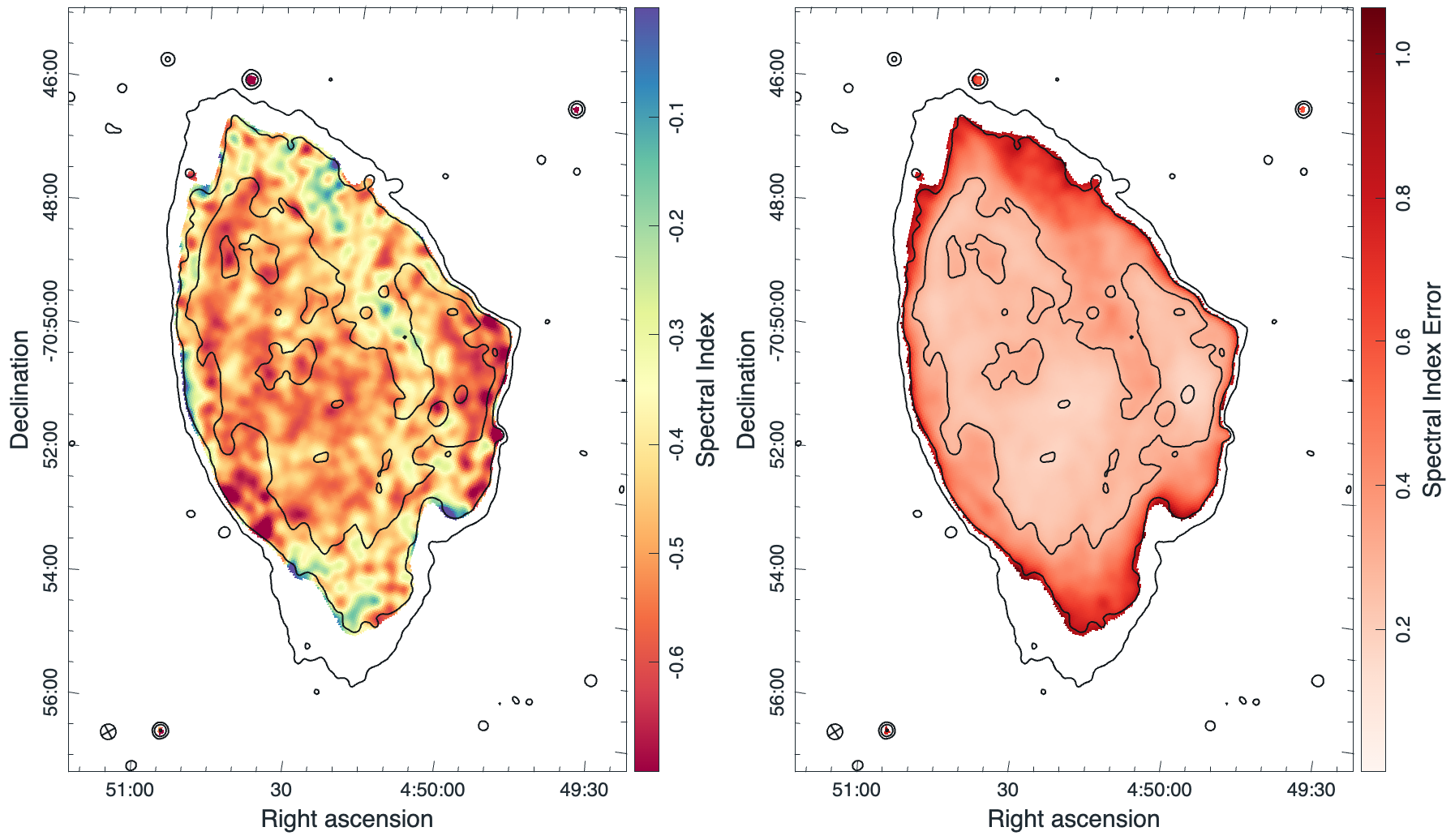}}
\caption{
Spectral index map and error map of \snr\ created using \ac{ASKAP} \ac{EMU} data at 888 and 944\,MHz, and MeerKAT data at 1.3\,GHz. Final images have synthesised beam sizes of 15\arcsec$\times$15\arcsec shown in the bottom left corners. The contours are radio-continuum contours from the MeerKAT 1.3\,GHz image at levels of 0.05, 0.15, 0.40, and 0.70 mJy beam$^{-1}$. {\bf Left} is spectral index map and {\bf right} is spectral index map error, generated as described in text.}
\label{fig:specindexmap}
\end{figure*}

We use the high resolution \ac{ASKAP} and MeerKAT radio data to generate a spectral index map. All images are convolved to the lowest resolution (\ac{ASKAP} 944\,MHz beam size\,=\,15\arcsec, pixel size\,=\,2\arcsec) before calculation. The images are then input into the \textsc{maths} function in the \textsc{MIRIAD} software which calculates the line of best fit through the data points for each pixel, which are then used to generate the final image (see Fig.~\ref{fig:specindexmap}, left). We estimate a spectral index error map using standard error propagation. For each frequency pair ($\nu_1, \nu_2$), the spectral index error is calculated as $\sigma_\alpha$ = 1/ln($\nu_1$/$\nu_2$)$\times\sqrt{(\sigma_1/S_1)^2+(\sigma_2/S_2)}$ where $\sigma_i$ is the \ac{RMS} noise in each image and $S_i$ is the flux density. We calculated the errors for all frequency pairs and averaged them to obtain the final uncertainty map. Both images were then cut to remove values with uncertainties $>$ 1.0. We measure an average value of $\alpha$\,=\,$-$0.46$\pm$0.36. The spectral index is mostly uniform with $\alpha\sim-0.4$ to $-0.5$, but there are two areas of slightly steeper spectral index in the western bulge and on the south-eastern edge which extent to $\alpha\sim-0.6$ to $-0.7$.

These values are significantly steeper than the overall value obtained from the full set of flux measurements (see Fig.~\ref{fig:specindexgraph}), and the estimated errors are quite high. This is likely due to the smaller frequency range used (888$-$1300\,MHz) when compared with the total range%
 (88$-$8850\,MHz). Therefore, the map values likely overestimate the spectral index steepness and the total value $\alpha\,=\,-0.26$ is more accurate. The images had similar native resolutions and were convolved to the same resolution before map generation, and thus the uniformity of the spectral index is likely accurate. Both the spectral index map and spectral index error map show a uniform distribution, indicating that there are no areas of significant spectral variability, such as a central \ac{PWN}.

\subsection{Polarisation}
\label{subsec:polarisation}

\begin{figure*}
    \centering
    \includegraphics[width=\textwidth]{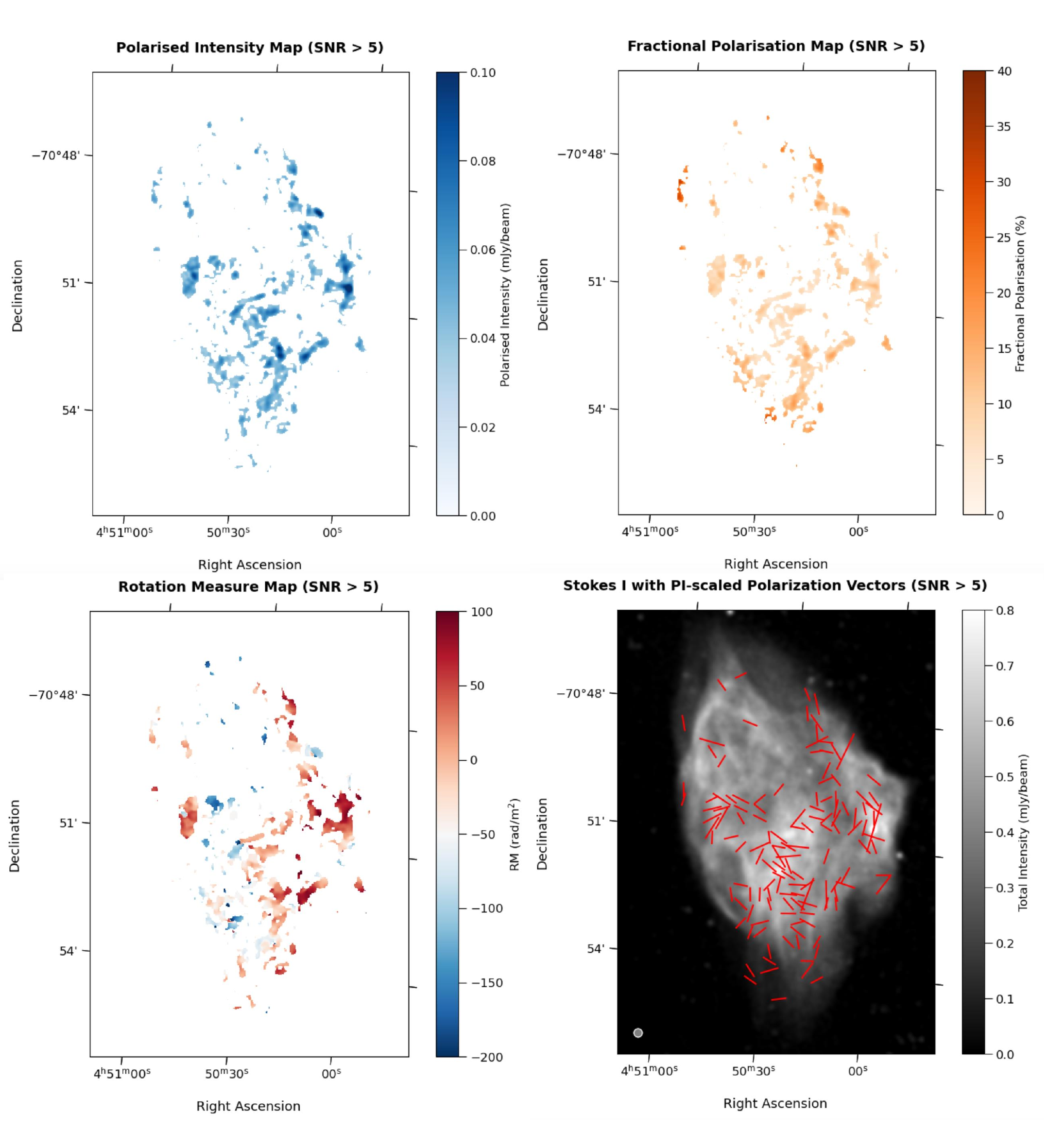}
    \caption{Polarisation and \ac{RM} output from RM synthesis analysis described in text. All images have been cut at a signal to noise ratio of 5$\sigma$. {\bf Top left} is polarised intensity, {\bf top right} is fractional polarisation, {\bf bottom left} is rotation measure (RM), and {\bf bottom right} is magnetic field vectors overlaid over the Stokes~I total intensity image. Background Stokes~I image is linearly scaled and has a beam size of 8\arcsec$\times$8\arcsec shown in the bottom left corner.}
    \label{fig:polarisation}
\end{figure*}

We conduct a polarisation and \ac{RM} analysis using the MeerKAT radio data.
This analysis was done using the Python \textsc{pyrmsynth} program\footnote{\url{https://mrbell.github.io/pyrmsynth/}}, which uses the \textsc{RMCLEAN} method as described in \citet{Heald2009}. The total polarised intensity is calculated as the magnitude of the polarisation, $PI=\sqrt{Q^2+U^2}$, where $PI$\,=\,polarised intensity, and $Q$ and $U$ are the flux density measurements of the respective Stokes parameters (Figure~\ref{fig:polarisation}, top left). The fractional polarisation is calculated as $PI/I$, where $I$ is the measured Stokes~I flux density (Figure~\ref{fig:polarisation}, top right). The \ac{RM} measurement follows the method of \citet{Heald2009}, which decomposes the complex polarisation function, $PI\,=\,Q+iU$ as a function of $\lambda^2$ into components at different Faraday depths (Fig.~\ref{fig:polarisation}, bottom left). This \ac{RM} is then used to de-rotate the polarisation vectors to measure the intrinsic electric field direction, which we then rotate by 90\D to map the intrinsic magnetic field %
(Figure~\ref{fig:polarisation}, bottom right). 

The results show relatively uniform polarisation values predominantly located in the southern half of the shell. We measure an average $PI$ of 0.05$\pm$0.01\,mJy\,beam$^{-1}$ and average fractional polarisation of 9$\pm$2\,\%.
We measure an average \ac{RM} of $-$16$\pm$11\,rad\,m$^{-2}$, ranging from $\sim-$200 to +100\,rad\,m$^{-2}$.
The \ac{RM} differs across the \ac{SNR}, with positive \ac{RM} values seen predominantly along the western edge with a smaller patch on the north-eastern side, and negative values observed predominantly in the centre, slightly offset to the east.

The magnetic field structure we observe appears highly disorganised. This is consistent with the low fractional polarisation ($\sim$10\%)~\citep{Reynolds2012} and the broad range of \ac{RM} values ($-$200 to +100\,rad\,m$^{-2}$) seen across the remnant. Such characteristics suggest either an intrinsically complex magnetic field structure, potentially indicating magnetic turbulence, or possible depolarisation effects along our line of sight. We note that at the MeerKAT observing frequency, there may be significant uncertainties in these magnetic field orientations due to the high absolute \ac{RM} values observed. Therefore, the interpretation of this magnetic field structure should be taken with caution.

\subsection{Magnetic field}
\label{subsec:magfield}

We can estimate the strength of the magnetic field in Veliki using an adapted equipartition calculation, following the method outlined in \citet{Filipovic2023}. We use the calculated values of spectral index $\alpha$ = 0.26\footnote{This paper primarily uses the notation of $S\propto\nu^\alpha$ which gives a spectral index value of $\alpha$ = $-$0.26. This equation uses the opposite, equally correct, notation of $S\propto\nu^{-\alpha}$, therefore flipping the sign of the radio spectral index value.} and flux density at 1\,GHz $S_{\text{1\,GHz}}$ = 0.66 Jy, and assume values of shock velocity $v$\,=\,170 \kms, speed of sound in the \ac{ISM} $c_s$ = 10\,\kms, 
filling factor $f$\,=\,0.25, compression ratio $r\approx$\,4 and injection parameter $\xi\approx$\,4, for strong non-modified shocks, and equal proton and electron temperatures $T_p = T_e$. The value of $\xi\approx$\,4 assumes a non-modified shock, and the assumed shock velocity in the ambient medium will give a compression ratio $r\approx$\,4, consistent with a strong shock. Assuming electron equipartition, we find $B\,=\,$20.0\,$\mu$G in an ambient field of 6.1 $\mu$G, and assuming proton equipartition we find $B\,=\,$68.9\,$\mu$G in an ambient field of 21.0$\,\mu$G. 

\section{Discussion}
\label{sec:discussion}

\subsection{\snr's size}
\label{subsec:size}

\snr\ was already one of the largest \acp{SNR} in the \ac{LMC}, and the newly visible filamentary structure extends its size even further to 150$\times$81\,pc. There is the possibility that these blowout structures may not represent the \ac{SNR} shell, but could potentially be leaked ionising radiation rather than physical matter. We observe a web-like or patchy structure in the radio emission over the \ac{SNR} (see Fig.~\ref{fig:SNR}) which may indicate a porous \ac{ISM}, expected in lower metallicity environments such as the \ac{LMC}~\citep{Dimaratos2015}. This porosity could facilitate the leakage of ionising radiation through channels of lower density material, naturally creating features that resemble physical blowouts. The spectral index appears to be flatter at the edges (see Fig.~\ref{fig:specindexmap}), however it is difficult to say this with certainty due to the cut used to generate the map, as the uncertainties were significantly higher in the filamentary areas. This leakage is a more likely scenario for the southern filaments, as the optical \SII/\Halpha\ shell does not extend to the outermost radio emission in this direction. Therefore, the southern filamentary structure may not be part of the physical structure. 

Even with this caveat however, \snr\ is still one of the largest in the \ac{LMC}. The only confirmed larger \ac{LMC} \ac{SNR} is MCSNR~J0507$-$6847 with a physical diameter of $\sim$150\,pc~\citep{Chu2000, Maitra2021, Zangrandi2024}, which is exclusively seen at X-ray frequencies. %
\snr's size is large when compared with other galaxies as well. The largest confirmed \ac{SMC} \ac{SNR} is J0056$-$7209~\citep{Haberl2012, Maggi2019} with a physical size of 99$\times$65\,pc~\citep{Maggi2019}.

A physical size comparison is more difficult with the Galactic \ac{SNR} population due to distance uncertainties. There are some Galactic \acp{SNR} that may reach the physical size of \snr, depending on which distance estimate is used. These include G278.94+1.35 (Diprotodon) which may reach up to 157$\times$154\,pc~\citep{Filipovic2024}, G65.1+0.6 which may reach up to 179.5$\times$179.5\,pc~\citep{Vukotic2019}, and G312.4$-$0.4 which may reach up to 154.8$\times$154.8\,pc~\citep{Vukotic2019}. The distances to these Galactic \acp{SNR} are debated in the literature however, and thus their physical sizes may be significantly smaller.

\begin{figure}
    \centering
    \includegraphics[width=\columnwidth]{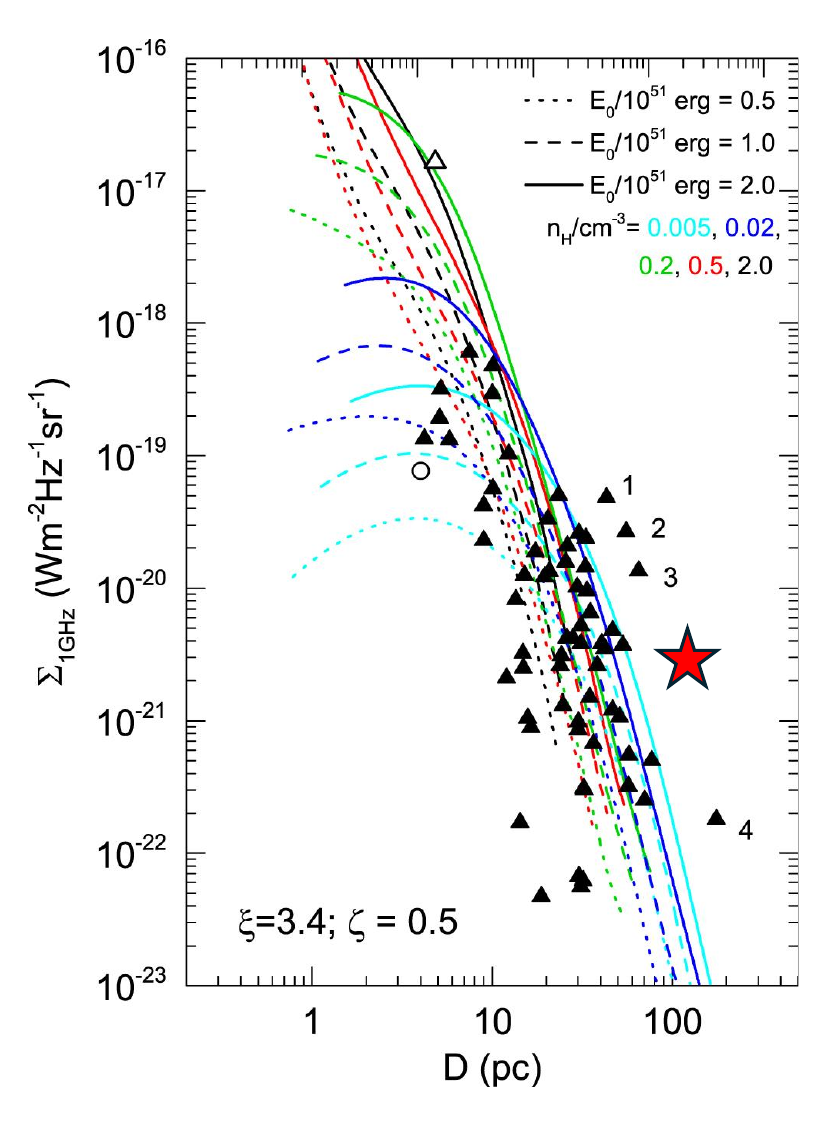}
    \caption{Radio surface brightness to diameter diagram for \acp{SNR} at frequency $\nu$\,=\,1\,GHz, adopted from \protect\citet[][their Fig.~3]{Pavlovic2018}, shown as black triangles. Different line colours represent different ambient densities, while different line types represent different explosion energies. The open circle is young Galactic \ac{SNR} G1.9+0.3~\citep{Luken2020}, and the open triangle represents Cassiopeia~A. The numbers represent \acp{SNR} (1): CTB~37A, (2): Kes~97, (3): CTB~37B, and (4): G65.1+0.6. The red star represents \snr\ at estimated surface brightness of 1.7$\times$10$^{-21}$\,W\,m$^{-1}$\,Hz$^{-2}$\,sr$^{-1}$ and diameter of 150\,pc (the diameter is taken as the major axis). The image shows evolutionary tracks for representative cases with injection parameter $\xi$\,=\,3.4 and nonlinear magnetic field damping parameter $\zeta$\,=\,0.5.}
    \label{fig:sigd}
\end{figure}

We also compare \snr's radio surface brightness with these similarly sized \acp{SNR}. We measure \snr\ as having $\Sigma\,=\,1.7\times10^{-21}$\,W\,m$^{-2}$\,Hz$^{-1}$\,sr$^{-1}$. Of these compared \acp{SNR}, only G312.4$-$0.4 is brighter at $\Sigma$\,=\,4.7$\times$10$^{-21}$\,W\,m$^{-2}$\,Hz$^{-1}$\,sr$^{-1}$~\citep{Vukotic2019}. The other \acp{SNR} compared are approximately an order of magnitude dimmer, with G278.94+1.35 (Diprotodon) having $\Sigma\,=\,1.3\times10^{-22}$\,W\,m$^{-2}$\,Hz$^{-1}$\,sr$^{-1}$~\citep{Filipovic2024}, and G65.1+0.6 having $\Sigma\,=\,2\times10^{-22}$\,W\,m$^{-2}$\,Hz$^{-1}$\,sr$^{-1}$~\citep{Vukotic2019}. %
We also use the $\Sigma$-D relationship of \citet{Pavlovic2018} to compare \snr's properties with those of a larger \ac{SNR} sample (see Figure~\ref{fig:sigd}, from their Fig.~3). \snr\ falls outside of the predicted \ac{SNR} evolutionary tracks, demonstrating that its combination of larger size and brighter radio emission is an outlier in the typical \ac{SNR} sample.

The compared \acp{SNR} are likely expanding in a rarefied environment, for example, in the case of MCSNR~J0507$-$6847~\citep{Chu2000, Maitra2021} and Diprotodon~\citep{Filipovic2024}, whose larger size is attributed to this rarefied ambient density and their older age. %
Conversely, \snr\ is predicted to be expanding into an ambient density of $\sim0.3-0.5$\,cm$^{-3}$~\citep{Williams2004}, and this higher density environment may help to explain the brighter radio emission.

\subsection{Environment}
\label{subsec:environment}

The \ac{LMC} environment has a lower metallicity than the \ac{MW}~\citep{Rolleston2002}, which is theoretically expected to impact the radiative cooling efficiency of \acp{SNR}. As radiative cooling is typically dominated by metal line emission, lower metallicity environments have reduced cooling efficiency~\citep{Thornton1998}. Thus, it would be expected for \ac{LMC} \acp{SNR} to remain in the adiabatic phase for longer than their respective Galactic counterparts. %
More recent \ac{LMC} surveys have also detected metallicity gradients within the \ac{LMC}, with the lowest metallicities observed in the south-west~\citep{Choudhury2021}, corresponding with \snr's location. %
\snr's properties display that it is now dominated by the swept-up \ac{ISM} rather than the inner ejecta, meaning that this delayed cooling may have a significant impact on the observed morphology. Thus, this environment may explain \snr's size and radio brightness if the onset of cooling were delayed.

We also note a molecular cloud PGCC~G282.62-35.35 with \ac{IR} and CO emission detected to the west-north-west of \snr, located just outside of the shell. We see no obvious radio-continuum signs that the \ac{SNR} is significantly interacting, such as limb brightening or distinct changes in morphology at this location, as seen in other \acp{SNR} with molecular cloud interaction (e.g. G166.0+4.3~\citep{Landecker1989}; Puppis A~\citep{Dubner1991}; G119.5+10.3~\citep{Pineault1997};  G18.8+0.3~\citep{Dubner1999}; N63A~\citep{Sano2019}; G46.8–0.3~\citep{Supan2022}; N49~\citep{Sano2023, Ghavam2024}). Veliki also appears in the analysis of the \ac{IR} properties of \acp{SNR} in the \ac{LMC} of \citet{Lakicevic2015}, where it is concluded that it is interacting, or about to interact, with surrounding clouds.
Using ratios of different {\it Herschel} IR data, they show that the dust temperatures and hydrogen column densities are consistent inside and outside of the \ac{SNR}, indicating that there is no current significant dust heating or sputtering. It is possible that their spatial correlation is due to a projection effect, the objects are simply in the same line of sight but at different physical locations in the \ac{LMC}. Conversely, it is possible that there is current interaction but it is not resulting in significant heating and sputtering, due to the slower radiative shocks. Once \snr's shocks have significantly slowed down and entered the radiative phase, it is possible that the interacting cloud material has settled down and entered into equilibrium inside and outside the \ac{SNR}, resulting in the observed \ac{IR} ratios~\citep{Lakicevic2015}. This scenario would also be consistent with the \OIII\ excess observed on the opposite rim compared to the molecular cloud, as this interaction may have caused the western edge to cool down faster due to the interaction.

Previous results also suggest expansion into an ambient density $\sim$0.3$-$0.5\,cm$^{-3}$~\citep{Williams2004} %
 which is well below molecular cloud densities~\citep{Fukui2009}. Therefore, it is possible that \snr\ is interacting with the molecular cloud, however the current data is not conclusive in this regard. Further observations, particularly in the $\gamma$-ray domain, would be useful for detecting and constraining any molecular cloud interaction.%

The integrated \HI\ data (see Fig.~\ref{fig:HICO}, panel (a)), shows a density gradient across the \ac{SNR}. Denser \HI\ material is seen on the western rim compared to the eastern, and the eastern rim shows a smoother radio-continuum structure. This density gradient is supported by \snr's elliptical morphology which is more constrained on the east and west edges with an observed blowout structure in the north and south. Such blowout structures are observed in other \acp{SNR} with such density gradients (e.g. SNR~J0455$-$6838~\citep{Crawford2008}; SNR~S147~\citep{Xiao2008}; G309.2$-$00.6~\citep{Gaensler1998}). We also observe an excess of \OIII\ emission on the eastern edge, corresponding with the lower density medium. This emission may be due to a faster, more energetic shock in this area, or it may be due to this rim cooling slower than the rest of the \ac{SNR}, both scenarios which are consistent with expansion into a lower density. These results strongly support a density gradient in the surrounding medium.

We also detect an expanding \HI\ shell in the $p$--$v$ \HI\ diagram at the \ac{LMC} velocity range (see Fig.~\ref{fig:HICO}, panel (b)). Such a cavity can be caused if the progenitor star's stellar winds shaped the surrounding environment. In the case of Veliki, this may have formed an axialsymmetric cavity into which the expansion occurred, explaining the barrel-like morphology and observed north-south outflows. This has been observed in some \acp{SNR} such as G290.1$-$0.8~\citep{Slane2002}, G166.0$+$4.3~\citep{Bocchino2009}, and W49B~\citep{Zhu2014}. For an \ac{SNR} as evolved as Veliki, the expansion likely would have already reached the boundary of the initial cavity, as shown by the spatial correlation within Veliki's shell, and thus the expanding \HI\ may be a combination of the initial progenitor's stellar wind and the expanding shockfront.

\subsection{Evolutionary phase}
\label{subsec:evolutionary phase}

\snr's optical properties are consistent with an almost fully radiative \ac{SNR}, with 
the \SII/\Halpha\ emission forming a filamentary shell across the entire \ac{SNR}. %
The \Halpha\ emission indicates areas where the post-shock gas has cooled sufficiently to enable recombination~\citep{Raymond1979, Vink2020}. The \SII\ emission acts as a tracer of radiative shocks~\citep{Vucetic2023}, and the strong \SII/\Halpha\ correlation indicates predominantly radiative shocks. 

One possible exception to this is the strong \OIII\ emission observed in the eastern shell. This emission may be caused by faster, more energetic shocks in this region~\citep{Dopita1996}. This faster expansion would likely be due to the external density gradient, and may indicate a region of non-radiative expansion. %
Older \acp{SNR} can display both radiative and non-radiative shock fronts due to their complex morphology~\citep{Vink2020}, even when the whole \ac{SNR} is predominantly radiative. This phenomenon has been studied in detail for some \acp{SNR}, such as RCW~86~\citep{Vink2006, Yamaguchi2016} and Cygnus Loop~\citep{Salvesen2009, Vucetic2023}. 

Conversely, the observed \OIII\ emission may be due to different cooling rates over the \ac{SNR}. Due to the lower density on this side of the \ac{SNR}, this scenario may be more likely than the stronger shock scenario. While the shell may expand faster within a lower density medium, if there is not sufficient ambient material present, then strong shocks capable of producing \OIII\ emission are less likely. On the other hand, expansion into a lower density will reduce the cooling rate, and thus this edge may be cooling more slowly than the rest of the \ac{SNR}. In this case, the \OIII\ emission would be due to higher temperatures in this area, and would not necessarily indicate non-radiative shocks.

Thus, the optical morphology indicates that \snr\ is either fully radiative with an area of higher temperatures on the eastern side, or predominantly radiative with a small area of non-radiative expansion. It should be noted that the previous X-ray models~\citep{Williams2004} indicate that the interior X-ray emission still has sufficient thermal pressure to drive expansion, suggesting that \snr\ may not be solely momentum-driven, and thus could potentially support areas of non-radiative expansion.

\subsubsection{Evolutionary modelling}
\label{subsubsec:evolution model}

We also apply the \ac{SNR} evolution model of \cite{2017AJ....153..239L} and \cite{2019AJ....158..149L} using the measured X-ray temperature and emission measure from \citet{Maggi2016} with the new mean radius measured from the high-resolution MeerKAT data.
Standard forward shock models, with various ejecta power-law indices and the cloudy \ac{ISM} model, appropriate to \ac{MM} \acp{SNR}, were explored. 
All models yielded explosion energies of $\simeq8.6\times10^{51}$ ergs, ages of $\simeq$43000 yr, and \ac{ISM} densities of $7\times10^{-2}$ cm$^{-3}$ (standard models) or 2 to $5\times10^{-2}$ cm$^{-3}$ (cloudy models). The estimated age ranges are in agreement with the previous results of \citet{Williams2004}, however the modelled ambient densities are approximately an order of magnitude lower. The previous measurements relied on measuring the \Halpha\ emission measure in the filamentary optical shell, while the model results give the intercloud density of the \ac{ISM} as described in~\citep{White1991}.  This may explain the difference in density estimates if the \Halpha\ filaments are representing radiative shocks as they propagate through the denser intercloud medium.

The high explosion energy is required to reproduce \snr's larger size with the measured high emission measure. This higher explosion energy scenario is also supported by \snr's location on the $\Sigma$-D evolutinary diagram (see Fig.~\ref{fig:sigd}). While \snr\ is an outlier in this distribution, it is offset to the higher explosion energy side.
Such high explosion energies are not uncommon, e.g. based on chemical evolution models of the Galaxy \citep{2006ApJ...653.1145K}, and models of \ac{SNR} populations \citep{2017ApJ...837...36L,2020ApJS..248...16L,2022ApJ...931...20L}.
The models also yield the transition times to the radiative phase, which are $\simeq$95,000 yr. This would suggest that \snr\ is not fully radiative despite the observed radiative shell. %
This discrepancy can be explained in the cloudy model if the shocks are propagating within the clouds. Propagation through this denser material could cause these shocks to become radiative before the intercloud shocks, and thus could explain how we can see a filamentary radiative shell prior to the transition time.  

This scenario would  result in a higher compression ratio in the clouds and would explain the optical filamentary structure as well as the higher-than-expected radio surface brightness. An issue with this scenario is the absence of non-radiative filaments over the \ac{SNR}. If these radiative shocks were attributed only to the propagation through the clouds, we would still expect to see the non-radiative filaments from the intercloud regions. This would likely result in a scenario where we see both radiative and non-radiative shocks anti-correlating with each other, a scenario not observed in \snr. The absence of any observed non-radiative shocks may argue against this scenario but does not rule it out.

\subsection{Spectral Index}
\label{subsec:spec index discussion}

The measured spectral index value of $\alpha\,=\,-0.26\pm0.02$ is flatter than expected for an \ac{SNR}. This value is flatter than the previously measured value of $\alpha$\,=\,$-$0.43$\pm$0.06~\citep{Cajko2009}, as well as flatter than the observed average for the \ac{LMC} population~\citep{Maggi2016, Bozzetto2017} and the theoretical value of $\alpha\,=\,-0.5$~\citep{Bell1978}. 

The observed spectral index values of \acp{SNR} are typically attributed to \ac{DSA} theory~\citep{Bell1978, Reynolds1992, Urosevic2014}. Standard \ac{DSA} theory predicts that the spectral index should steepen as an \ac{SNR} ages, as the particle acceleration becomes less efficient and thus the corresponding shock and compression ratio becomes weaker~\citep{Onic2013, Urosevic2014}. This is observed in the majority of evolved \acp{SNR} which exhibit a spectral index $-0.6<\alpha<-0.5$~\citep{Ferrand2012, Ranasinghe2023, Green2025}. However, there are several possible mechanisms that can act to flatten an \ac{SNR}'s spectral index.

\subsubsection{Morphology types with flatter indices}
\label{subsubsec:SNR types with flatter indices}

Some \ac{SNR} morphology types are naturally expected to have flatter spectral indices and this is observed in numerous \acp{SNR}~\citep{Ferrand2012, Ranasinghe2023, Green2025}. For example, composite \acp{SNR} host a central \ac{PWN}, and are expected to have a flatter overall spectral index due to the \ac{PWN} contribution %
(average $\alpha\,=\,-0.41$ for Galactic population~\citep{Ranasinghe2023}). %
We observe no obvious \ac{PWN} within the radio images (see Figure~\ref{fig:SNR}) nor any spectral variation (see Figure~\ref{fig:specindexmap}) that may indicate this possibility. There is also no observed radio or X-ray point source that may indicate a central pulsar. \snr\ appears as a typical shell-type \ac{SNR}, and thus we deem this scenario unlikely.

Another class of \acp{SNR} with theoretically predicted flatter spectral indices are the \ac{MM} (or filled-centre or thermal composite) remnants~\citep{Rho1998}. While \snr\ is classified as a \ac{MM} \ac{SNR}, their flatter spectral index (average $\alpha\,=\,-0.34$ for Galactic population~\citep{Ranasinghe2023}) is generally attributed to environmental interactions~\citep{Rho1998}. $\gamma$-ray observations could help constrain this possibility, particularly as they could determine potential molecular cloud interaction. For example, $\gamma$-ray observations consistent with hadronic emission from neutral pion decay would give support to the molecular cloud interaction scenario~\citep{Uchiyama2010}, but this data is not currently available. Future $\gamma$-ray observations could help constrain this scenario. As there is not sufficient evidence to determine interaction with a denser environment (see Section~\ref{subsec:environment}), and so this scenario is left as a possibility.%

\subsubsection{Second-order Fermi acceleration}
\label{subsubsec:second order fermi}

One suggested mechanism that may cause radio spectral index flattening is the combination of \ac{DSA} with second-order Fermi acceleration~\citep{Onic2013, Urosevic2014}. In this scenario, particles initially accelerated by the shock front undergo further stochastic acceleration through interactions with magnetic turbulence in the post-shock region~\citep{Liu2008, Cho2006, Petrosian2004}. %
This mechanism depends on magnetic field strengths and orientations, and is thus highly sensitive to magnetic field variations~\citep{Cho2006, Fan2010}. 

Thus, we would expect the magnetic field turbulence and varying shock conditions to correlate with spectral index variations, which is not observed in \snr. If second-order Fermi acceleration were the dominant mechanism, then we would expect to see variation in the spectral index map (see Fig.~\ref{fig:specindexmap}) which correlate with the magnetic field turbulence (rotation measure and fractional polarisation maps, see Fig.~\ref{fig:polarisation}). If the \OIII\ emission does represent non-radiative shocks, then we would also expect to see correlation with these varying shock conditions. As no such correlations are seen, this argues against second-order Fermi acceleration as a dominant mechanism.

\subsubsection{Thermal contribution}
\label{subsubsec:thermal contribution}

Some evolved \acp{SNR} can have a significant thermal bremsstrahlung contribution which flattens the spectral index as they sweep up significant amounts of interstellar material~\citep{Urosevic2005, Onic2012, Onic2013, Urosevic2014}. Additionally, \ac{MM} \acp{SNR} can have a significant thermal bremsstrahlung contribution, either attributed to denser environment interactions which is typical for \ac{MM} \acp{SNR}, or contribution from the X-ray filled interior~\citep{Onic2012, Vink2012}. There is no evidence for sufficiently dense environmental interactions following the arguments from previous sections (see Sections~\ref{subsec:environment} for prior discussion). It is also unlikely that any significant thermal contribution is attributed to the thermal X-ray emission, as this would cause spectral variation to correlate with the interior X-ray emission, which is not observed.

Thermal bremsstrahlung contribution is theoretically expected for evolutionarily old \acp{SNR}, particularly \acp{SNR} in the radiative phase of evolution. Radiative \acp{SNR} with a high compression ratio can have sufficient amounts of high density ionised material in the downstream region to generate significant thermal bremsstrahlung contribution. 
To test this scenario, we model the spectral index of \snr\ as a combination of thermal and non-thermal components using a two-component model similar to the method of~\citet{Onic2012}, where we define $\alpha_{\text{ total}}\,=\,\alpha_{\text{thermal}}+\alpha_{\text{non-thermal}}$. For a rough estimate, we restrict the expected values to $\alpha_{\text{thermal}}\,=\,-0.1$ and $\alpha_{\text{non-thermal}}\,=\,-0.5$. This model required a thermal contribution of 58.6\% at 1\,GHz and achieves a reduced $\chi^2$ value of 0.664, representing a relatively good fit for the data. This is a significant thermal contribution, but is still possible from older \acp{SNR}. For example, the analysis of \citet{Onic2012} show a thermal contribution for IC~443 of 3$-$57\%~\citep{Onic2012}. %

Some potential problems with this scenario are the lack of spectral curvature and the absence of a low-frequency spectral turnover in our observed spectrum (see Figure~\ref{fig:specindexgraph}). Thermal bremsstrahlung contribution is expected to begin to dominate at higher frequencies, and can result in a flattening at higher frequencies and thus a curved spectrum~\citep{Reynolds2008, Onic2012, Urosevic2014}, as observed in other \acp{SNR} (e.g. IC~443, 3C~391, 3C~396~\citep{Onic2012}; HB3~\citep{Tian2005, Urosevic2007, Onic2008}). We measure the flux densities over a broad frequency range (88$-$8850\,GHz) and observe no such spectral curvature. The absence of spectral curvature does not preclude thermal bremsstrahlung contribution however, as the amount of curvature can be very low and can easily be lost in the scatter of the data points. 

Low-frequency spectral turnovers are observed in some \acp{SNR}, and can be a sign of possible thermal bremsstrahlung emission at higher frequencies, such as in the cases of 3C~391 and IC~443~\citep{Brogan2005,Onic2012}. The absence of such a turnover in \snr's spectrum does not preclude thermal bremsstrahlung contribution however. In evolved \acp{SNR}, the thermal plasma can be more diffuse, thus having a lower optical depth and potentially shifting the absorption turnover to frequencies below our observational range ($<$88\,MHz). This thermal emission can still contribute at higher radio frequencies where bremsstrahlung dominates over absorption effects, thus resulting in an overall flatter spectral index with no observed spectral turnover. This is consistent with a more evolved \ac{SNR} such as \snr, where the plasma 
may be more diffuse than in younger \acp{SNR} such as 3C~391 and IC~443. Additional spectral index data points in the  microwave and infrared domain would significantly help to constrain the integrated radio spectral index and thus help to distinguish between synchrotron and thermal bremsstrahlung emission components.

Both spectral curvature and a low-frequency spectral turnover are seen in some \acp{SNR} with thermal contribution, however their absence does not preclude this possibility, particularly for an evolved \ac{SNR}. Therefore, thermal contribution is likely a contributing factor to the spectral index flattening.

\subsubsection{Higher compression ratio}
\label{subsubsec:higher compression ratio}

The final mechanism we consider is that of spectral index flattening caused by a high compression ratio. In standard \ac{DSA} theory, the energy spectrum of accelerated particles is affected by the shock compression ratio~\citep{Onic2013, Urosevic2014}. Typically, the relationship between the radio spectral index $\alpha$ and compression ratio $r$ is given by $\alpha\,=3/(2(r-1))$. Our measured spectral index of $\alpha\,=\,0.26\pm0.02$\footnote{This paper primarily uses the notation of $S\propto\nu^{\alpha}$ which gives a spectral index value of $\alpha\,=\,-0.26$. This compression ratio equation is derived from \ac{DSA} theory and thus uses the opposite, equally correct, notation of $S\propto\nu^{-\alpha}$, therefore flipping the sign of the radio spectral index value.} corresponds to a compression ratio of $r\sim6.8$, thus exceeding the standard limit of $r\,=\,4$ for strong adiabatic shocks in an ideal gas. Higher compression ratios are theoretically expected in fully radiative shocks, with $r\,=\,M_T^2$, where $M_T$ is the isothermal Mach number~\citep{Urosevic2014}. For example, for radiative shocks of $\sim$100$-$200\,km\,s$^{-1}$, the compression ratio can reach values as high as $\sim$100$-$400 (assuming a typical speed of sound in the \ac{ISM} of $v_c\sim10$\kms). Therefore, if \ac{DSA} theory is still accurate for fully radiative shocks, this is expected to significantly flatten the particle spectrum and naturally produce a flatter radio spectral index.
See the studies of \citet{Onic2013} and \citet{Urosevic2014} and references therein for more theoretical details.

Multiple lines of evidence support this scenario for \snr. Older \acp{SNR} such as \snr\ are expected to naturally have higher compression ratios as the shocks become radiative and the expansion slows down. Therefore, \snr's prominent radiative shell may indicate significant deceleration in the shell and a higher compression. The derived magnetic field is slightly higher than expected for older \acp{SNR}. For example, the Galactic \acp{SNR} Ancora and Diprotodon have predicted values of 7.7\,$\mu$G (for electron equipartition) or 41.7\,$\mu$G (for proton equipartition), and 13.7$-$19.6\,$\mu$G (depending on distance estimate used) respectively. The higher magnetic field observed in \snr\ can be explained by a higher compression ratio, as this compression would also compress, and thus strengthen, the swept-up magnetic field.

This scenario is also supported by the uniform spectral index. As the compression ratio increase is due to the radiative shell which encompasses the entire remnant, it would naturally result in a roughly equal flattening over the entire \ac{SNR}.

\subsection{What is Veliki?}
\label{subsec: summary}

Several mechanisms can be responsible for \snr's observed properties, particularly the bright radio surface brightness and flat spectral index. With the evidence available we determine that the most likely scenario is that of a fully radiative \ac{SNR}, with spectral flattening due to thermal bremsstrahlung contribution and a higher compression ratio.

This scenario is the simplest model to explain the observed properties. It posits that \snr\ is fully within the radiative phase, and these radiative shocks have sufficiently cooled and decelerated to increase compression. This higher compression ratio would provide high density ionised material in the downstream region, resulting in thermal bremsstrahlung contribution. It would also compress the swept-up magnetic field, increasing magnetic field strength and synchrotron emission in the shell. This combination would result in the high radio surface brightness.

\snr's larger size can be attributed to its expansion into the lower metallicity environment of the \ac{LMC}, where the cooling would likely have been delayed. Thus \snr\ would remain in the adiabatic phase for longer and expand to a larger size. The asymmetric morphology indicates that there is an external density gradient, with denser material along the east-west axis and less dense material in the north-south, resulting in the blowout and barrel-shaped structure. It is possible that \snr\ is also the result of a higher energy explosion, as predicted by the previous models (see Section.~\ref{subsec:evolutionary phase}), which would help explain the brighter radio surface brightness and larger size.

\section{Conclusions}
\label{sec:conclusion}

The new high-resolution radio observations of the \ac{LMC} \ac{SNR} \snr\ show previously unseen faint filamentary structures extending out of the north and south of the remnant. This increases the extent of \snr\ to 150$\times$81\,pc, making it one of the largest \acp{SNR} in the \ac{LMC}. Additionally, it has an unusually high radio surface-brightness, approximately an order of magnitude brighter than other \acp{SNR} or comparable size, and has one of the lowest average radio spectral indices ($\alpha\,=\,-0.26\pm0.02$). We observe a bright \SII/\Halpha\ shell over the remnant indicating predominantly radiative shocks, as well as higher temperatures or faster, possibly non-radiative, shocks on the eastern rim shown by an excess of \OIII\ emission.

We investigate several theoretical scenarios to explain these unusual properties. We determine that \snr\ is most likely a fully radiative \ac{SNR} and the flatter spectral index and brighter surface brightness is attributed to a higher compression ratio and possible thermal bremsstrahlung contamination. We also consider the alternative scenario of the observed properties resulting from expansion into a cloudy \ac{ISM}, resulting in a variable compression ratio and causing the shocks to become radiative and flattening the spectral index. 

\snr\ may also have had a higher explosion energy ($\sim$8.6$\times$10$^{51}$\,ergs), explaining the brighter radio surface brightness and larger size. The larger size can also be explained by expansion into the lower metallicity \ac{LMC} environment. Both the listed scenarios can potentially explain \snr's properties, and more observations are required to fully differentiate between them. In particular, observations that can better constrain \snr's environment, are vital to fully determine \snr's nature.

\section*{Acknowledgements}

The MeerKAT telescope is operated by the South African Radio Astronomy Observatory, which is a facility of the National Research Foundation, an agency of the Department of Science and Innovation.

This scientific work uses data obtained from Inyarrimanha Ilgari Bundara / the Murchison Radio-astronomy Observatory. We acknowledge the Wajarri Yamaji People as the Traditional Owners and native title holders of the Observatory site. CSIRO’s ASKAP radio telescope is part of the Australia Telescope National Facility\footnote{\label{foot:ATNF}\url{http://www.atnf.csiro.au}}. Operation of ASKAP is funded by the Australian Government with support from the National Collaborative Research Infrastructure Strategy. ASKAP uses the resources of the Pawsey Supercomputing Research Centre. Establishment of ASKAP, Inyarrimanha Ilgari Bundara, the CSIRO Murchison Radio-astronomy Observatory and the Pawsey Supercomputing Research Centre are initiatives of the Australian Government, with support from the Government of Western Australia and the Science and Industry Endowment Fund.

Support for the operation of the MWA is provided by the Australian Government (NCRIS), under a contract to Curtin University administered by Astronomy Australia Limited. 

The National Radio Astronomy Observatory is a facility of the
National Science Foundation, operated under a cooperative agreement
by Associated Universities, Inc.

DU and BA acknowledge the financial support provided by the Ministry of Science, Technological Development and Innovation of the Republic of Serbia through the contract 451-03-136/2025-03/200104  and for support through the joint project of the Serbian Academy of Sciences and Arts and Bulgarian Academy of Sciences  "Optical search for Galactic and extragalactic supernova remnants''. BA additionally acknowledges the funding provided by the Science Fund of the Republic of Serbia through project \#7337 ”Modeling Binary Systems That End in Stellar Mergers and Give Rise to Gravitational Waves” (MOBY).
HS acknowledges the financial support provided by the Japan Society for the Promotion of Science (JSPS) KAKENHI Grant Numbers JP20KK0309 and JP24H00246.
HS acknowledges the financial support provided by the Japan Society for the Promotion of Science (JSPS) KAKENHI Grant Numbers JP20KK0309 and JP24H00246.

\newcommand\eprint{in press }

\bibsep=0pt

\bibliographystyle{aa_url_saj}

{\small

\bibliography{0450}

\begin{thebibliography}{139}
\expandafter\ifx\csname natexlab\endcsname\relax\def\natexlab#1{#1}\fi

\bibitem[{{Acharyya} {et~al.}(2023){Acharyya}, {Adam}, {Aguasca-Cabot}, {Agudo}, {Aguirre-Santaella}, {Alfaro}, {Aloisio}, {Alves Batista}, {Amato}, {Ang{\"u}ner}, {Aramo}, {Arcaro}, {Asano}, {Aschersleben}, {Ashkar}, {Backes}, {Baktash}, {Balazs}, {Balbo}, {Ballet}, {Bamba}, {Baquero Larriva}, {Barbosa Martins}, {Barres de Almeida}, {Barrio}, {Bastieri}, {Batista}, {Batkovic}, {Baxter}, {Becerra Gonz{\'a}lez}, {Becker Tjus}, {Benbow}, {Bernardini}, {Bernardos Mart{\'\i}n}, {Bernete Medrano}, {Berti}, {Bertucci}, {Beshley}, {Bhattacharjee}, {Bhattacharyya}, {Bigongiari}, {Biland}, {Bissaldi}, {Bocchino}, {Bordas}, {Borkowski}, {Bottacini}, {B{\"o}ttcher}, {Bradascio}, {Brown}, {Bulgarelli}, {Burmistrov}, {Caroff}, {Carosi}, {Carqu{\'\i}n}, {Casanova}, {Cascone}, {Cassol}, {Cerruti}, {Chadwick}, {Chaty}, {Chen}, {Chiavassa}, {Chytka}, {Conforti}, {Cortina}, {Costa}, {Costantini}, {Cotter}, {Crestan}, {Cristofari}, {D'Ammando}, {Dalchenko}, {Dazzi}, {De Angelis}, {De Caprio}, {de Gouveia Dal Pino}, {De
  Martino}, {de Naurois}, {de Souza}, {del Valle}, {Delgado Giler}, {Delgado}, {della Volpe}, {Depaoli}, {Di Girolamo}, {Di Piano}, {Di Pierro}, {Di Tria}, {Di Venere}, {Diebold}, {Doro}, {Dumora}, {Dwarkadas}, {Eckner}, {Egberts}, {Emery}, {Escudero}, {Falceta-Goncalves}, {Fedorova}, {Fegan}, {Feng}, {Ferenc}, {Ferrand}, {Fiandrini}, {Filipovic}, {Fioretti}, {Foffano}, {Fontaine}, {Fukui}, {Gaggero}, {Galanti}, {Galaz}, {Gallozzi}, {Gammaldi}, {Garczarczyk}, {Gasbarra}, {Gasparrini}, {Ghalumyan}, {Giarrusso}, {Giavitto}, {Giglietto}, {Giordano}, {Giuliani}, {Glicenstein}, {Goldoni}, {Goulart Coelho}, {Granot}, {Green}, {Green}, {Grondin}, {Gueta}, {Hadasch}, {Hamal}, {Hassan}, {Hayashi}, {Heller}, {Hern{\'a}ndez Cadena}, {Hiroshima}, {Hnatyk}, {Hnatyk}, {Hofmann}, {Holder}, {Holler}, {Horan}, {Horvath}, {Hrabovsky}, {H{\"u}tten}, {Iarlori}, {Inada}, {Incardona}, {Inoue}, {Iocco}, {Jamrozy}, {Jin}, {Jung-Richardt}, {Jury{\v{s}}ek}, {Kantzas}, {Karas}, {Katagiri}, {Kerszberg}, {Kn{\"o}dlseder}, {Komin},
  {Kornecki}, {Kosack}, {Kowal}, {Kubo}, {Lamastra}, {Lapington}, {Lemoine-Goumard}, {Lenain}, {Leone}, {Leto}, {Leuschner}, {Lindfors}, {Lohse}, {Lombardi}, {Longo}, {L{\'o}pez-Coto}, {L{\'o}pez-Oramas}, {Loporchio}, {Luque-Escamilla}, {Macias}, {Majumdar}, {Mandat}, {Mangano}, {Manic{\`o}}, {Mariotti}, {Marquez}, {Marsella}, {Mart{\'\i}}, {Martin}, {Mart{\'\i}nez}, {Mazin}, {Menchiari}, {Meyer}, {Miceli}, {Miceli}, {Micha{\l}owski}, \& {Mitchell}}]{Acharyya2023}
{Acharyya}, A., {Adam}, R., {Aguasca-Cabot}, A., {et~al.} 2023, \href{https://ui.adsabs.harvard.edu/abs/2023MNRAS.523.5353A}{\mnras}, \href{https://ui.adsabs.harvard.edu/abs/2023MNRAS.523.5353A}{523, 5353}

\bibitem[{{Anderson} {et~al.}(2025){Anderson}, {Camilo}, {Faerber}, {Bietenholz}, {Bordiu}, {Bufano}, {Chibueze}, {Cotton}, {Ingallinera}, {Loru}, {Rigby}, {Riggi}, {Thompson}, {Trigilio}, {Umana}, \& {Williams}}]{Anderson2025}
{Anderson}, L.~D., {Camilo}, F., {Faerber}, T., {et~al.} 2025, \href{https://ui.adsabs.harvard.edu/abs/2025A&A...693A.247A}{\aap}, \href{https://ui.adsabs.harvard.edu/abs/2025A&A...693A.247A}{693, A247}

\bibitem[{{Ball} {et~al.}(2023){Ball}, {Kothes}, {Rosolowsky}, {West}, {Becker}, {Filipovi{\'c}}, {Gaensler}, {Hopkins}, {Koribalski}, {Landecker}, {Leahy}, {Marvil}, {Sun}, {Bufano}, {Carretti}, {Ingallinera}, {Van Eck}, \& {Willis}}]{Ball2023}
{Ball}, B.~D., {Kothes}, R., {Rosolowsky}, E., {et~al.} 2023, \href{https://ui.adsabs.harvard.edu/abs/2023MNRAS.524.1396B}{\mnras}, \href{https://ui.adsabs.harvard.edu/abs/2023MNRAS.524.1396B}{524, 1396}

\bibitem[{Bell(1978)}]{Bell1978}
Bell, A.~R. 1978, Monthly Notices of the Royal Astronomical Society, 182, 147

\bibitem[{{Blair} {et~al.}(2006){Blair}, {Ghavamian}, {Sankrit}, \& {Danforth}}]{Blair2006}
{Blair}, W.~P., {Ghavamian}, P., {Sankrit}, R., and {Danforth}, C.~W. 2006, \href{https://ui.adsabs.harvard.edu/abs/2006ApJS..165..480B}{\apjs}, \href{https://ui.adsabs.harvard.edu/abs/2006ApJS..165..480B}{165, 480}

\bibitem[{{Bocchino} {et~al.}(2009){Bocchino}, {Miceli}, \& {Troja}}]{Bocchino2009}
{Bocchino}, F., {Miceli}, M., and {Troja}, E. 2009, \href{https://ui.adsabs.harvard.edu/abs/2009A&A...498..139B}{\aap}, \href{https://ui.adsabs.harvard.edu/abs/2009A&A...498..139B}{498, 139}

\bibitem[{{Bozzetto} {et~al.}(2015){Bozzetto}, {Filipovic}, {Haberl}, {Sasaki}, {Kavanagh}, {Maggi}, {Urosevic}, \& {Sturm}}]{2015PKAS...30..149B}
{Bozzetto}, L.~M., {Filipovic}, M.~D., {Haberl}, F., {et~al.} 2015, \href{https://ui.adsabs.harvard.edu/abs/2015PKAS...30..149B}{Publication of Korean Astronomical Society}, \href{https://ui.adsabs.harvard.edu/abs/2015PKAS...30..149B}{30, 149}

\bibitem[{{Bozzetto} {et~al.}(2023){Bozzetto}, {Filipovi{\'c}}, {Sano}, {Alsaberi}, {Barnes}, {Boji{\v{c}}i{\'c}}, {Brose}, {Chomiuk}, {Crawford}, {Dai}, {Ghavam}, {Haberl}, {Hill}, {Hopkins}, {Ingallinera}, {Jarrett}, {Kavanagh}, {Koribalski}, {Kothes}, {Leahy}, {Lenc}, {Leonidaki}, {Maggi}, {Maitra}, {Matthew}, {Payne}, {Pennock}, {Points}, {Reid}, {Riggi}, {Rowell}, {Sasaki}, {Safi-Harb}, {van Loon}, {Tothill}, {Uro{\v{s}}evi{\'c}}, \& {Zangrandi}}]{Bozzetto2023}
{Bozzetto}, L.~M., {Filipovi{\'c}}, M.~D., {Sano}, H., {et~al.} 2023, \href{https://ui.adsabs.harvard.edu/abs/2023MNRAS.518.2574B}{\mnras}, \href{https://ui.adsabs.harvard.edu/abs/2023MNRAS.518.2574B}{518, 2574}

\bibitem[{{Bozzetto} {et~al.}(2017){Bozzetto}, {Filipovi{\'c}}, {Vukoti{\'c}}, {Pavlovi{\'c}}, {Uro{\v{s}}evi{\'c}}, {Kavanagh}, {Arbutina}, {Maggi}, {Sasaki}, {Haberl}, {Crawford}, {Roper}, {Grieve}, \& {Points}}]{Bozzetto2017}
{Bozzetto}, L.~M., {Filipovi{\'c}}, M.~D., {Vukoti{\'c}}, B., {et~al.} 2017, \href{https://ui.adsabs.harvard.edu/abs/2017ApJS..230....2B}{\apjs}, \href{https://ui.adsabs.harvard.edu/abs/2017ApJS..230....2B}{230, 2}

\bibitem[{{Bradley} {et~al.}(2025){Bradley}, {Smeaton}, {Tothill}, {Filipovi{\'c}}, {Becker}, {Hopkins}, {Koribalski}, {Lazarevi{\'c}}, {Leahy}, {Rowell}, {Velovi{\'c}}, \& {Uro{\v{s}}evi{\'c}}}]{Bradley2025}
{Bradley}, A.~C., {Smeaton}, Z.~J., {Tothill}, N.~F.~H., {et~al.} 2025, {arXiv e-prints, \href{https://ui.adsabs.harvard.edu/abs/2025arXiv250205299B}{}\eprint  \href{}{(DOI:  10.48550/arXiv.2502.05299)}, arXiv:2502.05299}

\bibitem[{{Brogan} {et~al.}(2005){Brogan}, {Lazio}, {Kassim}, \& {Dyer}}]{Brogan2005}
{Brogan}, C.~L., {Lazio}, T.~J., {Kassim}, N.~E., and {Dyer}, K.~K. 2005, \href{https://ui.adsabs.harvard.edu/abs/2005AJ....130..148B}{\aj}, \href{https://ui.adsabs.harvard.edu/abs/2005AJ....130..148B}{130, 148}

\bibitem[{{Burger-Scheidlin} {et~al.}(2024){Burger-Scheidlin}, {Brose}, {Mackey}, {Filipovi{\'c}}, {Goswami}, {Guillen}, {de O{\~n}a Wilhelmi}, \& {Sushch}}]{BurgerSchiedlin2024}
{Burger-Scheidlin}, C., {Brose}, R., {Mackey}, J., {et~al.} 2024, \href{https://ui.adsabs.harvard.edu/abs/2024A&A...684A.150B}{A\&A}, \href{https://ui.adsabs.harvard.edu/abs/2024A&A...684A.150B}{684, A150}

\bibitem[{{\v Cajko} {et~al.}(2009){\v Cajko}, {Crawford}, \& {Filipovi\'c}}]{Cajko2009}
{\v Cajko}, K.~O., {Crawford}, E.~J., and {Filipovi\'c}, M.~D. 2009, \href{https://ui.adsabs.harvard.edu/abs/2009SerAJ.179...55C}{Serbian Astronomical Journal}, \href{https://ui.adsabs.harvard.edu/abs/2009SerAJ.179...55C}{179, 55}

\bibitem[{{Cho} and {Lazarian}(2006){Cho} \& {Lazarian}}]{Cho2006}
{Cho}, J. and {Lazarian}, A. 2006, \href{https://ui.adsabs.harvard.edu/abs/2006ApJ...638..811C}{\apj}, \href{https://ui.adsabs.harvard.edu/abs/2006ApJ...638..811C}{638, 811}

\bibitem[{{Choudhury} {et~al.}(2021){Choudhury}, {de Grijs}, {Bekki}, {Cioni}, {Ivanov}, {van Loon}, {Miller}, {Niederhofer}, {Oliveira}, {Ripepi}, {Sun}, \& {Subramanian}}]{Choudhury2021}
{Choudhury}, S., {de Grijs}, R., {Bekki}, K., {et~al.} 2021, \href{https://ui.adsabs.harvard.edu/abs/2021MNRAS.507.4752C}{\mnras}, \href{https://ui.adsabs.harvard.edu/abs/2021MNRAS.507.4752C}{507, 4752}

\bibitem[{{Chu} {et~al.}(2000){Chu}, {Kim}, {Points}, {Petre}, \& {Snowden}}]{Chu2000}
{Chu}, Y.-H., {Kim}, S., {Points}, S.~D., {Petre}, R., and {Snowden}, S.~L. 2000, \href{https://ui.adsabs.harvard.edu/abs/2000AJ....119.2242C}{\aj}, \href{https://ui.adsabs.harvard.edu/abs/2000AJ....119.2242C}{119, 2242}

\bibitem[{{Clarke} {et~al.}(1976){Clarke}, {Little}, \& {Mills}}]{Clarke1976}
{Clarke}, J.~N., {Little}, A.~G., and {Mills}, B.~Y. 1976, \href{https://ui.adsabs.harvard.edu/abs/1976AuJPA..40....1C}{Australian Journal of Physics Astrophysical Supplement}, \href{https://ui.adsabs.harvard.edu/abs/1976AuJPA..40....1C}{40, 1}

\bibitem[{{Comrie} {et~al.}(2021){Comrie}, {Wang}, {Hsu}, {Moraghan}, {Harris}, {Pang}, {Pi{\'n}ska}, {Chiang}, {Simmonds}, {Chang}, {Jan}, \& {Lin}}]{Comrie2021}
{Comrie}, A., {Wang}, K.-S., {Hsu}, S.-C., {et~al.} 2021, \href{https://ui.adsabs.harvard.edu/abs/2021ascl.soft03031C}{{CARTA: Cube Analysis and Rendering Tool for Astronomy}}, Astrophysics Source Code Library, record ascl:2103.031

\bibitem[{{Cotton} {et~al.}(2024{\natexlab{a}}){Cotton}, {Filipovi{\'c}}, {Camilo}, {Indebetouw}, {Alsaberi}, {Anih}, {Baker}, {Bastian}, {Boji{\v{c}}i{\'c}}, {Carli}, {Cavallaro}, {Crawford}, {Dai}, {Haberl}, {Levin}, {Luken}, {Pennock}, {Rajabpour}, {Stappers}, {van Loon}, {Zijlstra}, {Buchner}, {Geyer}, {Goedhart}, \& {Serylak}}]{Cotton2024SMC}
{Cotton}, W.~D., {Filipovi{\'c}}, M.~D., {Camilo}, F., {et~al.} 2024{\natexlab{a}}, \href{https://ui.adsabs.harvard.edu/abs/2024MNRAS.529.2443C}{\mnras}, \href{https://ui.adsabs.harvard.edu/abs/2024MNRAS.529.2443C}{529, 2443}

\bibitem[{{Cotton} {et~al.}(2024{\natexlab{b}}){Cotton}, {Kothes}, {Camilo}, {Chandra}, {Buchner}, \& {Nyamai}}]{Cotton2024}
{Cotton}, W.~D., {Kothes}, R., {Camilo}, F., {et~al.} 2024{\natexlab{b}}, \href{https://ui.adsabs.harvard.edu/abs/2024ApJS..270...21C}{\apjs}, \href{https://ui.adsabs.harvard.edu/abs/2024ApJS..270...21C}{270, 21}

\bibitem[{{Crawford} {et~al.}(2008){Crawford}, {Filipovi\'c}, {de Horta}, {Stootman}, \& {Payne}}]{Crawford2008}
{Crawford}, E.~J., {Filipovi\'c}, M.~D., {de Horta}, A.~Y., {Stootman}, F.~H., and {Payne}, J.~L. 2008, \href{https://ui.adsabs.harvard.edu/abs/2008SerAJ.177...61C}{Serbian Astronomical Journal}, \href{https://ui.adsabs.harvard.edu/abs/2008SerAJ.177...61C}{177, 61}

\bibitem[{{Dimaratos} {et~al.}(2015){Dimaratos}, {Cormier}, {Bigiel}, \& {Madden}}]{Dimaratos2015}
{Dimaratos}, A., {Cormier}, D., {Bigiel}, F., and {Madden}, S.~C. 2015, \href{https://ui.adsabs.harvard.edu/abs/2015A&A...580A.135D}{\aap}, \href{https://ui.adsabs.harvard.edu/abs/2015A&A...580A.135D}{580, A135}

\bibitem[{{Dopita} and {Sutherland}(1996){Dopita} \& {Sutherland}}]{Dopita1996}
{Dopita}, M.~A. and {Sutherland}, R.~S. 1996, \href{https://ui.adsabs.harvard.edu/abs/1996ApJS..102..161D}{\apjs}, \href{https://ui.adsabs.harvard.edu/abs/1996ApJS..102..161D}{102, 161}

\bibitem[{{Dubner} {et~al.}(1999){Dubner}, {Giacani}, {Reynoso}, {Goss}, {Roth}, \& {Green}}]{Dubner1999}
{Dubner}, G., {Giacani}, E., {Reynoso}, E., {et~al.} 1999, \href{https://ui.adsabs.harvard.edu/abs/1999AJ....118..930D}{\aj}, \href{https://ui.adsabs.harvard.edu/abs/1999AJ....118..930D}{118, 930}

\bibitem[{{Dubner} {et~al.}(1991){Dubner}, {Braun}, {Winkler}, \& {Goss}}]{Dubner1991}
{Dubner}, G.~M., {Braun}, R., {Winkler}, P.~F., and {Goss}, W.~M. 1991, \href{https://ui.adsabs.harvard.edu/abs/1991AJ....101.1466D}{\aj}, \href{https://ui.adsabs.harvard.edu/abs/1991AJ....101.1466D}{101, 1466}

\bibitem[{{Fan} {et~al.}(2010){Fan}, {Liu}, \& {Fryer}}]{Fan2010}
{Fan}, Z., {Liu}, S., and {Fryer}, C.~L. 2010, \href{https://ui.adsabs.harvard.edu/abs/2010MNRAS.406.1337F}{\mnras}, \href{https://ui.adsabs.harvard.edu/abs/2010MNRAS.406.1337F}{406, 1337}

\bibitem[{{Ferrand} and {Safi-Harb}(2012){Ferrand} \& {Safi-Harb}}]{Ferrand2012}
{Ferrand}, G. and {Safi-Harb}, S. 2012, \href{https://ui.adsabs.harvard.edu/abs/2012AdSpR..49.1313F}{Advances in Space Research}, \href{https://ui.adsabs.harvard.edu/abs/2012AdSpR..49.1313F}{49, 1313}

\bibitem[{{Filipovi{\'c}} {et~al.}(2021){Filipovi{\'c}}, {Boji{\v{c}}i{\'c}}, {Grieve}, {Norris}, {Tothill}, {Shobhana}, {Rudnick}, {Prandoni}, {Andernach}, {Hurley-Walker}, {Alsaberi}, {Anderson}, {Collier}, {Crawford}, {For}, {Galvin}, {Haberl}, {Hopkins}, {Ingallinera}, {Kavanagh}, {Koribalski}, {Kothes}, {Leahy}, {Leverenz}, {Maggi}, {Maitra}, {Marvil}, {Pannuti}, {Park}, {Payne}, {Pennock}, {Riggi}, {Rowell}, {Sano}, {Sasaki}, {Staveley-Smith}, {Trigilio}, {Umana}, {Uro{\v{s}}evi{\'c}}, {van Loon}, \& {Vardoulaki}}]{2021MNRAS.507.2885F}
{Filipovi{\'c}}, M.~D., {Boji{\v{c}}i{\'c}}, I.~S., {Grieve}, K.~R., {et~al.} 2021, \href{https://ui.adsabs.harvard.edu/abs/2021MNRAS.507.2885F}{\mnras}, \href{https://ui.adsabs.harvard.edu/abs/2021MNRAS.507.2885F}{507, 2885}

\bibitem[{{Filipovi{\'c}} {et~al.}(2023){Filipovi{\'c}}, {Dai}, {Arbutina}, {Hurley-Walker}, {Brose}, {Becker}, {Sano}, {Uro{\v{s}}evi{\'c}}, {Jarrett}, {Hopkins}, {Alsaberi}, {Alsulami}, {Bordiu}, {Ball}, {Bufano}, {Burger-Scheidlin}, {Crawford}, {English}, {Haberl}, {Ingallinera}, {Kapinska}, {Kavanagh}, {Koribalski}, {Kothes}, {Lazarevi{\'c}}, {Mackey}, {Rowell}, {Leahy}, {Loru}, {Macgregor}, {Nicastro}, {Norris}, {Riggi}, {Sasaki}, {Stupar}, {Trigilio}, {Umana}, {Vernstrom}, \& {Vukoti{\'c}}}]{Filipovic2023}
{Filipovi{\'c}}, M.~D., {Dai}, S., {Arbutina}, B., {et~al.} 2023, \href{https://ui.adsabs.harvard.edu/abs/2023AJ....166..149F}{\aj}, \href{https://ui.adsabs.harvard.edu/abs/2023AJ....166..149F}{166, 149}

\bibitem[{{Filipovi{\'c}} {et~al.}(2008){Filipovi{\'c}}, {Haberl}, {Winkler}, {Pietsch}, {Payne}, {Crawford}, {de Horta}, {Stootman}, \& {Reaser}}]{Filipovic2008}
{Filipovi{\'c}}, M.~D., {Haberl}, F., {Winkler}, P.~F., {et~al.} 2008, \href{https://ui.adsabs.harvard.edu/abs/2008A&A...485...63F}{\aap}, \href{https://ui.adsabs.harvard.edu/abs/2008A&A...485...63F}{485, 63}

\bibitem[{{Filipovi\'c} {et~al.}(1998){Filipovi\'c}, {Haynes}, {White}, \& {Jones}}]{Filipovic1998a}
{Filipovi\'c}, M.~D., {Haynes}, R.~F., {White}, G.~L., and {Jones}, P.~A. 1998, \href{https://ui.adsabs.harvard.edu/abs/1998A&AS..130..421F}{\aaps}, \href{https://ui.adsabs.harvard.edu/abs/1998A&AS..130..421F}{130, 421}

\bibitem[{{Filipovi\'c} {et~al.}(1995{\natexlab{a}}){Filipovi\'c}, {Haynes}, {White}, {Jones}, {Klein}, \& {Wielebinski}}]{Filipovic1995}
{Filipovi\'c}, M.~D., {Haynes}, R.~F., {White}, G.~L., {et~al.} 1995{\natexlab{a}}, \href{https://ui.adsabs.harvard.edu/abs/1995A&AS..111..311F}{\aaps}, \href{https://ui.adsabs.harvard.edu/abs/1995A&AS..111..311F}{111, 311}

\bibitem[{{Filipovi\'c} {et~al.}(1995{\natexlab{b}}){Filipovi\'c}, {Haynes}, {White}, {Jones}, {Klein}, \& {Wielebinski}}]{Filipovic1995IV}
{Filipovi\'c}, M.~D., {Haynes}, R.~F., {White}, G.~L., {et~al.} 1995{\natexlab{b}}, \href{https://ui.adsabs.harvard.edu/abs/1995A&AS..111..311F}{\aaps}, \href{https://ui.adsabs.harvard.edu/abs/1995A&AS..111..311F}{111, 311}

\bibitem[{{Filipovi{\'c}} {et~al.}(2024){Filipovi{\'c}}, {Lazarevi{\'c}}, {Araya}, {Hurley-Walker}, {Kothes}, {Sano}, {Rowell}, {Martin}, {Fukui}, {Alsaberi}, {Arbutina}, {Ball}, {Bordiu}, {Brose}, {Bufano}, {Burger-Scheidlin}, {Anne Collins}, {Crawford}, {Dai}, {William Duchesne}, {Fuller}, {Hopkins}, {Ingallinera}, {Inoue}, {Jarrett}, {Koribalski}, {Leahy}, {Luken}, {Mackey}, {Macgregor}, {Norris}, {Payne}, {Riggi}, {Riseley}, {Sasaki}, {Smeaton}, {Sushch}, {Stupar}, {Umana}, {Uro{\v{s}}evi{\'c}}, {Velovi{\'c}}, {Vernstrom}, {Vukoti{\'c}}, \& {West}}]{Filipovic2024}
{Filipovi{\'c}}, M.~D., {Lazarevi{\'c}}, S., {Araya}, M., {et~al.} 2024, \href{https://ui.adsabs.harvard.edu/abs/2024PASA...41..112F}{\pasa}, \href{https://ui.adsabs.harvard.edu/abs/2024PASA...41..112F}{41, e112}

\bibitem[{{Filipovi{\'c}} {et~al.}(2022){Filipovi{\'c}}, {Payne}, {Alsaberi}, {Norris}, {Macgregor}, {Rudnick}, {Koribalski}, {Leahy}, {Ducci}, {Kothes}, {Andernach}, {Barnes}, {Boji{\v{c}}i{\'c}}, {Bozzetto}, {Brose}, {Collier}, {Crawford}, {Crocker}, {Dai}, {Galvin}, {Haberl}, {Heber}, {Hill}, {Hopkins}, {Hurley-Walker}, {Ingallinera}, {Jarrett}, {Kavanagh}, {Lenc}, {Luken}, {Mackey}, {Manojlovi{\'c}}, {Maggi}, {Maitra}, {Pennock}, {Points}, {Riggi}, {Rowell}, {Safi-Harb}, {Sano}, {Sasaki}, {Shabala}, {Stevens}, {van Loon}, {Tothill}, {Umana}, {Uro{\v{s}}evi{\'c}}, {Velovi{\'c}}, {Vernstrom}, {West}, \& {Wan}}]{Filipovic2022}
{Filipovi{\'c}}, M.~D., {Payne}, J.~L., {Alsaberi}, R.~Z.~E., {et~al.} 2022, \href{https://ui.adsabs.harvard.edu/abs/2022MNRAS.512..265F}{\mnras}, \href{https://ui.adsabs.harvard.edu/abs/2022MNRAS.512..265F}{512, 265}

\bibitem[{{Filipovic} {et~al.}(2025){Filipovic}, {Smeaton}, {Kothes}, {Mantovanini}, {Kostic}, {Leahy}, {Ahmad}, {Anderson}, {Araya}, {Ball}, {Becker}, {Bordiu}, {Bradley}, {Brose}, {Burger-Scheidlin}, {Dai}, {Duchesne}, {Galvin}, {Hopkins}, {Hurley-Walker}, {Koribalski}, {Lazarevic}, {Lundqvist}, {Mackey}, {Martin}, {McGee}, {Mitrasinovic}, {Payne}, {Riggi}, {Ross}, {Rowell}, {Rudnick}, {Sano}, {Sasaki}, {Soria}, {Urosevic}, {Vukotic}, \& {West}}]{Filipovic2025Arxiv}
{Filipovic}, M.~D., {Smeaton}, Z.~J., {Kothes}, R., {et~al.} 2025, {arXiv e-prints, \href{https://ui.adsabs.harvard.edu/abs/2025arXiv250504041F}{}\eprint  \href{}{(DOI:  10.48550/arXiv.2505.04041)}, arXiv:2505.04041}

\bibitem[{Filipovi{\'c} and Tothill(2021)Filipovi{\'c} \& Tothill}]{book1}
Filipovi{\'c}, M.~D. and Tothill, N. F.~H. 2021, Principles of Multimessenger Astronomy, 2514-3433 (IOP Publishing)

\bibitem[{{Filipovi\'c} {et~al.}(1996){Filipovi\'c}, {White}, {Haynes}, {Jones}, {Meinert}, {Wielebinski}, \& {Klein}}]{Filipovic1996}
{Filipovi\'c}, M.~D., {White}, G.~L., {Haynes}, R.~F., {et~al.} 1996, \href{https://ui.adsabs.harvard.edu/abs/1996A&AS..120...77F}{\aaps}, \href{https://ui.adsabs.harvard.edu/abs/1996A&AS..120...77F}{120, 77}

\bibitem[{{Filipovic} {et~al.}(1996){Filipovic}, {White}, {Jones}, {Haynes}, {Pietsch}, {Wielebinski}, \& {Klein}}]{1996ASPC..112...91F}
{Filipovic}, M.~D., {White}, G.~L., {Jones}, P.~A., {et~al.} 1996, in Astronomical Society of the Pacific Conference Series, Vol. 112, The History of the Milky Way and Its Satellite System, ed. A.~{Burkert}, D.~H. {Hartmann}, and S.~A. {Majewski}, 91

\bibitem[{{For} {et~al.}(2018){For}, {Staveley-Smith}, {Hurley-Walker}, {Franzen}, {Kapi{\'n}ska}, {Filipovi{\'c}}, {Collier}, {Wu}, {Grieve}, {Callingham}, {Bell}, {Bernardi}, {Bowman}, {Briggs}, {Cappallo}, {Deshpande}, {Dwarakanath}, {Gaensler}, {Greenhill}, {Hancock}, {Hazelton}, {Hindson}, {Johnston-Hollitt}, {Kaplan}, {Lenc}, {Lonsdale}, {McKinley}, {McWhirter}, {Mitchell}, {Morales}, {Morgan}, {Morgan}, {Oberoi}, {Offringa}, {Ord}, {Prabu}, {Procopio}, {Shankar}, {Srivani}, {Subrahmanyan}, {Tingay}, {Wayth}, {Webster}, {Williams}, {Williams}, \& {Zheng}}]{For2018}
{For}, B.~Q., {Staveley-Smith}, L., {Hurley-Walker}, N., {et~al.} 2018, \href{https://ui.adsabs.harvard.edu/abs/2018MNRAS.480.2743F}{\mnras}, \href{https://ui.adsabs.harvard.edu/abs/2018MNRAS.480.2743F}{480, 2743}

\bibitem[{{Fukui} {et~al.}(2009){Fukui}, {Kawamura}, {Wong}, {Murai}, {Iritani}, {Mizuno}, {Mizuno}, {Onishi}, {Hughes}, {Ott}, {Muller}, {Staveley-Smith}, \& {Kim}}]{Fukui2009}
{Fukui}, Y., {Kawamura}, A., {Wong}, T., {et~al.} 2009, \href{https://ui.adsabs.harvard.edu/abs/2009ApJ...705..144F}{\apj}, \href{https://ui.adsabs.harvard.edu/abs/2009ApJ...705..144F}{705, 144}

\bibitem[{{Gaensler} {et~al.}(1998){Gaensler}, {Green}, \& {Manchester}}]{Gaensler1998}
{Gaensler}, B.~M., {Green}, A.~J., and {Manchester}, R.~N. 1998, \href{https://ui.adsabs.harvard.edu/abs/1998MNRAS.299..812G}{\mnras}, \href{https://ui.adsabs.harvard.edu/abs/1998MNRAS.299..812G}{299, 812}

\bibitem[{{Galvin} and {Filipovi\'c}(2014){Galvin} \& {Filipovi\'c}}]{Galvin2014M31}
{Galvin}, T.~J. and {Filipovi\'c}, M.~D. 2014, \href{https://ui.adsabs.harvard.edu/abs/2014SerAJ.189...15G}{Serbian Astronomical Journal}, \href{https://ui.adsabs.harvard.edu/abs/2014SerAJ.189...15G}{189, 15}

\bibitem[{{Galvin} {et~al.}(2012){Galvin}, {Filipovi{\'c}}, {Crawford}, {Wong}, {Payne}, {De Horta}, {White}, {Tothill}, {Dra{\v{s}}kovi{\'c}}, {Pannuti}, {Grimes}, {Cahall}, {Millar}, \& {Laine}}]{Galvin2012}
{Galvin}, T.~J., {Filipovi{\'c}}, M.~D., {Crawford}, E.~J., {et~al.} 2012, \href{https://ui.adsabs.harvard.edu/abs/2012Ap&SS.340..133G}{\apss}, \href{https://ui.adsabs.harvard.edu/abs/2012Ap&SS.340..133G}{340, 133}

\bibitem[{{Galvin} {et~al.}(2014){Galvin}, {Filipovi{\'c}}, {Tothill}, {Crawford}, {O'Brien}, {Seymour}, {Pannuti}, {Kosakowski}, \& {Sharma}}]{Galvin2014Sculpt}
{Galvin}, T.~J., {Filipovi{\'c}}, M.~D., {Tothill}, N. F.~H., {et~al.} 2014, \href{https://ui.adsabs.harvard.edu/abs/2014Ap&SS.353..603G}{\apss}, \href{https://ui.adsabs.harvard.edu/abs/2014Ap&SS.353..603G}{353, 603}

\bibitem[{{Ghavam} {et~al.}(2024){Ghavam}, {Filipovi{\'c}}, {Alsaberi}, {Barnes}, {Crawford}, {Haberl}, {Kavanagh}, {Maggi}, {Payne}, {Rowell}, {Hidetoshi}, {Sasaki}, {Rajabpour}, {Tothill}, \& {Uro{\v{s}}evi{\'c}}}]{Ghavam2024}
{Ghavam}, M., {Filipovi{\'c}}, M.~D., {Alsaberi}, R., {et~al.} 2024, \href{https://ui.adsabs.harvard.edu/abs/2024PASA...41...89G}{\pasa}, \href{https://ui.adsabs.harvard.edu/abs/2024PASA...41...89G}{41, e089}

\bibitem[{{Goedhart} {et~al.}(2024){Goedhart}, {Cotton}, {Camilo}, {Thompson}, {Umana}, {Bietenholz}, {Woudt}, {Anderson}, {Bordiu}, {Buckley}, {Buemi}, {Bufano}, {Cavallaro}, {Chen}, {Chibueze}, {Egbo}, {Frank}, {Hoare}, {Ingallinera}, {Irabor}, {Kraan-Korteweg}, {Kurapati}, {Leto}, {Loru}, {Mutale}, {Obonyo}, {Plavin}, {Rajohnson}, {Rigby}, {Riggi}, {Seidu}, {Serra}, {Smart}, {Stappers}, {Steyn}, {Surnis}, {Trigilio}, {Williams}, {Abbott}, {Adam}, {Asad}, {Baloyi}, {Bauermeister}, {Bennet}, {Bester}, {Botha}, {Brederode}, {Buchner}, {Burger}, {Cheetham}, {Cloete}, {de Villiers}, {de Villiers}, {du Toit}, {Esterhuyse}, {Fanaroff}, {Fourie}, {Gamatham}, {Gatsi}, {Geyer}, {Gouws}, {Gumede}, {Heywood}, {Hokwana}, {Hoosen}, {Horn}, {Horrell}, {Hugo}, {Isaacson}, {J{\'o}zsa}, {Jonas}, {Jordaan}, {Joubert}, {Julie}, {Kapp}, {Kriek}, {Kriel}, {Krishnan}, {Kusel}, {Legodi}, {Lehmensiek}, {Lord}, {Macfarlane}, {Magnus}, {Magozore}, {Main}, {Malan}, {Manley}, {Marais}, {Maree}, {Martens}, {Maruping}, {McAlpine},
  {Merry}, {Mgodeli}, {Millenaar}, {Mokone}, {Monama}, {New}, {Ngcebetsha}, {Ngoasheng}, {Nicolson}, {Ockards}, {Oozeer}, {Passmoor}, {Patel}, {Peens-Hough}, {Perkins}, {Ramaila}, {Ratcliffe}, {Renil}, {Richter}, {Salie}, {Sambu}, {Schollar}, {Schwardt}, {Schwartz}, {Serylak}, {Siebrits}, {Sirothia}, {Slabber}, {Smirnov}, {Tiplady}, {van Balla}, {van der Byl}, {Van Tonder}, {Venter}, {Venter}, {Welz}, \& {Williams}}]{Goedhart2024}
{Goedhart}, S., {Cotton}, W.~D., {Camilo}, F., {et~al.} 2024, \href{https://ui.adsabs.harvard.edu/abs/2024MNRAS.531..649G}{\mnras}, \href{https://ui.adsabs.harvard.edu/abs/2024MNRAS.531..649G}{531, 649}

\bibitem[{{Green}(2025)}]{Green2025}
{Green}, D.~A. 2025, \href{https://ui.adsabs.harvard.edu/abs/2025JApA...46...14G}{Journal of Astrophysics and Astronomy}, \href{https://ui.adsabs.harvard.edu/abs/2025JApA...46...14G}{46, 14}

\bibitem[{{Guzman} {et~al.}(2019){Guzman}, {Whiting}, {Voronkov}, {Mitchell}, {Ord}, {Collins}, {Marquarding}, {Lahur}, {Maher}, {Van Diepen}, {Bannister}, {Wu}, {Lenc}, {Khoo}, \& {Bastholm}}]{Guzman2019}
{Guzman}, J., {Whiting}, M., {Voronkov}, M., {et~al.} 2019, \href{https://ui.adsabs.harvard.edu/abs/2019ascl.soft12003G}{{ASKAPsoft: ASKAP science data processor software}}

\bibitem[{{Haberl}(2014)}]{Haberl2014}
{Haberl}, F. 2014, in The X-ray Universe 2014, ed. J.-U. {Ness}, 4

\bibitem[{{Haberl} {et~al.}(2012){Haberl}, {Sturm}, {Filipovi{\'c}}, {Pietsch}, \& {Crawford}}]{Haberl2012}
{Haberl}, F., {Sturm}, R., {Filipovi{\'c}}, M.~D., {Pietsch}, W., and {Crawford}, E.~J. 2012, \href{https://ui.adsabs.harvard.edu/abs/2012A&A...537L...1H}{\aap}, \href{https://ui.adsabs.harvard.edu/abs/2012A&A...537L...1H}{537, L1}

\bibitem[{{Heald}(2009)}]{Heald2009}
{Heald}, G. 2009, in IAU Symposium, Vol. 259, Cosmic Magnetic Fields: From Planets, to Stars and Galaxies, ed. K.~G. {Strassmeier}, A.~G. {Kosovichev}, and J.~E. {Beckman}, 591--602

\bibitem[{{Hopkins} {et~al.}(2025){Hopkins}, {Kapinska}, {Marvil}, {Vernstrom}, {Collier}, {Norris}, {Gordon}, {Duchesne}, {Rudnick}, {Gupta}, {Carretti}, {Anderson}, {Dai}, {G{\"u}rkan}, {Parkinson}, {Prandoni}, {Riggi}, {Saraf}, {Ma}, {Filipovi{\'c}}, {Umana}, {Bahr-Kalus}, {Koribalski}, {Lenc}, {Ingallinera}, {Afonso}, {Ahmad}, {Ahmed}, {Alexander}, {Andernach}, {Asorey}, {Battisti}, {Bilicki}, {Botteon}, {Brown}, {Br{\"u}ggen}, {Cowley}, {Dage}, {Hale}, {Hardcastle}, {Kothes}, {Lazarevi{\'c}}, {Lin}, {Luken}, {Moss}, {Prathap}, {Rahman}, {Reiprich}, {Riseley}, {Salvato}, {Seymour}, {Shabala}, {Smith}, {Vaccari}, {van Loon}, {Wong}, {Alsaberi}, {Asher}, {Ball}, {Barbosa}, {Biava}, {Bradley}, {Carvajal}, {Crawford}, {Galvin}, {Huynh}, {Leahy}, {Matute}, {Moss}, {Pappalardo}, {Smeaton}, {Velovi{\'c}}, \& {Zafar}}]{2025arXiv250508271H}
{Hopkins}, A.~M., {Kapinska}, A., {Marvil}, J., {et~al.} 2025, {arXiv e-prints, \href{https://ui.adsabs.harvard.edu/abs/2025arXiv250508271H}{}\eprint  \href{}{(DOI:  10.48550/arXiv.2505.08271)}, arXiv:2505.08271}

\bibitem[{{Hotan} {et~al.}(2021){Hotan}, {Bunton}, {Chippendale}, {Whiting}, {Tuthill}, {Moss}, {McConnell}, {Amy}, {Huynh}, {Allison}, {Anderson}, {Bannister}, {Bastholm}, {Beresford}, {Bock}, {Bolton}, {Chapman}, {Chow}, {Collier}, {Cooray}, {Cornwell}, {Diamond}, {Edwards}, {Feain}, {Franzen}, {George}, {Gupta}, {Hampson}, {Harvey-Smith}, {Hayman}, {Heywood}, {Jacka}, {Jackson}, {Jackson}, {Jeganathan}, {Johnston}, {Kesteven}, {Kleiner}, {Koribalski}, {Lee-Waddell}, {Lenc}, {Lensson}, {Mackay}, {Mahony}, {McClure-Griffiths}, {McConigley}, {Mirtschin}, {Ng}, {Norris}, {Pearce}, {Phillips}, {Pilawa}, {Raja}, {Reynolds}, {Roberts}, {Roxby}, {Sadler}, {Shields}, {Schinckel}, {Serra}, {Shaw}, {Sweetnam}, {Troup}, {Tzioumis}, {Voronkov}, \& {Westmeier}}]{Hotan2021}
{Hotan}, A.~W., {Bunton}, J.~D., {Chippendale}, A.~P., {et~al.} 2021, \href{https://ui.adsabs.harvard.edu/abs/2021PASA...38....9H}{\pasa}, \href{https://ui.adsabs.harvard.edu/abs/2021PASA...38....9H}{38, e009}

\bibitem[{{Hurley-Walker} {et~al.}(2017){Hurley-Walker}, {Callingham}, {Hancock}, {Franzen}, {Hindson}, {Kapi{\'n}ska}, {Morgan}, {Offringa}, {Wayth}, {Wu}, {Zheng}, {Murphy}, {Bell}, {Dwarakanath}, {For}, {Gaensler}, {Johnston-Hollitt}, {Lenc}, {Procopio}, {Staveley-Smith}, {Ekers}, {Bowman}, {Briggs}, {Cappallo}, {Deshpande}, {Greenhill}, {Hazelton}, {Kaplan}, {Lonsdale}, {McWhirter}, {Mitchell}, {Morales}, {Morgan}, {Oberoi}, {Ord}, {Prabu}, {Shankar}, {Srivani}, {Subrahmanyan}, {Tingay}, {Webster}, {Williams}, \& {Williams}}]{HurleyWalker2017}
{Hurley-Walker}, N., {Callingham}, J.~R., {Hancock}, P.~J., {et~al.} 2017, \href{https://ui.adsabs.harvard.edu/abs/2017MNRAS.464.1146H}{\mnras}, \href{https://ui.adsabs.harvard.edu/abs/2017MNRAS.464.1146H}{464, 1146}

\bibitem[{{Johnston} {et~al.}(2008){Johnston}, {Taylor}, {Bailes}, {Bartel}, {Baugh}, {Bietenholz}, {Blake}, {Braun}, {Brown}, {Chatterjee}, {Darling}, {Deller}, {Dodson}, {Edwards}, {Ekers}, {Ellingsen}, {Feain}, {Gaensler}, {Haverkorn}, {Hobbs}, {Hopkins}, {Jackson}, {James}, {Joncas}, {Kaspi}, {Kilborn}, {Koribalski}, {Kothes}, {Landecker}, {Lenc}, {Lovell}, {Macquart}, {Manchester}, {Matthews}, {McClure-Griffiths}, {Norris}, {Pen}, {Phillips}, {Power}, {Protheroe}, {Sadler}, {Schmidt}, {Stairs}, {Staveley-Smith}, {Stil}, {Tingay}, {Tzioumis}, {Walker}, {Wall}, \& {Wolleben}}]{Johnston2008}
{Johnston}, S., {Taylor}, R., {Bailes}, M., {et~al.} 2008, \href{https://ui.adsabs.harvard.edu/abs/2008ExA....22..151J}{Experimental Astronomy}, \href{https://ui.adsabs.harvard.edu/abs/2008ExA....22..151J}{22, 151}

\bibitem[{{Kavanagh} {et~al.}(2020){Kavanagh}, {Sasaki}, {Breitschwerdt}, {de Avillez}, {Filipovi{\'c}}, {Galvin}, {Haberl}, {Hatzidimitriou}, {Henze}, {Plucinsky}, {Saeedi}, {Sokolovsky}, \& {Williams}}]{Kavanagh2020}
{Kavanagh}, P.~J., {Sasaki}, M., {Breitschwerdt}, D., {et~al.} 2020, \href{https://ui.adsabs.harvard.edu/abs/2020A&A...637A..12K}{\aap}, \href{https://ui.adsabs.harvard.edu/abs/2020A&A...637A..12K}{637, A12}

\bibitem[{{Khabibullin} {et~al.}(2023){Khabibullin}, {Churazov}, {Bykov}, {Chugai}, \& {Sunyaev}}]{Khabibullin2024}
{Khabibullin}, I.~I., {Churazov}, E.~M., {Bykov}, A.~M., {Chugai}, N.~N., and {Sunyaev}, R.~A. 2023, \href{https://ui.adsabs.harvard.edu/abs/2023MNRAS.521.5536K}{MNRAS}, \href{https://ui.adsabs.harvard.edu/abs/2023MNRAS.521.5536K}{521, 5536}

\bibitem[{{Kim} {et~al.}(2003){Kim}, {Staveley-Smith}, {Dopita}, {Sault}, {Freeman}, {Lee}, \& {Chu}}]{2003ApJS..148..473K}
{Kim}, S., {Staveley-Smith}, L., {Dopita}, M.~A., {et~al.} 2003, \href{https://ui.adsabs.harvard.edu/abs/2003ApJS..148..473K}{\apjs}, \href{https://ui.adsabs.harvard.edu/abs/2003ApJS..148..473K}{148, 473}

\bibitem[{{Kobayashi} {et~al.}(2006){Kobayashi}, {Umeda}, {Nomoto}, {Tominaga}, \& {Ohkubo}}]{2006ApJ...653.1145K}
{Kobayashi}, C., {Umeda}, H., {Nomoto}, K., {Tominaga}, N., and {Ohkubo}, T. 2006, \href{https://ui.adsabs.harvard.edu/abs/2006ApJ...653.1145K}{\apj}, \href{https://ui.adsabs.harvard.edu/abs/2006ApJ...653.1145K}{653, 1145}

\bibitem[{{Kothes} {et~al.}(2017){Kothes}, {Reich}, {Foster}, \& {Reich}}]{Kothes2017}
{Kothes}, R., {Reich}, P., {Foster}, T.~J., and {Reich}, W. 2017, \href{https://ui.adsabs.harvard.edu/abs/2017A&A...597A.116K}{A\&A}, \href{https://ui.adsabs.harvard.edu/abs/2017A&A...597A.116K}{597, A116}

\bibitem[{{Laki{\'c}evi{\'c}} {et~al.}(2015){Laki{\'c}evi{\'c}}, {van Loon}, {Meixner}, {Gordon}, {Bot}, {Roman-Duval}, {Babler}, {Bolatto}, {Engelbracht}, {Filipovi{\'c}}, {Hony}, {Indebetouw}, {Misselt}, {Montiel}, {Okumura}, {Panuzzo}, {Patat}, {Sauvage}, {Seale}, {Sonneborn}, {Temim}, {Uro{\v{s}}evi{\'c}}, \& {Zanardo}}]{Lakicevic2015}
{Laki{\'c}evi{\'c}}, M., {van Loon}, J.~T., {Meixner}, M., {et~al.} 2015, \href{https://ui.adsabs.harvard.edu/abs/2015ApJ...799...50L}{\apj}, \href{https://ui.adsabs.harvard.edu/abs/2015ApJ...799...50L}{799, 50}

\bibitem[{{Landecker} {et~al.}(1989){Landecker}, {Pineault}, {Routledge}, \& {Vaneldik}}]{Landecker1989}
{Landecker}, T.~L., {Pineault}, S., {Routledge}, D., and {Vaneldik}, J.~F. 1989, \href{https://ui.adsabs.harvard.edu/abs/1989MNRAS.237..277L}{\mnras}, \href{https://ui.adsabs.harvard.edu/abs/1989MNRAS.237..277L}{237, 277}

\bibitem[{{Lazarevi{\'c}} {et~al.}(2024){Lazarevi{\'c}}, {Filipovi{\'c}}, {Koribalski}, {Smeaton}, {Hopkins}, {Alsaberi}, {Velovi{\'c}}, {Ball}, {Kothes}, {Leahy}, \& {Ingallinera}}]{Lazarevic2024}
{Lazarevi{\'c}}, S., {Filipovi{\'c}}, M.~D., {Koribalski}, B.~S., {et~al.} 2024, \href{https://ui.adsabs.harvard.edu/abs/2024RNAAS...8..107L}{Research Notes of the American Astronomical Society}, \href{https://ui.adsabs.harvard.edu/abs/2024RNAAS...8..107L}{8, 107}

\bibitem[{{Leahy} {et~al.}(2019){Leahy}, {Wang}, {Lawton}, {Ranasinghe}, \& {Filipovi{\'c}}}]{2019AJ....158..149L}
{Leahy}, D., {Wang}, Y., {Lawton}, B., {Ranasinghe}, S., and {Filipovi{\'c}}, M. 2019, \href{https://ui.adsabs.harvard.edu/abs/2019AJ....158..149L}{\aj}, \href{https://ui.adsabs.harvard.edu/abs/2019AJ....158..149L}{158, 149}

\bibitem[{{Leahy}(2017)}]{2017ApJ...837...36L}
{Leahy}, D.~A. 2017, \href{https://ui.adsabs.harvard.edu/abs/2017ApJ...837...36L}{\apj}, \href{https://ui.adsabs.harvard.edu/abs/2017ApJ...837...36L}{837, 36}

\bibitem[{{Leahy} and {Filipovi{\'c}}(2022){Leahy} \& {Filipovi{\'c}}}]{2022ApJ...931...20L}
{Leahy}, D.~A. and {Filipovi{\'c}}, M.~D. 2022, \href{https://ui.adsabs.harvard.edu/abs/2022ApJ...931...20L}{\apj}, \href{https://ui.adsabs.harvard.edu/abs/2022ApJ...931...20L}{931, 20}

\bibitem[{{Leahy} {et~al.}(2020){Leahy}, {Ranasinghe}, \& {Gelowitz}}]{2020ApJS..248...16L}
{Leahy}, D.~A., {Ranasinghe}, S., and {Gelowitz}, M. 2020, \href{https://ui.adsabs.harvard.edu/abs/2020ApJS..248...16L}{\apjs}, \href{https://ui.adsabs.harvard.edu/abs/2020ApJS..248...16L}{248, 16}

\bibitem[{{Leahy} and {Williams}(2017){Leahy} \& {Williams}}]{2017AJ....153..239L}
{Leahy}, D.~A. and {Williams}, J.~E. 2017, \href{https://ui.adsabs.harvard.edu/abs/2017AJ....153..239L}{\aj}, \href{https://ui.adsabs.harvard.edu/abs/2017AJ....153..239L}{153, 239}

\bibitem[{{Liu} {et~al.}(2008){Liu}, {Fan}, {Fryer}, {Wang}, \& {Li}}]{Liu2008}
{Liu}, S., {Fan}, Z.-H., {Fryer}, C.~L., {Wang}, J.-M., and {Li}, H. 2008, \href{https://ui.adsabs.harvard.edu/abs/2008ApJ...683L.163L}{\apjl}, \href{https://ui.adsabs.harvard.edu/abs/2008ApJ...683L.163L}{683, L163}

\bibitem[{{Luken} {et~al.}(2020){Luken}, {Filipovi{\'c}}, {Maxted}, {Kothes}, {Norris}, {Allison}, {Blackwell}, {Braiding}, {Brose}, {Burton}, {De Horta}, {Galvin}, {Harvey-Smith}, {Hurley-Walker}, {Leahy}, {Ralph}, {Roper}, {Rowell}, {Sushch}, {Uro{\v{s}}evi{\'c}}, \& {Wong}}]{Luken2020}
{Luken}, K.~J., {Filipovi{\'c}}, M.~D., {Maxted}, N.~I., {et~al.} 2020, \href{https://ui.adsabs.harvard.edu/abs/2020MNRAS.492.2606L}{MNRAS}, \href{https://ui.adsabs.harvard.edu/abs/2020MNRAS.492.2606L}{492, 2606}

\bibitem[{{Maggi} {et~al.}(2019){Maggi}, {Filipovi{\'c}}, {Vukoti{\'c}}, {Ballet}, {Haberl}, {Maitra}, {Kavanagh}, {Sasaki}, \& {Stupar}}]{Maggi2019}
{Maggi}, P., {Filipovi{\'c}}, M.~D., {Vukoti{\'c}}, B., {et~al.} 2019, \href{https://ui.adsabs.harvard.edu/abs/2019A&A...631A.127M}{\aap}, \href{https://ui.adsabs.harvard.edu/abs/2019A&A...631A.127M}{631, A127}

\bibitem[{{Maggi} {et~al.}(2016){Maggi}, {Haberl}, {Kavanagh}, {Sasaki}, {Bozzetto}, {Filipovi{\'c}}, {Vasilopoulos}, {Pietsch}, {Points}, {Chu}, {Dickel}, {Ehle}, {Williams}, \& {Greiner}}]{Maggi2016}
{Maggi}, P., {Haberl}, F., {Kavanagh}, P.~J., {et~al.} 2016, \href{https://ui.adsabs.harvard.edu/abs/2016A&A...585A.162M}{\aap}, \href{https://ui.adsabs.harvard.edu/abs/2016A&A...585A.162M}{585, A162}

\bibitem[{{Maitra} {et~al.}(2021){Maitra}, {Haberl}, {Maggi}, {Kavanagh}, {Vasilopoulos}, {Sasaki}, {Filipovi{\'c}}, \& {Udalski}}]{Maitra2021}
{Maitra}, C., {Haberl}, F., {Maggi}, P., {et~al.} 2021, \href{https://ui.adsabs.harvard.edu/abs/2021MNRAS.504..326M}{\mnras}, \href{https://ui.adsabs.harvard.edu/abs/2021MNRAS.504..326M}{504, 326}

\bibitem[{{Mathewson} and {Clarke}(1973){Mathewson} \& {Clarke}}]{Mathewson1973}
{Mathewson}, D.~S. and {Clarke}, J.~N. 1973, \href{https://ui.adsabs.harvard.edu/abs/1973ApJ...180..725M}{\apj}, \href{https://ui.adsabs.harvard.edu/abs/1973ApJ...180..725M}{180, 725}

\bibitem[{{Mathewson} {et~al.}(1985){Mathewson}, {Ford}, {Tuohy}, {Mills}, {Turtle}, \& {Helfand}}]{Mathewson1985}
{Mathewson}, D.~S., {Ford}, V.~L., {Tuohy}, I.~R., {et~al.} 1985, \href{https://ui.adsabs.harvard.edu/abs/1985ApJS...58..197M}{\apjs}, \href{https://ui.adsabs.harvard.edu/abs/1985ApJS...58..197M}{58, 197}

\bibitem[{{McGee} {et~al.}(1972){McGee}, {Brooks}, \& {Batchelor}}]{McGee1972}
{McGee}, R.~X., {Brooks}, J.~W., and {Batchelor}, R.~A. 1972, \href{https://ui.adsabs.harvard.edu/abs/1972AuJPh..25..581M}{Australian Journal of Physics}, \href{https://ui.adsabs.harvard.edu/abs/1972AuJPh..25..581M}{25, 581}

\bibitem[{{Meixner} {et~al.}(2006){Meixner}, {Gordon}, {Indebetouw}, {Hora}, {Whitney}, {Blum}, {Reach}, {Bernard}, {Meade}, {Babler}, {Engelbracht}, {For}, {Misselt}, {Vijh}, {Leitherer}, {Cohen}, {Churchwell}, {Boulanger}, {Frogel}, {Fukui}, {Gallagher}, {Gorjian}, {Harris}, {Kelly}, {Kawamura}, {Kim}, {Latter}, {Madden}, {Markwick-Kemper}, {Mizuno}, {Mizuno}, {Mould}, {Nota}, {Oey}, {Olsen}, {Onishi}, {Paladini}, {Panagia}, {Perez-Gonzalez}, {Shibai}, {Sato}, {Smith}, {Staveley-Smith}, {Tielens}, {Ueta}, {van Dyk}, {Volk}, {Werner}, \& {Zaritsky}}]{Meixner2006}
{Meixner}, M., {Gordon}, K.~D., {Indebetouw}, R., {et~al.} 2006, \href{https://ui.adsabs.harvard.edu/abs/2006AJ....132.2268M}{\aj}, \href{https://ui.adsabs.harvard.edu/abs/2006AJ....132.2268M}{132, 2268}

\bibitem[{{Millar} {et~al.}(2012){Millar}, {White}, \& {Filipovi\'c}}]{Millar2012}
{Millar}, W.~C., {White}, G.~L., and {Filipovi\'c}, M.~D. 2012, \href{https://ui.adsabs.harvard.edu/abs/2012SerAJ.184...19M}{Serbian Astronomical Journal}, \href{https://ui.adsabs.harvard.edu/abs/2012SerAJ.184...19M}{184, 19}

\bibitem[{{Millar} {et~al.}(2011){Millar}, {White}, {Filipovi{\'c}}, {Payne}, {Crawford}, {Pannuti}, \& {Staggs}}]{Millar2011}
{Millar}, W.~C., {White}, G.~L., {Filipovi{\'c}}, M.~D., {et~al.} 2011, \href{https://ui.adsabs.harvard.edu/abs/2011Ap&SS.332..221M}{\apss}, \href{https://ui.adsabs.harvard.edu/abs/2011Ap&SS.332..221M}{332, 221}

\bibitem[{{Norris} {et~al.}(2011){Norris}, {Hopkins}, {Afonso}, {Brown}, {Condon}, {Dunne}, {Feain}, {Hollow}, {Jarvis}, {Johnston-Hollitt}, {Lenc}, {Middelberg}, {Padovani}, {Prandoni}, {Rudnick}, {Seymour}, {Umana}, {Andernach}, {Alexander}, {Appleton}, {Bacon}, {Banfield}, {Becker}, {Brown}, {Ciliegi}, {Jackson}, {Eales}, {Edge}, {Gaensler}, {Giovannini}, {Hales}, {Hancock}, {Huynh}, {Ibar}, {Ivison}, {Kennicutt}, {Kimball}, {Koekemoer}, {Koribalski}, {L{\'o}pez-S{\'a}nchez}, {Mao}, {Murphy}, {Messias}, {Pimbblet}, {Raccanelli}, {Randall}, {Reiprich}, {Roseboom}, {R{\"o}ttgering}, {Saikia}, {Sharp}, {Slee}, {Smail}, {Thompson}, {Urquhart}, {Wall}, \& {Zhao}}]{Norris2011}
{Norris}, R.~P., {Hopkins}, A.~M., {Afonso}, J., {et~al.} 2011, \href{https://ui.adsabs.harvard.edu/abs/2011PASA...28..215N}{PASA}, \href{https://ui.adsabs.harvard.edu/abs/2011PASA...28..215N}{28, 215}

\bibitem[{{Norris} {et~al.}(2021){Norris}, {Marvil}, {Collier}, {Kapi{\'n}ska}, {O'Brien}, {Rudnick}, {Andernach}, {Asorey}, {Brown}, {Br{\"u}ggen}, {Crawford}, {English}, {Rahman}, {Filipovi{\'c}}, {Gordon}, {G{\"u}rkan}, {Hale}, {Hopkins}, {Huynh}, {HyeongHan}, {James Jee}, {Koribalski}, {Lenc}, {Luken}, {Parkinson}, {Prandoni}, {Raja}, {Reiprich}, {Riseley}, {Shabala}, {Sheil}, {Vernstrom}, {Whiting}, {Allison}, {Anderson}, {Ball}, {Bell}, {Bunton}, {Galvin}, {Gupta}, {Hotan}, {Jacka}, {Macgregor}, {Mahony}, {Maio}, {Moss}, {Pandey-Pommier}, \& {Voronkov}}]{Norris2021}
{Norris}, R.~P., {Marvil}, J., {Collier}, J.~D., {et~al.} 2021, \href{https://ui.adsabs.harvard.edu/abs/2021PASA...38...46N}{PASA}, \href{https://ui.adsabs.harvard.edu/abs/2021PASA...38...46N}{38, e046}

\bibitem[{{O'Brien} {et~al.}(2013){O'Brien}, {Filipovi{\'c}}, {Crawford}, {Tothill}, {Collier}, {De Horta}, {Wong}, {Dra{\v{s}}kovi{\'c}}, {Payne}, {Pannuti}, {Napier}, {Griffith}, {Staggs}, \& {Kotu{\v{s}}}}]{OBrien2013}
{O'Brien}, A.~N., {Filipovi{\'c}}, M.~D., {Crawford}, E.~J., {et~al.} 2013, \href{https://ui.adsabs.harvard.edu/abs/2013Ap&SS.347..159O}{\apss}, \href{https://ui.adsabs.harvard.edu/abs/2013Ap&SS.347..159O}{347, 159}

\bibitem[{{Oni{\'c}}(2013)}]{Onic2013}
{Oni{\'c}}, D. 2013, \href{https://ui.adsabs.harvard.edu/abs/2013Ap&SS.346....3O}{\apss}, \href{https://ui.adsabs.harvard.edu/abs/2013Ap&SS.346....3O}{346, 3}

\bibitem[{{Oni\'c} and {Uro\v sevi\'c}(2008){Oni\'c} \& {Uro\v sevi\'c}}]{Onic2008}
{Oni\'c}, D. and {Uro\v sevi\'c}, D. 2008, \href{https://ui.adsabs.harvard.edu/abs/2008SerAJ.177...67O}{Serbian Astronomical Journal}, \href{https://ui.adsabs.harvard.edu/abs/2008SerAJ.177...67O}{177, 67}

\bibitem[{{Oni{\'c}} {et~al.}(2012){Oni{\'c}}, {Uro{\v{s}}evi{\'c}}, {Arbutina}, \& {Leahy}}]{Onic2012}
{Oni{\'c}}, D., {Uro{\v{s}}evi{\'c}}, D., {Arbutina}, B., and {Leahy}, D. 2012, \href{https://ui.adsabs.harvard.edu/abs/2012ApJ...756...61O}{\apj}, \href{https://ui.adsabs.harvard.edu/abs/2012ApJ...756...61O}{756, 61}

\bibitem[{{Pannuti} {et~al.}(2011){Pannuti}, {Schlegel}, {Filipovi{\'c}}, {Payne}, {Petre}, {Harrus}, {Staggs}, \& {Lacey}}]{Pannuti2011}
{Pannuti}, T.~G., {Schlegel}, E.~M., {Filipovi{\'c}}, M.~D., {et~al.} 2011, \href{https://ui.adsabs.harvard.edu/abs/2011AJ....142...20P}{\aj}, \href{https://ui.adsabs.harvard.edu/abs/2011AJ....142...20P}{142, 20}

\bibitem[{{Pannuti} {et~al.}(2015){Pannuti}, {Swartz}, {Laine}, {Schlegel}, {Lacey}, {Moffitt}, {Sharma}, {Lackey-Stewart}, {Kosakowski}, {Filipovi{\'c}}, \& {Payne}}]{Pannuti2015}
{Pannuti}, T.~G., {Swartz}, D.~A., {Laine}, S., {et~al.} 2015, \href{https://ui.adsabs.harvard.edu/abs/2015AJ....150...91P}{\aj}, \href{https://ui.adsabs.harvard.edu/abs/2015AJ....150...91P}{150, 91}

\bibitem[{{Pavlovi{\'c}} {et~al.}(2018){Pavlovi{\'c}}, {Uro{\v{s}}evi{\'c}}, {Arbutina}, {Orlando}, {Maxted}, \& {Filipovi{\'c}}}]{Pavlovic2018}
{Pavlovi{\'c}}, M.~Z., {Uro{\v{s}}evi{\'c}}, D., {Arbutina}, B., {et~al.} 2018, \href{https://ui.adsabs.harvard.edu/abs/2018ApJ...852...84P}{Astrophys. J.}, \href{https://ui.adsabs.harvard.edu/abs/2018ApJ...852...84P}{852, 84}

\bibitem[{{Payne} {et~al.}(2008){Payne}, {White}, \& {Filipovi{\'c}}}]{Payne2008}
{Payne}, J.~L., {White}, G.~L., and {Filipovi{\'c}}, M.~D. 2008, \href{https://ui.adsabs.harvard.edu/abs/2008MNRAS.383.1175P}{\mnras}, \href{https://ui.adsabs.harvard.edu/abs/2008MNRAS.383.1175P}{383, 1175}

\bibitem[{{Payne} {et~al.}(2007){Payne}, {White}, {Filipovi{\'c}}, \& {Pannuti}}]{Payne2007}
{Payne}, J.~L., {White}, G.~L., {Filipovi{\'c}}, M.~D., and {Pannuti}, T.~G. 2007, \href{https://ui.adsabs.harvard.edu/abs/2007MNRAS.376.1793P}{\mnras}, \href{https://ui.adsabs.harvard.edu/abs/2007MNRAS.376.1793P}{376, 1793}

\bibitem[{{Pennock} {et~al.}(2021){Pennock}, {van Loon}, {Filipovi{\'c}}, {Andernach}, {Haberl}, {Kothes}, {Lenc}, {Rudnick}, {White}, {Agliozzo}, {Ant{\'o}n}, {Boji{\v{c}}i{\'c}}, {Bomans}, {Collier}, {Crawford}, {Hopkins}, {Jeganathan}, {Kavanagh}, {Koribalski}, {Leahy}, {Maggi}, {Maitra}, {Marvil}, {Micha{\l}owski}, {Norris}, {Oliveira}, {Payne}, {Sano}, {Sasaki}, {Staveley-Smith}, \& {Vardoulaki}}]{Pennock2021}
{Pennock}, C.~M., {van Loon}, J.~T., {Filipovi{\'c}}, M.~D., {et~al.} 2021, \href{https://ui.adsabs.harvard.edu/abs/2021MNRAS.506.3540P}{\mnras}, \href{https://ui.adsabs.harvard.edu/abs/2021MNRAS.506.3540P}{506, 3540}

\bibitem[{{Petrosian} and {Liu}(2004){Petrosian} \& {Liu}}]{Petrosian2004}
{Petrosian}, V. and {Liu}, S. 2004, \href{https://ui.adsabs.harvard.edu/abs/2004ApJ...610..550P}{\apj}, \href{https://ui.adsabs.harvard.edu/abs/2004ApJ...610..550P}{610, 550}

\bibitem[{Pietrzy{\'n}ski {et~al.}(2019)Pietrzy{\'n}ski, Graczyk, Gallenne, Gieren, Thompson, Pilecki, Karczmarek, G{\'o}rski, Suchomska, Taormina, {et~al.}}]{Pietrzynski2019}
Pietrzy{\'n}ski, G., Graczyk, D., Gallenne, A., {et~al.} 2019, \href{https://ui.adsabs.harvard.edu/abs/2019Natur.567..200P}{Nature}, \href{https://ui.adsabs.harvard.edu/abs/2019Natur.567..200P}{567, 200}

\bibitem[{{Pilbratt} {et~al.}(2010){Pilbratt}, {Riedinger}, {Passvogel}, {Crone}, {Doyle}, {Gageur}, {Heras}, {Jewell}, {Metcalfe}, {Ott}, \& {Schmidt}}]{Pilbratt2010}
{Pilbratt}, G.~L., {Riedinger}, J.~R., {Passvogel}, T., {et~al.} 2010, \href{https://ui.adsabs.harvard.edu/abs/2010A&A...518L...1P}{\aap}, \href{https://ui.adsabs.harvard.edu/abs/2010A&A...518L...1P}{518, L1}

\bibitem[{{Pineault} {et~al.}(1997){Pineault}, {Landecker}, {Swerdlyk}, \& {Reich}}]{Pineault1997}
{Pineault}, S., {Landecker}, T.~L., {Swerdlyk}, C.~M., and {Reich}, W. 1997, \href{https://ui.adsabs.harvard.edu/abs/1997A&A...324.1152P}{\aap}, \href{https://ui.adsabs.harvard.edu/abs/1997A&A...324.1152P}{324, 1152}

\bibitem[{{Planck Collaboration} {et~al.}(2016){Planck Collaboration}, {Ade}, {Aghanim}, {Arnaud}, {Ashdown}, {Aumont}, {Baccigalupi}, {Banday}, {Barreiro}, {Bartolo}, {Battaner}, {Benabed}, {Beno{\^\i}t}, {Benoit-L{\'e}vy}, {Bernard}, {Bersanelli}, {Bielewicz}, {Bonaldi}, {Bonavera}, {Bond}, {Borrill}, {Bouchet}, {Boulanger}, {Bucher}, {Burigana}, {Butler}, {Calabrese}, {Catalano}, {Chamballu}, {Chiang}, {Christensen}, {Clements}, {Colombi}, {Colombo}, {Combet}, {Couchot}, {Coulais}, {Crill}, {Curto}, {Cuttaia}, {Danese}, {Davies}, {Davis}, {de Bernardis}, {de Rosa}, {de Zotti}, {Delabrouille}, {D{\'e}sert}, {Dickinson}, {Diego}, {Dole}, {Donzelli}, {Dor{\'e}}, {Douspis}, {Ducout}, {Dupac}, {Efstathiou}, {Elsner}, {En{\ss}lin}, {Eriksen}, {Falgarone}, {Fergusson}, {Finelli}, {Forni}, {Frailis}, {Fraisse}, {Franceschi}, {Frejsel}, {Galeotta}, {Galli}, {Ganga}, {Giard}, {Giraud-H{\'e}raud}, {Gjerl{\o}w}, {Gonz{\'a}lez-Nuevo}, {G{\'o}rski}, {Gratton}, {Gregorio}, {Gruppuso}, {Gudmundsson}, {Hansen}, {Hanson},
  {Harrison}, {Helou}, {Henrot-Versill{\'e}}, {Hern{\'a}ndez-Monteagudo}, {Herranz}, {Hildebrandt}, {Hivon}, {Hobson}, {Holmes}, {Hornstrup}, {Hovest}, {Huffenberger}, {Hurier}, {Jaffe}, {Jaffe}, {Jones}, {Juvela}, {Keih{\"a}nen}, {Keskitalo}, {Kisner}, {Knoche}, {Kunz}, {Kurki-Suonio}, {Lagache}, {Lamarre}, {Lasenby}, {Lattanzi}, {Lawrence}, {Leonardi}, {Lesgourgues}, {Levrier}, {Liguori}, {Lilje}, {Linden-V{\o}rnle}, {L{\'o}pez-Caniego}, {Lubin}, {Mac{\'\i}as-P{\'e}rez}, {Maggio}, {Maino}, {Mandolesi}, {Mangilli}, {Marshall}, {Martin}, {Mart{\'\i}nez-Gonz{\'a}lez}, {Masi}, {Matarrese}, {Mazzotta}, {McGehee}, {Melchiorri}, {Mendes}, {Mennella}, {Migliaccio}, {Mitra}, {Miville-Desch{\^e}nes}, {Moneti}, {Montier}, {Morgante}, {Mortlock}, {Moss}, {Munshi}, {Murphy}, {Naselsky}, {Nati}, {Natoli}, {Netterfield}, {N{\o}rgaard-Nielsen}, {Noviello}, {Novikov}, {Novikov}, {Oxborrow}, {Paci}, {Pagano}, {Pajot}, {Paladini}, {Paoletti}, {Pasian}, {Patanchon}, {Pearson}, {Pelkonen}, {Perdereau}, {Perotto}, {Perrotta},
  {Pettorino}, {Piacentini}, {Piat}, {Pierpaoli}, {Pietrobon}, {Plaszczynski}, {Pointecouteau}, {Polenta}, {Pratt}, {Pr{\'e}zeau}, {Prunet}, {Puget}, {Rachen}, {Reach}, {Rebolo}, {Reinecke}, {Remazeilles}, {Renault}, {Renzi}, {Ristorcelli}, {Rocha}, {Rosset}, {Rossetti}, {Roudier}, {Rubi{\~n}o-Mart{\'\i}n}, {Rusholme}, {Sandri}, {Santos}, {Savelainen}, {Savini}, {Scott}, {Seiffert}, {Shellard}, {Spencer}, {Stolyarov}, \& {Sudiwala}}]{Planck2016}
{Planck Collaboration}, {Ade}, P.~A.~R., {Aghanim}, N., {et~al.} 2016, \href{https://ui.adsabs.harvard.edu/abs/2016A&A...594A..28P}{\aap}, \href{https://ui.adsabs.harvard.edu/abs/2016A&A...594A..28P}{594, A28}

\bibitem[{{Ranasinghe} and {Leahy}(2023){Ranasinghe} \& {Leahy}}]{Ranasinghe2023}
{Ranasinghe}, S. and {Leahy}, D. 2023, \href{https://ui.adsabs.harvard.edu/abs/2023ApJS..265...53R}{\apjs}, \href{https://ui.adsabs.harvard.edu/abs/2023ApJS..265...53R}{265, 53}

\bibitem[{{Raymond}(1979)}]{Raymond1979}
{Raymond}, J.~C. 1979, \href{https://ui.adsabs.harvard.edu/abs/1979ApJS...39....1R}{\apjs}, \href{https://ui.adsabs.harvard.edu/abs/1979ApJS...39....1R}{39, 1}

\bibitem[{{Reynolds}(2008)}]{Reynolds2008}
{Reynolds}, S.~P. 2008, \href{https://ui.adsabs.harvard.edu/abs/2008ARA&A..46...89R}{\araa}, \href{https://ui.adsabs.harvard.edu/abs/2008ARA&A..46...89R}{46, 89}

\bibitem[{{Reynolds} and {Ellison}(1992){Reynolds} \& {Ellison}}]{Reynolds1992}
{Reynolds}, S.~P. and {Ellison}, D.~C. 1992, \href{https://ui.adsabs.harvard.edu/abs/1992ApJ...399L..75R}{\apjl}, \href{https://ui.adsabs.harvard.edu/abs/1992ApJ...399L..75R}{399, L75}

\bibitem[{{Reynolds} {et~al.}(2012){Reynolds}, {Gaensler}, \& {Bocchino}}]{Reynolds2012}
{Reynolds}, S.~P., {Gaensler}, B.~M., and {Bocchino}, F. 2012, \href{https://ui.adsabs.harvard.edu/abs/2012SSRv..166..231R}{\ssr}, \href{https://ui.adsabs.harvard.edu/abs/2012SSRv..166..231R}{166, 231}

\bibitem[{{Rho} and {Petre}(1998){Rho} \& {Petre}}]{Rho1998}
{Rho}, J. and {Petre}, R. 1998, \href{https://ui.adsabs.harvard.edu/abs/1998ApJ...503L.167R}{\apjl}, \href{https://ui.adsabs.harvard.edu/abs/1998ApJ...503L.167R}{503, L167}

\bibitem[{{Rolleston} {et~al.}(2002){Rolleston}, {Trundle}, \& {Dufton}}]{Rolleston2002}
{Rolleston}, W.~R.~J., {Trundle}, C., and {Dufton}, P.~L. 2002, \href{https://ui.adsabs.harvard.edu/abs/2002A&A...396...53R}{\aap}, \href{https://ui.adsabs.harvard.edu/abs/2002A&A...396...53R}{396, 53}

\bibitem[{{Salvesen} {et~al.}(2009){Salvesen}, {Raymond}, \& {Edgar}}]{Salvesen2009}
{Salvesen}, G., {Raymond}, J.~C., and {Edgar}, R.~J. 2009, \href{https://ui.adsabs.harvard.edu/abs/2009ApJ...702..327S}{\apj}, \href{https://ui.adsabs.harvard.edu/abs/2009ApJ...702..327S}{702, 327}

\bibitem[{{Sano} {et~al.}(2019){Sano}, {Matsumura}, {Nagaya}, {Yamane}, {Alsaberi}, {Filipovi{\'c}}, {Tachihara}, {Fujii}, {Tokuda}, {Tsuge}, {Yoshiike}, {Onishi}, {Kawamura}, {Minamidani}, {Mizuno}, {Yamamoto}, {Inutsuka}, {Inoue}, {Maxted}, {Rowell}, {Sasaki}, \& {Fukui}}]{Sano2019}
{Sano}, H., {Matsumura}, H., {Nagaya}, T., {et~al.} 2019, \href{https://ui.adsabs.harvard.edu/abs/2019ApJ...873...40S}{\apj}, \href{https://ui.adsabs.harvard.edu/abs/2019ApJ...873...40S}{873, 40}

\bibitem[{{Sano} {et~al.}(2023){Sano}, {Yamane}, {van Loon}, {Furuya}, {Fukui}, {Alsaberi}, {Bamba}, {Enokiya}, {Filipovi{\'c}}, {Indebetouw}, {Inoue}, {Kawamura}, {Laki{\'c}evi{\'c}}, {Law}, {Mizuno}, {Murase}, {Onishi}, {Park}, {Plucinsky}, {Rho}, {Richards}, {Rowell}, {Sasaki}, {Seok}, {Sharda}, {Staveley-Smith}, {Suzuki}, {Temim}, {Tokuda}, {Tsuge}, \& {Tachihara}}]{Sano2023}
{Sano}, H., {Yamane}, Y., {van Loon}, J.~T., {et~al.} 2023, \href{https://ui.adsabs.harvard.edu/abs/2023ApJ...958...53S}{\apj}, \href{https://ui.adsabs.harvard.edu/abs/2023ApJ...958...53S}{958, 53}

\bibitem[{{Slane} {et~al.}(2002){Slane}, {Smith}, {Hughes}, \& {Petre}}]{Slane2002}
{Slane}, P., {Smith}, R.~K., {Hughes}, J.~P., and {Petre}, R. 2002, \href{https://ui.adsabs.harvard.edu/abs/2002ApJ...564..284S}{\apj}, \href{https://ui.adsabs.harvard.edu/abs/2002ApJ...564..284S}{564, 284}

\bibitem[{{Smeaton} {et~al.}(2024){Smeaton}, {Filipovi{\'c}}, {Koribalski}, {Lazarevi{\'c}}, {Alsaberi}, {Becker}, {Dage}, {Gordon}, {Hopkins}, {Kothes}, {Leahy}, \& {Mitras̆inovi{\'c}}}]{Smeaton2024}
{Smeaton}, Z.~J., {Filipovi{\'c}}, M.~D., {Koribalski}, B.~S., {et~al.} 2024, \href{https://ui.adsabs.harvard.edu/abs/2024RNAAS...8..158S}{Research Notes of the American Astronomical Society}, \href{https://ui.adsabs.harvard.edu/abs/2024RNAAS...8..158S}{8, 158}

\bibitem[{{Smith} {et~al.}(2000){Smith}, {Leiton}, \& {Pizarro}}]{Smith2000}
{Smith}, C., {Leiton}, R., and {Pizarro}, S. 2000, in Astronomical Society of the Pacific Conference Series, Vol. 221, Stars, Gas and Dust in Galaxies: Exploring the Links, ed. D.~{Alloin}, K.~{Olsen}, and G.~{Galaz}, 83

\bibitem[{{Supan} {et~al.}(2022){Supan}, {Fischetto}, \& {Castelletti}}]{Supan2022}
{Supan}, L., {Fischetto}, G., and {Castelletti}, G. 2022, \href{https://ui.adsabs.harvard.edu/abs/2022A&A...664A..89S}{\aap}, \href{https://ui.adsabs.harvard.edu/abs/2022A&A...664A..89S}{664, A89}

\bibitem[{Theil(1950)}]{Theil1950}
Theil, H. 1950, Indagationes mathematicae, 12, 173

\bibitem[{{Thornton} {et~al.}(1998){Thornton}, {Gaudlitz}, {Janka}, \& {Steinmetz}}]{Thornton1998}
{Thornton}, K., {Gaudlitz}, M., {Janka}, H.~T., and {Steinmetz}, M. 1998, \href{https://ui.adsabs.harvard.edu/abs/1998ApJ...500...95T}{\apj}, \href{https://ui.adsabs.harvard.edu/abs/1998ApJ...500...95T}{500, 95}

\bibitem[{{Tian} and {Leahy}(2005){Tian} \& {Leahy}}]{Tian2005}
{Tian}, W.~W. and {Leahy}, D. 2005, \href{https://ui.adsabs.harvard.edu/abs/2005A&A...436..187T}{\aap}, \href{https://ui.adsabs.harvard.edu/abs/2005A&A...436..187T}{436, 187}

\bibitem[{{Tingay} {et~al.}(2013){Tingay}, {Goeke}, {Bowman}, {Emrich}, {Ord}, {Mitchell}, {Morales}, {Booler}, {Crosse}, {Wayth}, {Lonsdale}, {Tremblay}, {Pallot}, {Colegate}, {Wicenec}, {Kudryavtseva}, {Arcus}, {Barnes}, {Bernardi}, {Briggs}, {Burns}, {Bunton}, {Cappallo}, {Corey}, {Deshpande}, {Desouza}, {Gaensler}, {Greenhill}, {Hall}, {Hazelton}, {Herne}, {Hewitt}, {Johnston-Hollitt}, {Kaplan}, {Kasper}, {Kincaid}, {Koenig}, {Kratzenberg}, {Lynch}, {Mckinley}, {Mcwhirter}, {Morgan}, {Oberoi}, {Pathikulangara}, {Prabu}, {Remillard}, {Rogers}, {Roshi}, {Salah}, {Sault}, {Udaya-Shankar}, {Schlagenhaufer}, {Srivani}, {Stevens}, {Subrahmanyan}, {Waterson}, {Webster}, {Whitney}, {Williams}, {Williams}, \& {Wyithe}}]{Tingay2013}
{Tingay}, S.~J., {Goeke}, R., {Bowman}, J.~D., {et~al.} 2013, \href{https://ui.adsabs.harvard.edu/abs/2013PASA...30....7T}{\pasa}, \href{https://ui.adsabs.harvard.edu/abs/2013PASA...30....7T}{30, e007}

\bibitem[{{Tramacere} {et~al.}(2025){Tramacere}, {Campana}, {Massaro}, {Bocchino}, {Miceli}, \& {Orlando}}]{Tramacere2025}
{Tramacere}, A., {Campana}, R., {Massaro}, E., {et~al.} 2025, \href{https://ui.adsabs.harvard.edu/abs/2025A&A...697A.200T}{\aap}, \href{https://ui.adsabs.harvard.edu/abs/2025A&A...697A.200T}{697, A200}

\bibitem[{{Turtle} and {Amy}(1991){Turtle} \& {Amy}}]{Turtle1991}
{Turtle}, A.~J. and {Amy}, S.~W. 1991, in IAU Symposium, Vol. 148, The Magellanic Clouds, ed. R.~{Haynes} and D.~{Milne}, 114

\bibitem[{{Uchiyama} {et~al.}(2010){Uchiyama}, {Blandford}, {Funk}, {Tajima}, \& {Tanaka}}]{Uchiyama2010}
{Uchiyama}, Y., {Blandford}, R.~D., {Funk}, S., {Tajima}, H., and {Tanaka}, T. 2010, \href{https://ui.adsabs.harvard.edu/abs/2010ApJ...723L.122U}{\apjl}, \href{https://ui.adsabs.harvard.edu/abs/2010ApJ...723L.122U}{723, L122}

\bibitem[{{Uro{\v{s}}evi{\'c}}(2014)}]{Urosevic2014}
{Uro{\v{s}}evi{\'c}}, D. 2014, \href{https://ui.adsabs.harvard.edu/abs/2014Ap&SS.354..541U}{\apss}, \href{https://ui.adsabs.harvard.edu/abs/2014Ap&SS.354..541U}{354, 541}

\bibitem[{{Uro{\v{s}}evi{\'c}} and {Pannuti}(2005){Uro{\v{s}}evi{\'c}} \& {Pannuti}}]{Urosevic2005}
{Uro{\v{s}}evi{\'c}}, D. and {Pannuti}, T.~G. 2005, \href{https://ui.adsabs.harvard.edu/abs/2005APh....23..577U}{Astroparticle Physics}, \href{https://ui.adsabs.harvard.edu/abs/2005APh....23..577U}{23, 577}

\bibitem[{{Uro{\v{s}}evi{\'c}} {et~al.}(2007){Uro{\v{s}}evi{\'c}}, {Pannuti}, \& {Leahy}}]{Urosevic2007}
{Uro{\v{s}}evi{\'c}}, D., {Pannuti}, T.~G., and {Leahy}, D. 2007, \href{https://ui.adsabs.harvard.edu/abs/2007ApJ...655L..41U}{\apjl}, \href{https://ui.adsabs.harvard.edu/abs/2007ApJ...655L..41U}{655, L41}

\bibitem[{{Vink}(2012)}]{Vink2012}
{Vink}, J. 2012, \href{https://ui.adsabs.harvard.edu/abs/2012A&ARv..20...49V}{\aapr}, \href{https://ui.adsabs.harvard.edu/abs/2012A&ARv..20...49V}{20, 49}

\bibitem[{{Vink}(2020)}]{Vink2020}
{Vink}, J. 2020, {Physics and Evolution of Supernova Remnants}

\bibitem[{{Vink} {et~al.}(2006){Vink}, {Bleeker}, {van der Heyden}, {Bykov}, {Bamba}, \& {Yamazaki}}]{Vink2006}
{Vink}, J., {Bleeker}, J., {van der Heyden}, K., {et~al.} 2006, \href{https://ui.adsabs.harvard.edu/abs/2006ApJ...648L..33V}{\apjl}, \href{https://ui.adsabs.harvard.edu/abs/2006ApJ...648L..33V}{648, L33}

\bibitem[{Virtanen {et~al.}(2020)Virtanen, Gommers, Oliphant, Haberland, Reddy, Cournapeau, Burovski, Peterson, Weckesser, Bright, {van der Walt}, Brett, Wilson, Millman, Mayorov, Nelson, Jones, Kern, Larson, Carey, Polat, Feng, Moore, {VanderPlas}, Laxalde, Perktold, Cimrman, Henriksen, Quintero, Harris, Archibald, Ribeiro, Pedregosa, {van Mulbregt}, \& {SciPy 1.0 Contributors}}]{Virtanen2020}
Virtanen, P., Gommers, R., Oliphant, T.~E., {et~al.} 2020, \href{https://rdcu.be/b08Wh}{Nature Methods}, \href{https://rdcu.be/b08Wh}{17, 261}

\bibitem[{{Vukoti{\'c}} {et~al.}(2019){Vukoti{\'c}}, {{\'C}iprijanovi{\'c}}, {Vu{\v{c}}eti{\'c}}, {Oni{\'c}}, \& {Uro{\v{s}}evi{\'c}}}]{Vukotic2019}
{Vukoti{\'c}}, B., {{\'C}iprijanovi{\'c}}, A., {Vu{\v{c}}eti{\'c}}, M.~M., {Oni{\'c}}, D., and {Uro{\v{s}}evi{\'c}}, D. 2019, \href{https://ui.adsabs.harvard.edu/abs/2019SerAJ.199...23S}{Serbian Astronomical Journal}, \href{https://ui.adsabs.harvard.edu/abs/2019SerAJ.199...23S}{199, 23}

\bibitem[{{Vu{\v{c}}eti{\'c}} {et~al.}(2023){Vu{\v{c}}eti{\'c}}, {Milanovi{\'c}}, {Uro{\v{s}}evi{\'c}}, {Raymond}, {Oni{\'c}}, {Milo{\v{s}}evi{\'c}}, \& {Petrov}}]{Vucetic2023}
{Vu{\v{c}}eti{\'c}}, M., {Milanovi{\'c}}, N., {Uro{\v{s}}evi{\'c}}, D., {et~al.} 2023, \href{https://ui.adsabs.harvard.edu/abs/2023SerAJ.207....9V}{Serbian Astronomical Journal}, \href{https://ui.adsabs.harvard.edu/abs/2023SerAJ.207....9V}{207, 9}

\bibitem[{{Wayth} {et~al.}(2015){Wayth}, {Lenc}, {Bell}, {Callingham}, {Dwarakanath}, {Franzen}, {For}, {Gaensler}, {Hancock}, {Hindson}, {Hurley-Walker}, {Jackson}, {Johnston-Hollitt}, {Kapi{\'n}ska}, {McKinley}, {Morgan}, {Offringa}, {Procopio}, {Staveley-Smith}, {Wu}, {Zheng}, {Trott}, {Bernardi}, {Bowman}, {Briggs}, {Cappallo}, {Corey}, {Deshpande}, {Emrich}, {Goeke}, {Greenhill}, {Hazelton}, {Kaplan}, {Kasper}, {Kratzenberg}, {Lonsdale}, {Lynch}, {McWhirter}, {Mitchell}, {Morales}, {Morgan}, {Oberoi}, {Ord}, {Prabu}, {Rogers}, {Roshi}, {Shankar}, {Srivani}, {Subrahmanyan}, {Tingay}, {Waterson}, {Webster}, {Whitney}, {Williams}, \& {Williams}}]{Wayth2015}
{Wayth}, R.~B., {Lenc}, E., {Bell}, M.~E., {et~al.} 2015, \href{https://ui.adsabs.harvard.edu/abs/2015PASA...32...25W}{\pasa}, \href{https://ui.adsabs.harvard.edu/abs/2015PASA...32...25W}{32, e025}

\bibitem[{{White} and {Long}(1991){White} \& {Long}}]{White1991}
{White}, R.~L. and {Long}, K.~S. 1991, \href{https://ui.adsabs.harvard.edu/abs/1991ApJ...373..543W}{\apj}, \href{https://ui.adsabs.harvard.edu/abs/1991ApJ...373..543W}{373, 543}

\bibitem[{{Williams} {et~al.}(2004){Williams}, {Chu}, {Dickel}, {Gruendl}, {Shelton}, {Points}, \& {Smith}}]{Williams2004}
{Williams}, R.~M., {Chu}, Y.~H., {Dickel}, J.~R., {et~al.} 2004, \href{https://ui.adsabs.harvard.edu/abs/2004ApJ...613..948W}{\apj}, \href{https://ui.adsabs.harvard.edu/abs/2004ApJ...613..948W}{613, 948}

\bibitem[{{Wong} {et~al.}(2011){Wong}, {Hughes}, {Ott}, {Muller}, {Pineda}, {Bernard}, {Chu}, {Fukui}, {Gruendl}, {Henkel}, {Kawamura}, {Klein}, {Looney}, {Maddison}, {Mizuno}, {Paradis}, {Seale}, \& {Welty}}]{2011ApJS..197...16W}
{Wong}, T., {Hughes}, A., {Ott}, J., {et~al.} 2011, \href{https://ui.adsabs.harvard.edu/abs/2011ApJS..197...16W}{\apjs}, \href{https://ui.adsabs.harvard.edu/abs/2011ApJS..197...16W}{197, 16}

\bibitem[{{Wong} {et~al.}(2017){Wong}, {Hughes}, {Tokuda}, {Indebetouw}, {Bernard}, {Onishi}, {Wojciechowski}, {Bandurski}, {Kawamura}, {Roman-Duval}, {Cao}, {Chen}, {Chu}, {Cui}, {Fukui}, {Montier}, {Muller}, {Ott}, {Paradis}, {Pineda}, {Rosolowsky}, \& {Sewi{\l}o}}]{2017ApJ...850..139W}
{Wong}, T., {Hughes}, A., {Tokuda}, K., {et~al.} 2017, \href{https://ui.adsabs.harvard.edu/abs/2017ApJ...850..139W}{\apj}, \href{https://ui.adsabs.harvard.edu/abs/2017ApJ...850..139W}{850, 139}

\bibitem[{{Wright} {et~al.}(1994){Wright}, {Griffith}, {Burke}, \& {Ekers}}]{Wright1994}
{Wright}, A.~E., {Griffith}, M.~R., {Burke}, B.~F., and {Ekers}, R.~D. 1994, \href{https://ui.adsabs.harvard.edu/abs/1994ApJS...91..111W}{\apjs}, \href{https://ui.adsabs.harvard.edu/abs/1994ApJS...91..111W}{91, 111}

\bibitem[{{Xiao} {et~al.}(2008){Xiao}, {F{\"u}rst}, {Reich}, \& {Han}}]{Xiao2008}
{Xiao}, L., {F{\"u}rst}, E., {Reich}, W., and {Han}, J.~L. 2008, \href{https://ui.adsabs.harvard.edu/abs/2008A&A...482..783X}{\aap}, \href{https://ui.adsabs.harvard.edu/abs/2008A&A...482..783X}{482, 783}

\bibitem[{{Yamaguchi} {et~al.}(2016){Yamaguchi}, {Katsuda}, {Castro}, {Williams}, {Lopez}, {Slane}, {Smith}, \& {Petre}}]{Yamaguchi2016}
{Yamaguchi}, H., {Katsuda}, S., {Castro}, D., {et~al.} 2016, \href{https://ui.adsabs.harvard.edu/abs/2016ApJ...820L...3Y}{\apjl}, \href{https://ui.adsabs.harvard.edu/abs/2016ApJ...820L...3Y}{820, L3}

\bibitem[{{Yew} {et~al.}(2018){Yew}, {Filipovi{\'c}}, {Roper}, {Collier}, {Crawford}, {Jarrett}, {Tothill}, {O'Brien}, {Pavlovi{\'c}}, {Pannuti}, {Galvin}, {Kapi{\'n}ska}, {Cluver}, {Banfield}, {Schlegel}, {Maxted}, \& {Grieve}}]{Yew2018}
{Yew}, M., {Filipovi{\'c}}, M.~D., {Roper}, Q., {et~al.} 2018, \href{https://ui.adsabs.harvard.edu/abs/2018PASA...35...15Y}{\pasa}, \href{https://ui.adsabs.harvard.edu/abs/2018PASA...35...15Y}{35, e015}

\bibitem[{{Yew} {et~al.}(2021){Yew}, {Filipovi{\'c}}, {Stupar}, {Points}, {Sasaki}, {Maggi}, {Haberl}, {Kavanagh}, {Parker}, {Crawford}, {Vukoti{\'c}}, {Uro{\v{s}}evi{\'c}}, {Sano}, {Seitenzahl}, {Rowell}, {Leahy}, {Bozzetto}, {Maitra}, {Leverenz}, {Payne}, {Park}, {Alsaberi}, \& {Pannuti}}]{Yew2021}
{Yew}, M., {Filipovi{\'c}}, M.~D., {Stupar}, M., {et~al.} 2021, \href{https://ui.adsabs.harvard.edu/abs/2021MNRAS.500.2336Y}{\mnras}, \href{https://ui.adsabs.harvard.edu/abs/2021MNRAS.500.2336Y}{500, 2336}

\bibitem[{{Zangrandi} {et~al.}(2024){Zangrandi}, {Jurk}, {Sasaki}, {Knies}, {Filipovi{\'c}}, {Haberl}, {Kavanagh}, {Maitra}, {Maggi}, {Saeedi}, {Bernreuther}, {Koribalski}, {Points}, \& {Staveley-Smith}}]{Zangrandi2024}
{Zangrandi}, F., {Jurk}, K., {Sasaki}, M., {et~al.} 2024, \href{https://ui.adsabs.harvard.edu/abs/2024A&A...692A.237Z}{\aap}, \href{https://ui.adsabs.harvard.edu/abs/2024A&A...692A.237Z}{692, A237}

\bibitem[{{Zhu} {et~al.}(2014){Zhu}, {Tian}, \& {Zuo}}]{Zhu2014}
{Zhu}, H., {Tian}, W.~W., and {Zuo}, P. 2014, \href{https://ui.adsabs.harvard.edu/abs/2014ApJ...793...95Z}{\apj}, \href{https://ui.adsabs.harvard.edu/abs/2014ApJ...793...95Z}{793, 95}

\end{thebibliography}
}

\clearpage

{\ }

\clearpage

{\ }

\newpage

\begin{strip}

{\ }

\vskip-20mm

\naslov{Istra{\zz}iva{\nj}e {\dz}inovskog ostatka supernove u Velikom Magelanovom Oblaku: Veliki (J0450.4$-$7050)}

\authors{
Z. J. Smeaton$^1$,
{\rbf M. D. Filipovi\cc$^1$},
R. Z. E. Alsaberi$^{2,1}$,
{\rbf B. Arbutina$^{3}$},
W. D. Cotton$^{4, 5}$,}
\authors{
E. J. Crawford$^{1}$,
A. M. Hopkins$^6$, 
R. Kothes$^7$,
D. Leahy$^8$,
J. L. Payne$^1$,
N. Rajabpour$^1$,
}
\authors{
H. Sano$^{2, 9}$,
M. Sasaki$^{10}$, 
{\rbf D. Uro{\ss}evi\cc$^3$},
{\rbf i} J. Th. van Loon$^{11}$
}

\vskip3mm

\address{$^1$Western Sydney University, Locked Bag 1797, Penrith South DC, NSW 2751, Australia}
\address{$^2$Faculty of Engineering, Gifu University, 1-1 Yanagido, Gifu 501-1193, Japan}
\address{$^{3}${\rit Katedra za Astronomiju, Matematichki Fakultet, Univer{z}iteta u Beogradu, Studen{t}{s}ki trg 16, 11000 Beograd, Srbija}}
\address{$^4$National Radio Astronomy Observatory, 520 Edgemont Road, Charlottesville, VA 22903, USA}
\address{$^5$South African Radio Astronomy Observatory
Liesbeek House, River Park, Gloucester Road
Cape Town, 7700, South Africa}
\address{$^6$School of Mathematical and Physical Sciences, 12 Wally’s Walk, Macquarie University, NSW 2109, Australia}
\address{$^7$Dominion Radio Astrophysical Observatory, Herzberg Astronomy \& Astrophysics, National Research Council Canada, P.O. Box 248, Penticton}
\address{$^8$Department of Physics and Astronomy, University of Calgary, Calgary, Alberta, T2N IN4, Canada}
\address{$^9$Center for Space Research and Utilization Promotion (c-SRUP), Gifu University, 1-1 Yanagido, Gifu 501-1193, Japan}
\address{$^{10}$Dr Karl Remeis Observatory, Erlangen Centre for Astroparticle Physics, Friedrich-Alexander-Universit\"{a}t Erlangen-N\"{u}rnberg, Sternwartstra{\ss}e 7, 96049 Bamberg, Germany}
\address{$^{11}$Lennard-Jones Laboratories, Keele University, ST5 5BG, UK}

\Email{19594271@student.westernsydney.edu.au}

\vskip3mm

\centerline{{\rrm UDK} \udc}

\vskip1mm

\centerline{\rit Uredjivaqki prilog}

\vskip.7cm

\baselineskip=3.8truemm

\begin{multicols}{2}

{
\rrm

Predstavljamo pregled visoke rezolucije u radio-kontinuumu i multifrekvencijsku analizu ostatka supernove (OSN) J0450.4$-$7050 u Velikom Magelanovom oblaku (VMO) kome smo dali nadimak Veliki. 
Ova posmatranja visoke rezolucije otkrivaju da je ostatak ve{\cc}ih dimenzija u odnosu na prethodna merenja, {\ss}to qini J0450$-$7050 jednim od najve{\cc}ih poznatih OSN. Dodatno, detektovali smo povr{\ss}inski sjaj ve{\cc}i
od oqekivanog i neobiqno blag spektralni indeks  ($\alpha\,=\,-0.26\pm0.02$), sa vrlo malo varijacija du{\zz} ostatka. Posmatrana je sjajna ljuska u liniji $\mathrm{H}\alpha$ {\ss}to ukazuje na znaqajno \mbox{hladjenje}, ali i eksces emisije 
u liniji $\mathrm{[OIII]}$ na istoqnom delu udara, sugeri{\ss}u{\cc}i ve{\cc}u brzinu delova udarnog talasa. Razmotrili smo nekoliko teorijskih scenarija emisije i radio-evolucije J0450$-$7050 u tipiqnom okru{\zz}enju VMO i zakljuqili da se verovatno radi o starom OSN sa
ve{\cc}om kompresijom na udarnom talasu koja daje bla{\zz}i netermalni spektar, u kombinaciji sa termalnim (zakoqnim) \mbox{zraqenjem} koje ima odredjeni doprinos.

}

\end{multicols}

\end{strip}

\end{document}